\documentclass[epjH]{svjour}
\usepackage{graphicx,wrapfig,lipsum,amssymb}
\usepackage{color}
\usepackage{url}
\usepackage{booktabs}
\newcommand{\beq}{\begin{equation}}
\newcommand{\eeq}{\end{equation}}
\newcommand{\beqa}{\begin{eqnarray}}
\newcommand{\eeqa}{\end{eqnarray}}
\def\la{\lower.5ex\hbox{$\; \buildrel < \over \sim \;$}}
\def\ga{\lower.5ex\hbox{$\; \buildrel > \over \sim \;$}}

\begin{document}
\title{Robert Dicke and the naissance of experimental gravity physics,  1957-1967}
\author{P.~J.~E. Peebles\thanks{\email{pjep@princeton.edu}} }%
\institute{Joseph Henry Laboratories\\ Princeton University, Princeton NJ USA}
\abstract{
The experimental study of gravity became much more active in the late 1950s, a change pronounced enough be termed the birth, or naissance, of experimental gravity physics. I present a review of developments in this subject since 1915, through the broad range of new approaches that commenced in the late 1950s, and up to the transition of experimental gravity physics to what might be termed a normal and accepted part of physical science in the late 1960s. This review shows the importance of advances in technology, here as in all branches of natural science. The role of contingency is illustrated by Robert Dicke's decision in the mid-1950s to change directions in mid-career, to lead a research group dedicated to the experimental study of gravity. The review also shows the power of nonempirical evidence. Some in the 1950s felt that general relativity theory is so logically sound as to be scarcely worth the testing. But Dicke and others argued that a poorly tested theory is only that, and that other nonempirical arguments, based on Mach's Principle and Dirac's Large Numbers hypothesis, suggested it would be worth looking for a better theory of gravity. I conclude by offering lessons from this history, some peculiar to the study of gravity physics during the naissance, some of more general relevance. The central lesson, which is familiar but not always well advertised, is that physical theories can be empirically established, sometimes with surprising results. 
} 
\maketitle

\section{Introduction}\label{sec:intro}
This is an examination of how the experimental study of gravity grew in the late 1950s and through the 1960s. The subject was a small science then that can be examined in some detail in the space of this paper. It offers a particularly clear illustration of the importance of ideas as well as technology in the origins of lines of research, which in this case have grown into Big Science. 

In the mid-1950s the experimental exploration of gravity physics was generally considered uninteresting. This was in part because there seemed to be little that could be done, apart from incremental improvements of the three classical tests of general relativity. But a serious contributing factor was that influential scientists accepted general relativity theory as a compellingly logical extension from classical electromagnetism to the theory of gravity. (An excellent example is the presentation in Landau and Lifshitz 1951, in the first English translation of the 1948 edition of {\it The Theory of Fields}.) I take this broad  acceptance of Einstein's general relativity, at a time when it had little empirical basis, to be an argument from the ``nonempirical evidence'' that respected scientists found the theory to be a logically compelling extension of what had come before. The term is borrowed from Dawid (2015), who argued for the merits of ``non-empirical theory assessment'' in present-day string theory, where the prospects for empirical assessment seem to be even more remote than it seemed to be for gravity physics in the 1950s. In Dawid's (2016) words, 
\begin{quotation}
\noindent By the term Ônon-empirical confirmationÕ I denote confirmation of a
theory by observations that lie beyond the theory's intended domain:
unlike in the case of empirical confirmation, the confirming
observations are not predicted by the theory they confirm.
Non-empirical confirmation resembles empirical confirmation, however,
in being based on observations about the world beyond the theory and
its endorsers. Main examples of non-empirical confirmation are based
on the observations that scientists have not succeeded in finding
serious alternatives to a given theory, that in some sense comparable
theories in the research field have turned out predictively successful
once tested, and that a theory provides explanations that had not been
aimed at during the theory's construction.
\end{quotation}
My use of the term ``nonempirical evidence,'' in a second theme of this paper, is meant to be in line with this statement, but I have ventured to include considerations of the vague but commonly applied criterion of elegance, or simplicity. I take as prototype for this criterion Einstein's (1945, p. 127) comment about the cosmological constant, $\rm\Lambda$, in the appendix for the second edition of {\it The Meaning of Relativity}:
\begin{quotation}
\noindent The introduction of the ``cosmologic member'' into the equations of gravity, though possible from the point of view of relativity, is to be rejected from the point of view of logical economy.
\end{quotation}
The term $\rm\Lambda$ in Einstein's field equation remained unpopular among influential scientists in the 1950s, and increasingly so through the 1990s (as reviewed in Sec.~\ref{sec:fractal}). But despite its inelegance $\rm\Lambda$ was eventually added to the standard and accepted theory under the pressure of experimental advances. We may of course counter this example of the force of empirical evidence with a prime illustration of  the successful application of nonempirical evidence: General relativity theory now passes demanding empirical tests. 

The change in thinking about the possibilities of probes of gravity physics that led to the experimental establishment of general relativity is part of what Will (1986) termed the renaissance of general relativity, and Blum, Lalli, and Renn (2015) termed its reinvention. Both names are appropriate for the subject as a whole, but on the empirical side the connotation of revival is inappropriate, because not a lot had happened earlier. There  was a paradigm shift in community opinion, in the sense of Kuhn's (1962) {\it Structure of Scientific Revolutions}: In the 1950s it was generally accepted that there is little of interest to do in experimental gravity physics, as one sees in the heavy ratio of theory to experiment in the international conferences in the 1950s reviewed in Section~\ref{sec:GRinthe50s}. In the 1960s new directions in experimental programs were becoming a familiar and accepted part of science. But since there was not a shift in standard and accepted fundamental physics I avoid the word ``revolution'' and use instead the term, ``the naissance of experimental gravity.'' 

I take this naissance to have started in 1957, at the Chapel Hill Conference on {\it The Role of Gravitation in Physics} (DeWitt 1957), where Dicke (1957a) emphasized the sparse experimental exploration of gravity physics, the promise of new technology that could help improve the situation, and experiments to this end in progress in his group. Others may have been thinking along similar lines, but Dicke was alone in making a clear statement of the situation in print and starting a long-term broad-based program of empirical  investigations of gravity. I take the naissance to have lasted about a decade before maturing into a part of normal science. 

Ideas are important to the general advance of science, but they may play a particularly big role in the origins of a research activity. A clear illustration is the ideas Dicke and others found for new lines of research from old arguments associated with Ernst Mach and Paul A. M. Dirac. This situation was important enough to the development of modern gravity physics to be reviewed in some detail, in Section~\ref{sec:Mach} of this paper. 

Section~\ref{sec:GRinthe50s} reviews the state of thinking about general relativity as a physical science at the start of the naissance, as revealed by the proceedings of international conferences in the late 1950s and early 1960s. Two of the leading actors in this search for ``The Role of Gravitation in Physics"  were John Archibald Wheeler and Robert Henry Dicke; their thinking is discussed in Section~\ref{Sec:WheelerDicke}. Section~\ref{sec:Mach} reviews ideas that motivated Dicke and others, and Section~\ref{sec:GravityGroup} presents a consideration of the style of exploration of these and other ideas in what became known as the Gravity Research Group. Activities in the group are  exemplified by accounts of the two experiments in progress that Dicke mentioned at the Chapel Hill Conference. Section~\ref{sec:tests} presents more brief reviews of other significant advances in experimental gravity physics, from 1915 through to the nominal end of the naissance. The great lines of research that have grown out of this early work have shown us that the theory Einstein completed a century ago fits an abundance of experimental and observational evidence on scales ranging from the laboratory to the Solar system and on out to the observable universe. This is a striking success for assessment from nonempirical evidence, but I hope not to be considered an anticlimax. As noted, we have been forced to add Einstein's (1917) cosmological constant, despite its general lack of appeal. Empirical evidence can be surprising. Lessons to be drawn from this and other aspects of the history are discussed in Section~\ref{sec:lessons}.\footnote{This paper was written in the culture of physics, and certainly could be made more complete by broader considerations, perhaps examined in the culture of the histroy of science. For example, in Section~\ref{sec:militaryfunding} I discuss  support for research during the naissance by agencies that are more normally associated with the military. My  thoughts in this section on why the agencies were doing this are only schematic; much more is known, as discussed for example in Wilson and Kaiser (2014), and I expect they and other historians are better equipped to follow this interesting story.  Industries also supported postwar research in gravity physics. Howard Forward is listed as a Hughes Aircraft Company Staff Doctoral Fellow in the paper by Forward,  Zipoy, Weber, et al. (1961) on modes of oscillation of the Earth, which figured in the search for gravitational waves (Sec.~\ref{sec:gravitationalwaves}). George Gamow's tour as a consultant to General Dynamics in the 1950s is celebrated for the missed chance for Gamow and Hoyle to hit on the thermal sea of radiation left from the hot early universe (as recalled by Hoyle 1981). I do not know why there were such connections between industry and gravity physics; as graduate students we used to joke that the aircraft companies were hoping to find antigravity. The Hughes Fellowships still exist, but I understand now focus on more practical training. During the naissance, nontenured faculty in physics at Princeton University typically had appointments halftime teaching and halftime research, the latter supported by a funding agency, often military. It allowed many more junior faculty than positions  available for tenure. The junior faculty were supposed to benefit from experience and contacts that could lead to jobs elsewhere, usually successfully. And research greatly benefitted from many active young people. This comfortable and productive arrangement at elite universities ended during the Vietnam War,  possibly a casualty of resentment of the particularly loud protests of the draft at elite universities that perhaps were least seriously afflicted by the draft. Perhaps there was something to our joke that Senator Mansfield  sought to prevent the military from corrupting our young minds. But I am not capable of judging the truth of this matter, or the actual effect on curiosity-driven research. In Section~\ref{Sec:WheelerDicke}, I mention Wheeler and Dicke's peaceful coexistence with their quite different philosophies of research in gravitation. Historians might see room for closer examination of this situation and, I expect, many other aspects of how empirical gravity physics grew.}

\section{General relativity and experimental gravity physics in the 1950s} \label{sec:GRinthe50s}
The modest state of experimental research in gravity physics in the mid-1950s is illustrated by the proceedings of the July 1955 Berne Conference, {\it Jubilee of Relativity Theory}  (Mercier and Kervaire 1956), on the occasion of the $50^{\rm th}$ anniversary of special relativity theory, the $40^{\rm th}$ for general relativity. Of the 34 papers in the proceedings there is just one on the fundamental empirical basis for general relativity: Trumpler's (1956) review of the modest advances in two of the original tests of general relativity, measurements of the deflection of light by the Sun and of the gravitational redshift of light from stars. Baade presented an important development, his correction to the extragalactic distance scale. He did not contribute to the proceedings, but the record of discussions of his report includes Robertson's (1956) comment on the possible need for ``the disreputable $\rm\Lambda$.'' Einstein's $\rm\Lambda$ became part of established gravity physics, but this happened a half-century later.

The experimental situation was improving, though slowly at first. At the March 1957 Chapel Hill {\it Conference on The Role of Gravitation in Physics} (DeWitt 1957; DeWitt and Rickles 2011), Dicke (1957a) stressed the contrast between the scant experimental work in gravity physics and the dense tests and experimental applications of quantum physics, and he outlined a research program he had commenced a few years earlier aimed at improving the situation. The only other commentary in the proceedings on empirical advances was Lilley's (1957) review of radio astronomy. Of immediate interest was the possibility of distinguishing between the Steady State and relativistic cosmological models by counts of radio sources as a function of flux density.  The counts proved to be faulty for this purpose, but they have proved to be of lasting interest for their near isotropy, an early hint to the large-scale homogeneity of the observable universe (Sec.~\ref{Sec:cosmology}).

At the June 1959 Royaumont Conference on {\it Les Th\'eories Relativistes de la Gravitation} (Lichnerowicz and Tonnelat 1962), Weber (1962) presented an analysis of how to build a gravitational wave detector. This was the only experimental paper among the 46 in the proceedings, and one may wonder how edifying the more technical aspects were to an audience that likely was almost entirely theorists. But Weber was introducing a new direction in the experimental investigation of general relativity theory. 

We see a strikingly abrupt change of emphasis to the search for empirical probes of gravity in the July 1961 NASA  {\it Conference on Experimental Tests of Theories of Relativity} (Roman 1961). Discussions of projects to which NASA could contribute included measurements of relativistic timekeeping in artificial satellites; tracking of artificial satellite orbits, including a satellite that shields a test mass from atmospheric drag and light pressure by jets that keep the enclosed test mass centered; the design conditions for a test of the relativistic Lense-Thirring inertial frame-dragging effect; and a search for detection of gravitational waves on a quieter site, the Moon. 

The general lack of interest in experimental general relativity and gravity physics in the 1950s was at least in part a result of competition from many interesting things to do in other branches of physics, from elementary particle physics to biophysics. But ongoing advances of technology were offering new possibilities for better probes of relativity. In some cases a particular technical advance motivated a specific experiment. For example, by 1955 Townes's group had a working ammonia beam maser (Gordon, Zeiger, and Townes 1955); M\o ller (1957) acknowledged ``stimulating discussions'' with Townes  ``on problems of general relativity in connection with the maser;'' and a year after that  Townes's group published a variant of the Kennedy-Thorndike aether drift experiment (Cedarholm, Bland, Havens, and Townes 1958; as discussed below in Sec.~\ref{sec:preferredmotion}). And a year after publication of the M{\"o}ssbauer (1958) effect, Pound and Rebka (1959) announced their plan to use it for an attempt at a  laboratory detection of the gravitational redshift, and Pound and Rebka (1960) announced the detection a year after that (as reviewed in Sec.~\ref{sec:gravredshift}). Dicke followed this pattern, but with a particular difference: he was systematically casting about for experiments that may help improve the empirical basis for gravity physics. 

\begin{figure}[t]
\begin{center}
\includegraphics[angle=0,width=5.in]{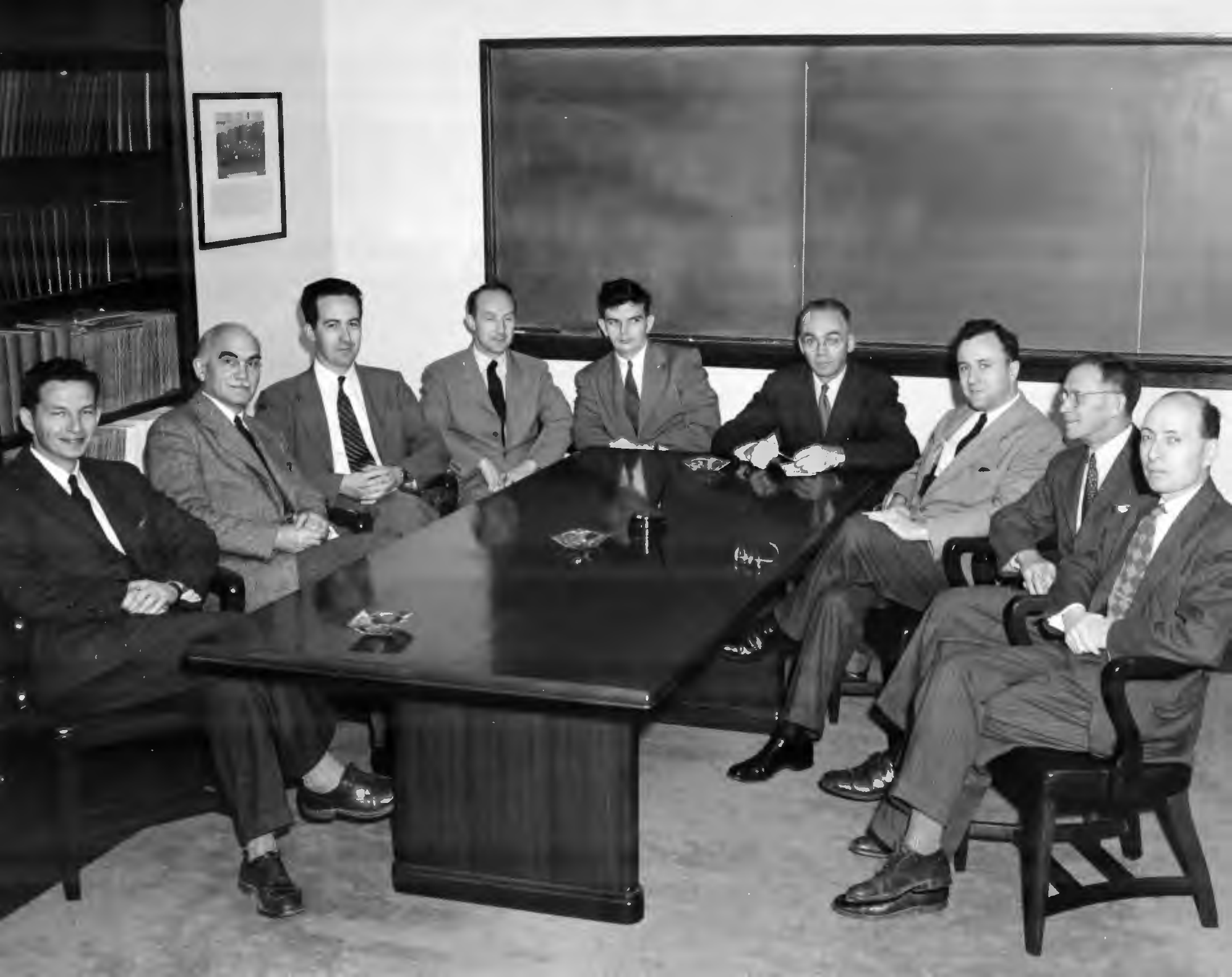} 
\caption{\label{Fig:faculty}  Members of the senior faculty in the Department of Physics, Palmer Physical Laboratory, Princeton University, in about 1950: from the left Rubby Sherr, Allen Shenstone, Donald Hamilton,  Eric Rogers, Robert Dicke, Walker Bleakney, John Wheeler,  Rudolf Ladenburg, and Eugene Wigner. }
\end{center}
\end{figure}
\section{Wheeler and Dicke on the role of gravitation in physics}\label{Sec:WheelerDicke}
At the 1957 Chapel Hill ``Conference on The Role of Gravitation in Physics'' John Archibald Wheeler
spoke on the need to better understand the physical meaning of general relativity, which he accepted as the unquantized theory of gravity. Robert Henry Dicke spoke on the need to better establish the experimental basis for gravity physics, and perhaps find a unquantized theory that is even better than general relativity. The two are shown in Figure~\ref{Fig:faculty} with other members of the senior faculty of the Department of Physics at Princeton University in about 1950. The photograph was taken a few years before both decided to turn to research in relativity and gravitation, when they were in mid-career: Wheeler was 44 and Dicke 39 in 1955. 

Wheeler's research interests had been in theoretical nuclear, particle, and atomic physics. In his autobiography (Wheeler and Ford 1998, p. 228) Wheeler recalled that his notes from 1952 revealed that he had learned ``from Shenstone 1/2 hour ago that I can teach relativity next year'' which was ``my first step into territory that would grip my imagination and command my research attention for the rest of my life.'' (Allen Shenstone, then chair of physics, is second from the left in Figure~\ref{Fig:faculty}.) By the end of the 1950s Wheeler was leading an active research group in theoretical general relativity, with participation by students, postdocs, faculty, and a steady stream of visitors.

Dicke recalled (in Lightman and Brawer 1990, p. 204) that during sabbatical leave at Harvard in 1954-55 he was thinking about the E\"otv\"os experiment, which tests whether the acceleration of gravity may depend on the nature of the test particle. The constraint was remarkably tight, but Dicke saw that it could be done even better with the much better technology he could use. Bill Hoffmann (2016), who was a graduate student then, recalled that Dicke returned from Harvard ``all fired up about gravity experiments.'' The quite abrupt change in direction of his active research career to gravity from what might be summarily termed quantum optics is illustrated by the list in Appendix A of the research topics of his graduate students before and after Dicke turned to gravity physics.\footnote{Dicke's early life, the story of how he joined the faculty at Princeton University, and what he did when he got there, are reviewed in Happer, Peebles, and Wilkinson 1999.}

\begin{figure}
\begin{center} 
\includegraphics[angle=270,width=2.5in]{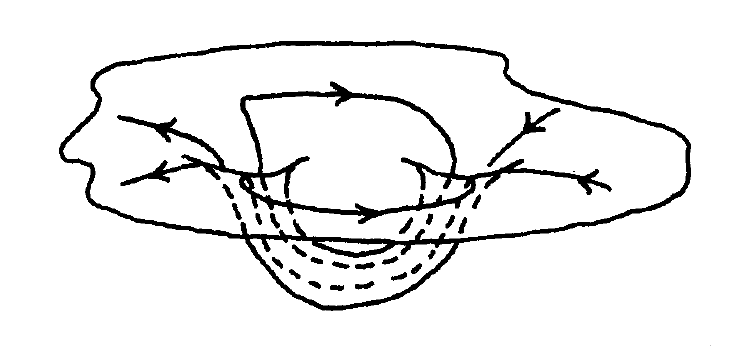} 
\caption{\label{fig:wormhole}  Wheeler's sketch of a wormhole that is charged without charge, and Dicke's table of orders of magnitude of physical parameters,  from the 1957 Chapel Hill Conference.}
\end{center}
\end{figure}
\begin{figure}[h]
\begin{center} 
\includegraphics[angle=270,width=4.in]{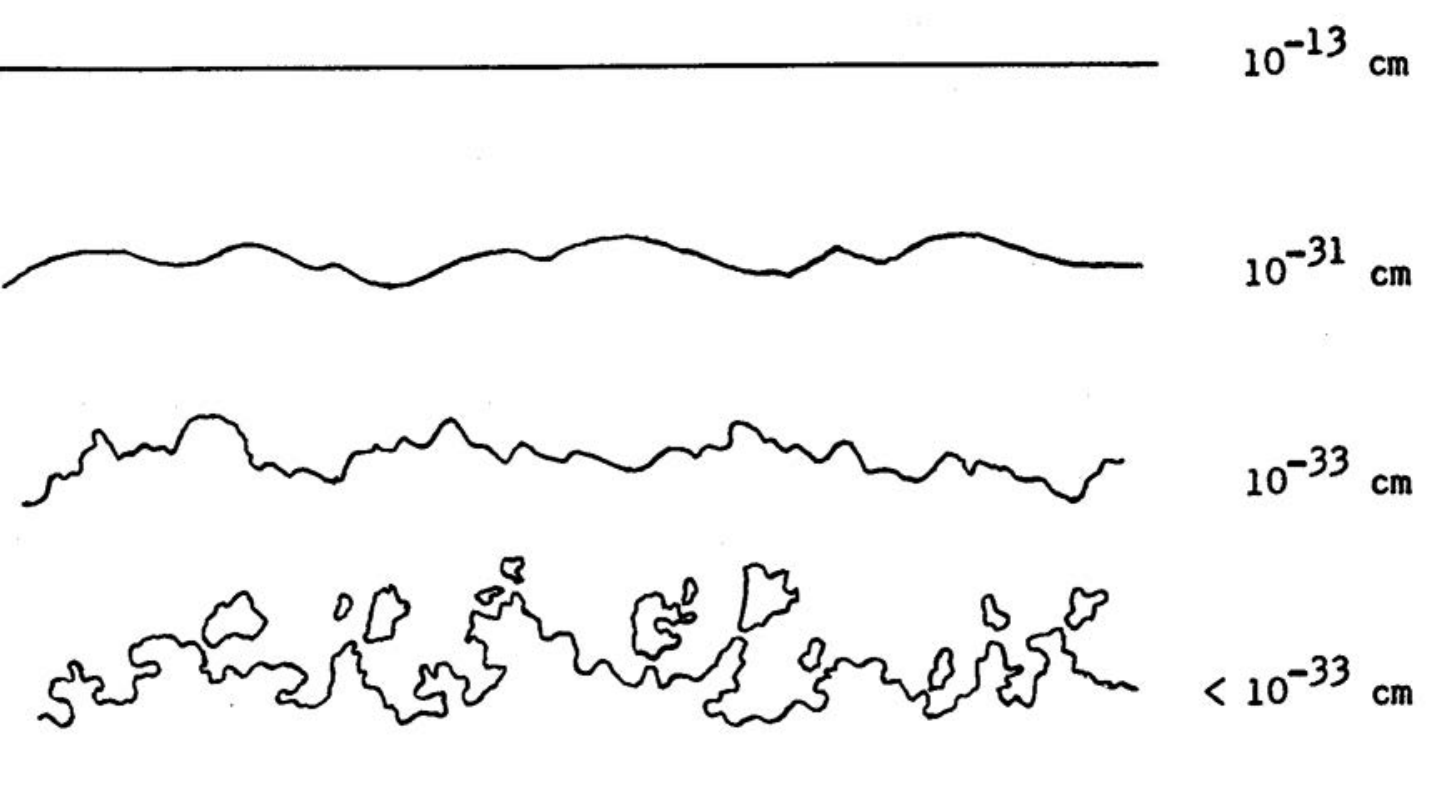} 
\caption{\label{fig:spacetimefoam}  Wheeler's sketch of spacetime foam,  from discussion at the 1957 Chapel Hill Conference.}
\end{center}
\end{figure}

We have samples of what Wheeler and Dicke were thinking as they turned to investigations of relativity and gravity physics from the proceedings of the Chapel Hill Conference. The title of Wheeler's (1957) paper, {\it The Present Position of Classical Relativity Theory and Some of its Problems}, reflects the separation of topics at the conference to ``unquantized general relativity" and ``quantized general relativity.'' The latter was as fascinating, challenging and widely debated then as it is now. Figure~\ref{fig:wormhole} from Wheeler's paper shows an example of his thinking in classical relativity: a wormhole in space-time threaded by electric flux, giving us ``charge without charge." He also spoke of ``mass without mass," in his concept of a geon produced by the nonlinear interaction of electromagnetic and spacetime curvature fields. Wheeler's thinking about quantized general relativity is illustrated by a report of his contribution to a discussion at the Chapel Hill Conference:
\begin{quotation}
\noindent WHEELER envisions ``foam-like structure'' for the vacuum, arising from these fluctuations of
the metric. He compared our observation of the vacuum with the view of
an aviator flying over the ocean. At high altitudes the ocean looks smooth,
but begins to show roughness as the aviator descends. In the ease of the
vacuum, WHEELER believes that if we look at it on a sufficiently small
scale it may even change its topological connectedness, thus [in the illustration in Fig.~\ref{fig:spacetimefoam}]:
\end{quotation}

Wheeler's imaginative approach to physics, and his modes of instruction of graduate students, produced great science and great generations of scientists. His inspiring style is seen in his opening paragraph in the Chapel Hill proceedings:
\begin{quotation}
\noindent We are here to consider an extraordinary topic, one that ranges from
the infinitely large to the infinitely small. We want to find what general
relativity and gravitation physics have to do with the description of nature.
This task imposes a heavy burden of judgement and courage on us, for
never before has theoretical physics had to face such wide subject matter,
assisted by so comprehensive a theory but so little tested by experiment.
\end{quotation}

Wheeler's part in the renaissance of the theoretical side of general relativity deserves a study that would expand on his recollections in Wheeler and Ford (1998), but only a few brief points may be noted here. The presence of singularly active research groups in gravity theory and experiment in the same physics department, commencing in the 1950s, cannot have been planned: Wheeler and Dicke turned to relativity and gravity from other research directions well after arriving at Princeton. The presence of Einstein  at the Princeton Institute for Advanced Study, and the relativist Howard Percy Robertson, who was based at Princeton University from 1929 to 1947, may have set a tradition that had some influence. Perhaps that is exemplified by Wheeler's feeling, which I heard expressed on occasion, that a philosophically satisfying universe has closed space sections, as Einstein had argued in the 1920s (and is reviewed in Sec.~\ref{sec:Mach}). Wheeler took an active interest in what was happening in Dicke's group. But Wheeler spent far more time studying the physical significance of general relativity theory as Einstein had written it down, and the  possible approaches to its quantization.

Dicke's thinking about his change of direction of research is illustrated by these quotes from his 1957 Chapel Hill paper,  {\it The Experimental Basis of Einstein's Theory} (Dicke 1957a, p. 5):
\begin{quotation}
\noindent It is unfortunate to note that the situation with respect to the experimental
checks of general relativity theory is not much better than it was a few years
after the theory was discovered --- say in 1920. This is in striking contrast
to the situation with respect to quantum theory, where we have literally
thousands of experimental checks.

\ldots

\noindent Professor Wheeler has already discussed the three famous checks of
general relativity; this is really very flimsy evidence on which to hang a
theory.

\ldots

\noindent It is a great challenge to the experimental physicist to
try to improve this situation; to try to devise new experiments and refine
old ones to give new checks on the theory. We have been accustomed to
thinking that gravity can play no role in laboratory-scale experiments; that
the gradients are too small, and that all gravitational effects are equivalent
to a change of frame of reference. Recently I have been changing my views
about this.
\end{quotation}   
In the second of these quotes Dicke was referring to Wheeler's summary comments on the classical three tests of general relativity: the orbit of the planet Mercury, the gravitational deflection of light passing near the Sun, and the gravitational redshift of light from stars. As it happens, the redshift test Wheeler mentioned, the measured shifts in the spectra of two white dwarf stars in binary systems, proved to be accurate in one case but quite erroneous in the other (as discussed in Sec.~\ref{sec:gravredshift}). Wheeler and Dicke might instead have referred to the measured wavelength shifts of solar absorption lines (St.~John 1928), but this test was problematic because the measured shifts varied across the face of the Sun and varied with the depth of origin of the lines in the solar atmosphere, largely results of turbulence. The measured solar redshifts were roughly in line with general relativity, however, tending to  scatter around the predicted value by no more than about 25\%, so one might say that in 1957 general relativity had passed about two and a half tests. 

At the Chapel Hill Conference Dicke mentioned work in progress in his group, largely in the discussion, beginning with this exchange:
\begin{quotation}
\noindent BERGMANN: What is the status of the experiments which it is rumored
are being done at Princeton?\medskip

\noindent DICKE: There are two experiments being started now. One is an improved
measurement of ``g''  to detect possible annual variations. This is coming
nicely, and I think we can improve earlier work by a factor of ten. This is
done by using a very short pendulum, without knife edges, just suspended
by a quartz fiber, oscillating at a high rate of around 30 cycles/sec. instead
of the long slow pendulum. The other experiment is a repetition of the
E\"otv\"os experiment. We put the whole system in a vacuum to get rid
of Brownian motion disturbances; we use better geometry than E\"otv\"os
used; and instead of looking for deflections, the apparatus would be in an
automatic feed-back loop such that the position is held fixed by feeding in
external torque to balance the gravitational torque. This leads to rapid
damping, and allows you to divide time up so that you don't need to
average over long time intervals, but can look at each separate interval of
time. This is being instrumented; we are worrying about such questions
as temperature control of the room right now, because we'd like stability
of the temperature to a thousandth of a degree, which is a bit difficult for
the whole room.

\ldots\smallskip

\noindent We have been working on an atomic clock, with which we will
be able to measure variations in the moon's rotation rate. Astronomical
observations are accurate enough so that, with a good atomic clock, it
should be possible in three years' time to detect variations in ``g'' of the
size of the effects we have been considering. We are working on a rubidium
clock, which we hope may be good to one part in~$10^{10}$.
\end{quotation}

In these comments Dicke mentioned three experimental projects. Tracking the motion of the Moon is reviewed in Section~\ref{sec:LLR}. The pendulum and E\"otv\"os experiments are discussed in Section~\ref{sec:GravityGroup}, for the purpose of illustrating how Dicke's group operated. 

\begin{figure}[h]
\begin{center} 
\includegraphics[angle=0,width=4.5in]{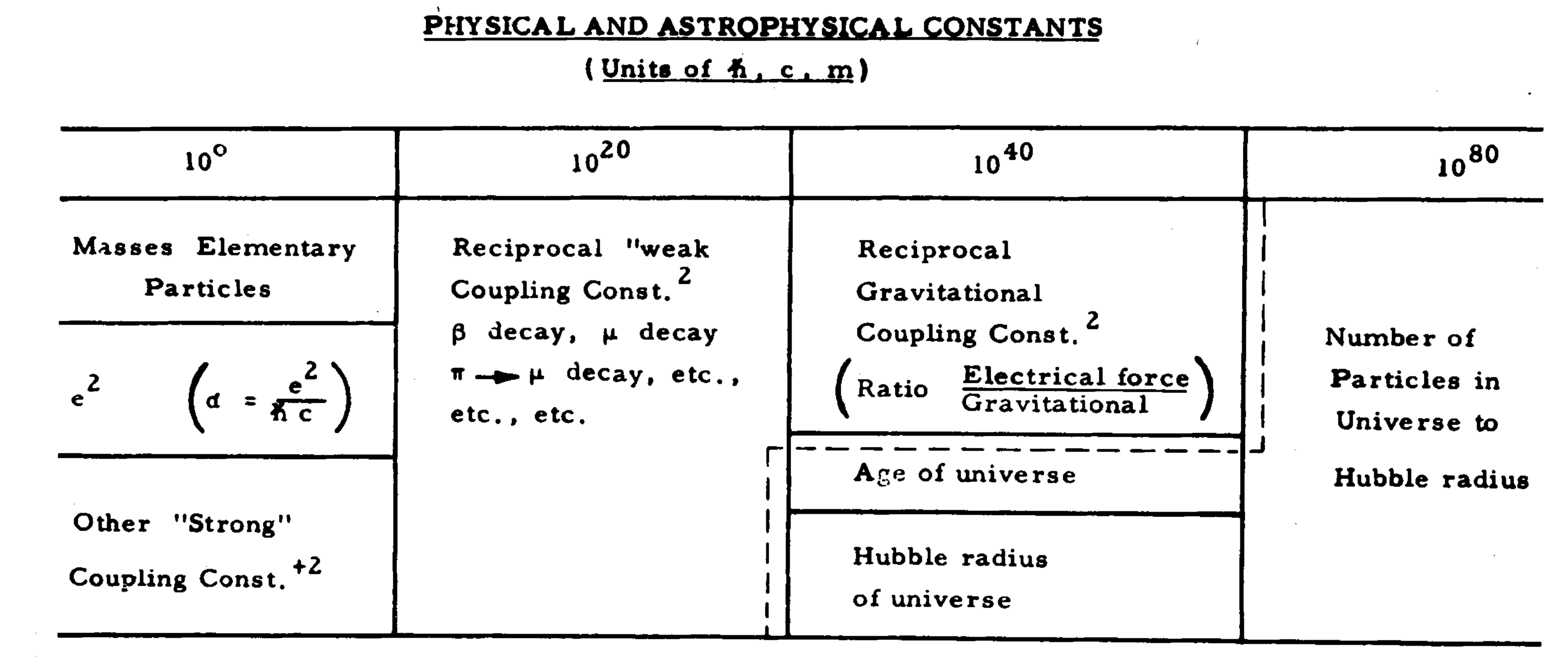} 
\caption{\label{fig:Dicketable} Dicke's table of orders of magnitude of physical parameters, from his paper at the 1957 Chapel Hill Conference.}
\end{center}
\end{figure}
In the third of the above quotes Dicke mentioned that ``I have been changing my views.'' Though not stated at this point in the discussion, it seems likely that his new views included the comments earlier in the paper about the table in Figure~\ref{fig:Dicketable} (copied from his Chapel Hill paper). The table was meant to motivate the idea that the strengths of the gravitational and weak interactions may evolve as the universe expands. This idea, and a more detailed discussion of the table, was presented in Dicke (1957b). We may suppose that Dicke's change of views also included his interest in Mach's Principle, for although he did not mention Mach in the Chapel Hill proceedings he advanced arguments along Machian ideas  in Dicke (1957c). The next section reviews these ideas, their influence on Dicke and others, and their role in shaping the progress of experimental gravity physics. 

\begin{figure}
\begin{center}
\includegraphics[angle=270,width=4.in]{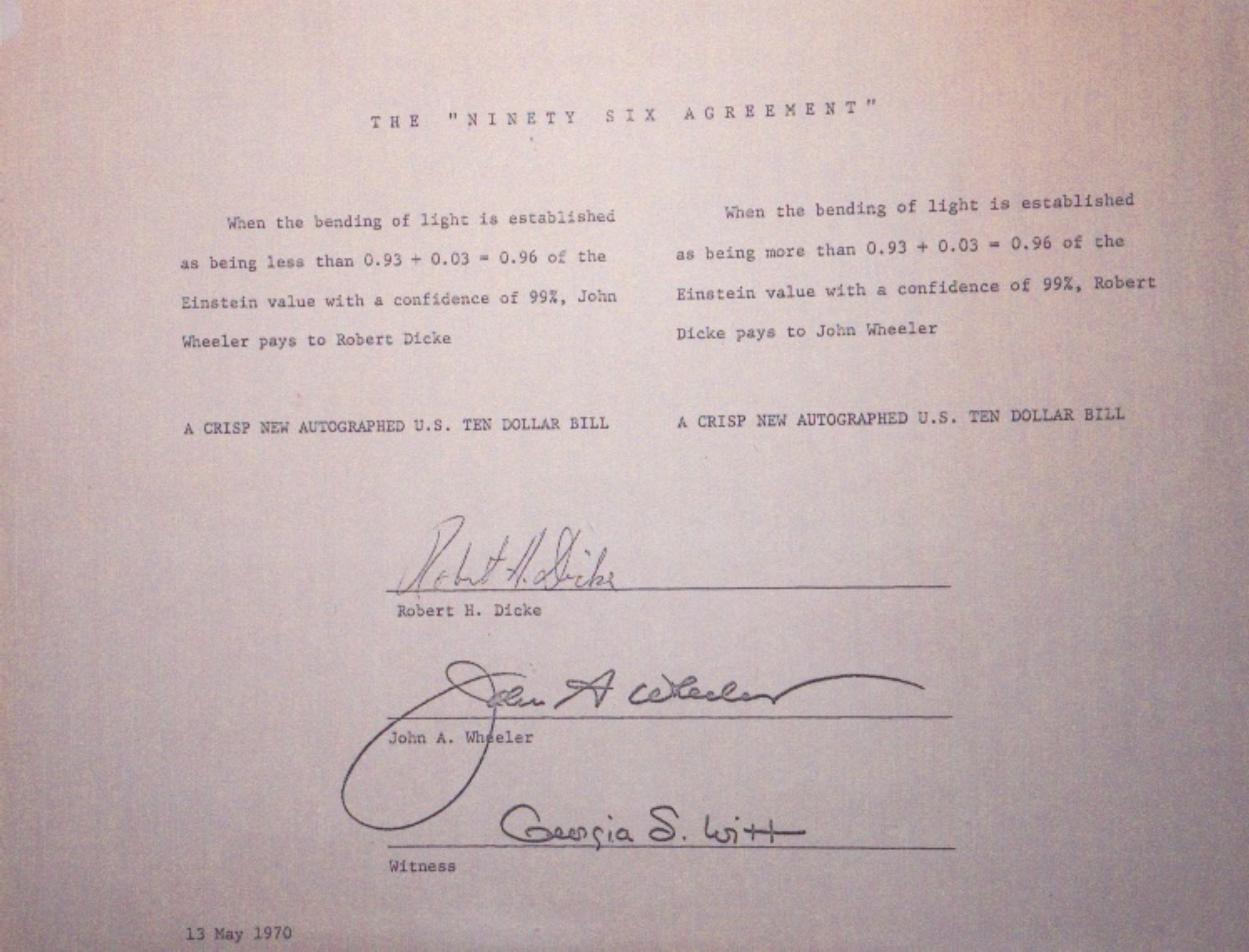} 
\caption{\label{Fig:WDbet}  A wager on the tests of gravity theories.}
\end{center}
\end{figure}
While Dicke was actively casting about for evidence that might lead to a better gravity theory than general relativity, Wheeler was not at all inclined to question the validity of Einstein's relativity.  Their differences are illustrated by the 1970 document\footnote{The witness, Georgia Witt, was Wheeler's secretary. The document is discolored, as if it had been tacked to a wall for a long time. But I was only recently made aware of it by Martin McHugh, who found it in the Robert Henry Dicke Papers, Box 15, Folder W, Department of Rare Books and Special Collections, Princeton University Library.} shown in Figure~\ref{Fig:WDbet}. At the time Dicke felt reasonably sure that the scalar-tensor gravity theory discussed in Section~\ref{sec:STgravity} fits the evidence better than general relativity. The details are reviewed in footnote \ref{footnote:wager} in Section~\ref{sec:testingBD}; here we need only note a few points. The scalar-tensor theory predicts that the precession of the orbit of Mercury is smaller than in general relativity.  However, Dicke (1964) had pointed out that if the interior of the Sun were rotating about as rapidly as the gas giant planets Jupiter and Saturn, the rotation of the solar surface being slowed by drag by the wind of plasma blowing away from the Sun, then the oblate solar mass distribution would contribute to the Newtonian precession, leaving a smaller non-Newtonian residual, consistent with the scalar-tensor theory. The Dicke and Goldenberg (1967) measurement of the shape of the Sun agreed with this idea. This interpretation required that the gravitational deflection of light by the Sun is no more than about 0.93 times the general relativity prediction. The added 0.03 in the document is not explained; it may be an indication of caution on Dicke's part. But my impression was that Dicke was confident of the argument because he had a new measurement---the shape of the Sun---that seemed to agree with two old considerations---Mach's principle and Dirac's large numbers in the context of the scalar-tensor gravity theory. Wheeler rested his case on the elegant simplicity of Einstein's general relativity theory, and he certainly gave the impression of confidence that the theory would continue to pass the improving tests. The network of tests is much tighter now, and it continues to agree with Wheeler's confidence in Einstein's theory. It might be noted, however, the evidence does require close to flat space sections, contrary to Einstein's and Wheeler's preference for a closed universe. 

In my recollection Wheeler and Dicke were comfortable with their differences, but I do not know of any examples of joint research or more than casual exchanges of ideas. (An exception is that Wheeler guided Dicke's work on his 1939 Princeton undergraduate senior thesis, ``A Logical Development of Quantum Mechanics and the Raman effect in the Atom.'')

I recall that other senior members of the Princeton physics department in the 1960s felt that the path to a deeper fundamental theory would be through quantum particle physics. They respected Dicke and Wheeler as excellent physicists who had made curious career choices. 

\section{Mach's Principle, Dirac's Large Numbers, and scalar-tensor gravity theory}\label{sec:Mach}
Considerations along lines discussed in this section motivated Einstein's development of general relativity and his introduction of basic elements of modern cosmology. They also motivated Pascual Jordan in his creation of the scalar-tensor gravity theory, and Dicke in his development of empirical gravity physics. These are loosely specified ideas that have never been part of the broadly accepted belief system in physics and astronomy, but they are important for an understanding of how empirical gravity physics grew, and they continue to attract interest, for evolving reasons. 

\subsection{Mach's Principle}\label{sec:machsprincile}
Just as motion may be considered meaningful only relative to the rest of the matter in the universe, it is logical to some to conjecture that inertial motion is meaningful only relative to what the rest of the matter is doing. The idea has a long history; it is most familiar now from the discussion in Ernst Mach's book, {\it The Science of Mechanics} (Mach 1893; p. 284 in the 1960 edition of the English translation). Mach commented on Newton's point, that if the surface of the water in a bucket is curved then the bucket is observed to be rotating relative to the distant stars. This could be taken to mean that rotation has an absolute meaning, but Mach and others questioned that. In Mach's words (translated from the German),  
\begin{quotation} 
\noindent No one is competent to say how the experiment would turn out if the sides of the vessel increased in thickness and mass till they were ultimately several leagues thick. The one experiment only lies before us, and our business is, to bring it into accord with the other facts known to us, and not with the arbitrary fictions of our imagination.
\end{quotation} 
Mach's admonition is in line with the theme of this paper: experiments matter (though Mach's positivist-empiricist philosophy was overly strict about imagination: Einstein's great imagination led to general relativity theory, which now proves to fit many facts later derived from experiments). The quote could be read to argue that the rotation of a very massive bucket might be expected to drag the motion of an inertial frame defined by local measurements, including the behavior of the surface of the water in the bucket. This would be an elegant anticipation of the relativistic Lense-Thirring inertial frame dragging near a rotating mass concentration such as the Earth, as follows. It is simplest to imagine a spherical shell of mass $M$ and radius $R$ rotating at angular velocity  $\rm\Omega$. The general relativity prediction, in lowest approximation, is that an inertial frame inside the shell precesses relative to distant matter at angular velocity, $\omega$, which is in order of magnitude
\beq
\omega\sim{GM\over Rc^2}\rm\Omega.\label{eq:LenseThirrng}
\eeq 
The successful test of this relativistic prediction applied outside the solid Earth is discussed in Section~\ref{sec:LenseThirrng}.

If in equation~(\ref{eq:LenseThirrng}) we replace $M$ by the observable mass of the universe in the relativistic Friedman-Lema\^\i tre cosmology that is expanding at about escape velocity, and we replace $R$  by the Hubble length (the distance at which the linear relation between the distances of galaxies and their redshifts extrapolates to apparent recession velocity equal to the velocity of light), then the factor $GM/Rc^2$ is of order unity, so $\omega\sim\rm\Omega$. Brill and Cohen (1966) examined this situation. It might invite one to imagine that, if the observable universe were said to be rotating, then local inertial frames would rotate with it, which is to say that the rotation would be meaningless. We may consider this qualitative argument to be one way to express Mach's Principle.

The broader physical implication of Mach's massive rotating bucket argument are still debated. Einstein, in his 1921 lectures on {\it The Meaning of Relativity} (Einstein 1923), noted that in general relativity a galaxy in otherwise empty asymptotically flat space-time could rotate, with all the usual effects of rotation, but it would be rotation relative to an otherwise empty universe. Einstein argued that, if this situation were allowed, it would mean that ``Mach was wholly wrong in his thought that inertia, as well as gravitation, depends upon a kind of mutual action between bodies''  and that ``from the standpoint of epistemology it is more satisfying to have the mechanical properties of space completely determined by matter, and this is the case only in a space-bounded  universe'' (Einstein 1923, p. 119). In subsequent editions the sentence ends ``only in a closed universe.'' Wheeler continued to argue for the philosophical appeal of this argument for a closed near homogeneous relativistic universe (as one sees in Misner, Thorne, and Wheeler 1973, \S 21.12). Following Einstein (1923),  some take Mach's Principle to be that a philosophically acceptable universe is described by a solution of Einstein's field equation in which inertial motion everywhere is determined solely by the distribution and motion of matter everywhere. 

The considerations presented in the 1921 lectures may account for Einstein's (1917) bold proposal that, apart from local fluctuations, the universe is homogenous. There was no empirical evidence of this at the time. His 1917 argument for homogeneity is difficult to follow, but the 1921 reasoning seems clear: homogeneity would prohibit the phenomenon of non-Machian inertia in an asymptotically flat space that contains only a single concentration of matter. The concept of large-scale homogeneity (which might be stated to be that the universe is a spatially stationary and isotropic random process) has come to be known as Einstein's Cosmological Principle. It was very influential in the development of cosmology well before there was any observational evidence in support of homogeneity. The network of well-checked cosmological tests we have now make a persuasive case for Einstein's Cosmological Principle from Mach's Principle (Sec.~\ref{Sec:GR-CMB}), which may be counted as a notable example of the power of nonempirical evidence --- when it is right.

One may debate whether the application of the Cosmological Principle really makes general relativity theory Machian. The issue does not seem to have long interested Einstein, but Machian ideas continued to fascinate others. For example, general relativity theory requires the same consistency of inertial motion defined by local experiments and observations of distant matter in a universe that satisfies the Cosmological Principle but has an arbitrarily small mean mass density. Dicke's thinking, as I recall it, was that this means Einstein's theory is unsatisfactory, that we need a better one that would make the distinction between inertial and noninertial motion meaningless in a universe that is empty apart from some test particles with arbitrarily small masses. 

Examples of thoughts along such directions are to be found in Sciama (1953, 1964), Brans and Dicke (1961),  Lynden-Bell (2010), and articles in the book, {\it Mach's Principle: From Newton's Bucket to Quantum Gravity} (Barbour and Pfister 1995). The sense of the thinking may be captured in Sciama's (1953) proposal that ``if local phenomena are strongly coupled to the universe as a whole, then local observations can give us information about the universe as  a whole.'' Such ideas, perhaps drawn from interpretations of Mach's principle, perhaps drawn from some other holistic concept of physical reality, inspired many of the gravity experiments discussed in Sections~\ref{sec:GravityGroup} and~\ref{sec:tests}.

\subsection{Dirac's Large Numbers Hypothesis}\label{sec:LNH}
Another  line of thought that was influential during the naissance started with the observation that the ratio of electrostatic to gravitational forces of attraction of a proton and electron is an exceedingly large number. The ratio of the expansion time $t$ in the relativistic Friedman-Lema\^\i tre cosmology to the characteristic time $e^2/m_ec^3$ defined by atomic parameters also is an exceedingly large number. And the two numbers have the same order of magnitude:
 \beq
{e^2\over G m_pm_e} \sim t{m_ec^3\over e^2} \sim 10^{40} \sim N. \label{eq:N}
\eeq
Here $m_e$ and $m_p$ are the masses of the electron and proton, $e$ is the magnitude of their charge, and $c$ is the velocity of light. The number $n_p$ of protons in the observable part of the universe, in the relativistic model, is really large, on the  order of $n_p\sim N^2$. Dirac (1937) mentioned Eddington's discussion of these large numbers (without citation; he may have meant Eddington 1936 p. 272). Dirac's Large Numbers Hypothesis (LNH) was that these numbers are very large because $N$ has been increasing for a very long time, and that the parameters in equation~(\ref{eq:N}) have been changing in such a way as to preserve rough equality of the two ratios. If physical units may be chosen such that the atomic parameters in equation~(\ref{eq:N}) are constant, or close to it, then preservation of the approximate numerical agreement of the ratios would require that as the universe expands, and the expansion time $t$ increases, the strength of the gravitational interaction decreases, as
\beq
G\sim t^{-1}.\label{eq:Gt}
\eeq 
This would mean that the number of protons is increasing as $n_p\sim N^2 \sim t^2$. Dirac was not clear about the manner of increase of $n_p$; he mentioned particle production in stellar interiors. The thought turned instead to the idea that $n_p$ is  the number of protons in the observable part of the universe, which has been increasing (in the standard cosmology and variants). 

\subsection{The weak and strong equivalence principles}\label{sec:WeakStrongEP}
Dicke argued that Dirac likely was on the right, Machian, track: As the universe evolves physical ``constants'' evolve under the influence of the evolving concentrations of matter. But my observation was that Dicke was even more attracted to ideas that suggested interesting experiments from which something of value might be uncovered. At the Chapel Hill Conference Dicke (1957a) mentioned experiments in progress motivated by the idea that Dirac's LNH 
\begin{quotation}
\noindent would imply
that the gravitational coupling constant varies with time. Hence it
might also well vary with position; hence gravitational energy might contribute
to weight in a different way from other energy, and the principle
of equivalence might be violated, or at least be only approximately true.
However, it is just at this point that the E\"otv\"os experiment is not accurate
enough to say anything; it says the Òstrong interactionsÓ are all right (as
regards the principle of equivalence), but it is the Òweak interactionsÓ we
are questioning. 

\ldots

\noindent Assuming that the gravitational binding energy of a body contributes
anomalously to its weight (e.g., does not contribute or contributes too
much), a large body would have a gravitational acceleration different from
that of a small one. A first possible effect is the slight difference between
the effective weight of an object when it is on the side of the earth toward
the sun and when it is on the side away from the sun.
\end{quotation}

The comment about a difference of gravitational accelerations of a large body and a small one is worth noting. It may be related to Dicke's (1962a) later remark: If the strength of the gravitational interaction were a function of position, then the gravitational binding energy of a massive body, such as the planet Jupiter, would be a function of position, and the gradient of the energy would be a force that would cause the orbit to differ from that of a low mass particle with negligible gravitational binding energy. Finzi (1962) discussed the same effect, in connection with the orbits of white dwarf stars. Nordtvedt (1968) introduced the formal analysis of the effect. 

The principle of equivalence Dicke mentioned at the Chapel Hill Conference has come to be termed the Strong Equivalence Principle: the prediction in general relativity theory that what is happening  on Earth is quite unaffected by the disposition of all exterior mass, in our Solar System, galaxy, and the rest of the universe, apart from tidal fields and the determination of local inertial motion. The Weak Equivalence Principle is that the gravitational acceleration of a test particle is independent of its nature. It is tested by the E\"otv\"os experiment, as Dicke noted. Early discussions of the distinction are in Dicke (1957a; 1959b, pp. 3-4; 1962a, pp. 15-31). 

The pendulum experiment Dicke mentioned at the Chapel Hill Conference, to check for variation of the gravitational acceleration $g$ at a fixed point on Earth, was a search for a possible violation of the Strong Principle. Maybe, Dicke suggested, the strength $G$ of gravity varies as the Earth moves around the Sun and moves relative to the rest of the mass of the universe, producing an annual variation of $g$, or maybe $G$ is decreasing as the universe expands, producing a  secular decrease of $g$. At Chapel Hill Dicke also mentioned precision tracking of the motion of the Moon. He did not  explain, but in later papers (notably Hoffmann, Krotkov, and Dicke 1960) stated the purpose to be to check for Dirac's LNH expressed in equation~(\ref{eq:Gt}), again in violation of the Strong Principle. Section~\ref{sec:LLR} reviews how tracking the Moon grew into a demanding test of general relativity with a tight constraint on the LNH, including the variation of $g$  that the pendulum experiments were meant to probe. 

\subsection{The scalar-tensor gravity theory}\label{sec:STgravity}
Pascual Jordan took Dirac's LNH seriously (as Schucking 1999 described). Jordan (1937, 1949) reviewed Eddington's (1936) and Dirac's (1937) considerations of the large numbers and, with Dirac, contemplated the idea that $n_p$ is growing because matter is being created, maybe in stars. Jordan and M\"oller (1947) and Jordan (1948) took the LNH as motivation for replacing Newton's gravitational constant $G$ by a scalar field in a scalar-tensor gravity theory in which the evolution of the scalar field could agree with the conjecture that the strength of the gravitational interaction is decreasing as the universe expands. In his book, {\it Schwerkraft und Weltall}, Jordan (1952) took note of Teller's (1948) point, that a larger $G$ in the past would imply a hotter Sun, which if too hot would violate evidence of early life on Earth, a serious constraint. Also in {\it Schwerkraft und Weltall}, the setup of the scalar-tensor theory required local violation of energy-momentum conservation if the strength of the gravitational interaction were evolving, again leading Jordan to contemplate generation of matter (p. 143), perhaps in stars, perhaps causing the masses of stars to increase as $G$ decreases. In {\it Schwerkraft und Weltall} and {\it Die Expansion der Erde} (Jordan 1966) Jordan considered the growing evidence for continental drift, which he pointed out might be caused by a decreasing value of $G$ that relieved stresses that allowed the continents to move. In the preface to {\it Die Expansion der Erde}, Jordan wrote that ``I must report that R. Dicke has independently arrived at similar hypothetical consequences'' (as expressed in the English translation in Jordan 1971). Jordan (1966) may have been referring to Dicke~(1961a) or~(1962b), where he discussed issues of continental drift. 

Fierz (1956) wrote down the special case of Jordan's approach to a scalar-tensor theory that preserves standard local physics, eliminating the violation of local energy conservation that Jordan was thinking might be relevant for stellar evolution. In the notation of Brans and Dicke (1961), Fierz's action (with units chosen so $c=1$) is
\beq
S = \int\sqrt{-g}\,d^4x\left[ \phi R + 16\pi L +w\,\phi_{,i}\phi^{,i}/\phi\right].\label{eq:BD}
\eeq
Newton's gravitational constant $G$ is replaced by $\phi^{-1}$, where $\phi$ is a scalar field. Since $\phi$ does not enter the action $L$ for matter and radiation, local physics is standard, local energy is conserved, and the gravitational acceleration of a test particle is independent of its nature, consistent with the E\"otv\"os experiment. The presence of $\phi$ in the denominator of the gradient energy density term makes the units consistent. The source term for $\phi$ in equation~(\ref{eq:BD}) is $w^{-1}R$, where the Ricci tensor $R$ is a measure of spacetime curvature. Brans (1961) showed that this can be reduced to
\beq
\Box\,\phi = 8\pi T/(3 + 2w),\label{eq:BD-T}
\eeq
where $T$ is the trace of the matter stress-energy tensor. Thus we see that the strength of the coupling of the scalar field to the rest of physics scales as about $w^{-1}$, where the constant $w$ is a free parameter. The larger the choice of $w$ the smaller the source of variation of $\phi$. That is, at large $w$ equation~(\ref{eq:BD}) may approach the Einstein-Hilbert action of general relativity with constant $\phi$ and $G=1/\phi$. 

Carl Brans (PhD Princeton 1961) arrived in Princeton in 1957 as a graduate student intending to work on mathematical analyses of spacetime structure. He recalled that Charles Misner (whose PhD in 1957 was directed by Wheeler) suggested that he talk to Dicke, who was looking for someone to turn Mach's Principle and Dirac's LNH into a gravity theory. Dicke told Brans about Sciama's (1953) schematic model for how this might be done; Brans then independently hit on the scalar-tensor action in the form of equation~(\ref{eq:BD}), with preservation of local energy-momentum conservation (Brans 2008; 2016). Brans later learned that Jordan and Fierz had done it first. 

In his earliest publications on gravity physics Dicke (1957b, c) referred to Jordan (1952) and Fierz (1956). If at the time Dicke had recognized the significance of the theory in these papers it would have been in character for him to have told Brans about them rather than Sciama, but we can only speculate about this. In any case, we see that Dicke's references to Jordan's research on scalar-tensor gravity theory and its physical implications were consistently brief though complete. Jordan's theory was cited in Dicke (1957b, c); Brans and Dicke (1961) and Dicke (1962c), on their considerations of the scalar-tensor theory; and in Dicke (1964) and Dicke and Goldenberg (1967), on the search for a solar oblateness that might allow more room for the scalar-tensor parameter $w$ in equation~(\ref{eq:BD}). Jordan does not seem to have been overly troubled by this manner of acknowledgement of his work, as indicated by their correspondence.\footnote{Quotations from the Robert Henry Dicke Papers, Box 4, Folder 4, Department of Rare Books and Special Collections, Princeton University Library.} In a letter to Dicke dated 2.7.1966 Jordan wrote
\begin{quotation}
\noindent I tried to study comprehensively the whole field of empirical facts which might be suited to allow a test of Dirac's hypothesis; and I have now the impression (or I hope to be justified to think so) that all relevant empirical facts are in best accord with theoretical expectation, and the whole picture is quite convincing. I shall be very anxious, as soon as the book is published, to learn what you think of it. (Some points are a little deviating from what you preferred to assume; I come to the result of a rather great value of $\dot\kappa/\kappa\sim 10^{-9}/$\,year \ldots\ I intend to be in USA for about a month \ldots\ Naturally I should be extremely glad to have the occasion to visit you. 
\end{quotation}
Here $\dot\kappa/\kappa$ is Jordan's notation for the fractional rate of change of the strength of gravity;  the notation used here is $\dot G/G$. Dicke's reply, dated 7 July 1966, after a paragraph welcoming Jordan's plan to visit the USA, was
\begin{quotation}
\noindent I was pleased to learn that you are publishing a new book. I think that perhaps you would agree with me that the implications for geophysics and astrophysics of a time rate of change of the gravitational  interaction is one of the most fascinating questions that one could consider. I always have my mind open looking for some new fragment of information that could have a bearing on this question. I am curious to know how you could have a time rate of change of gravitation as great as $10^{-9}$ per year and am looking forward to reading about it in your book.
\end{quotation}
At the time we felt reasonably sure that the evolution of $G$ could not be faster than about a tenth of Jordan's value (Sec.~\ref{Sec:Gdot}). Jordan and Dicke disagreed about aspects of the science, as here, but I know of no indication of disagreements between the two beyond this normal course of events in research.

For a more complete review of the historical development of scalar-tensor theories see Goenner (2012). For other reviews of Jordan's and Dicke's thinking about an evolving $G$ and its possible effects in geophysics and cosmology see Kragh (2015a,b; 2016).

\section{The Gravity Research Group}\label{sec:GravityGroup}
The experimental advances in gravity physics in the 1950s and 1960s grew out of independent work in many laboratories. But since Dicke was the central actor it is appropriate to give particular attention to his approach. The examples discussed here are drawn from the projects he mentioned at the Chapel Hill Conference: a repetition of the classical E\"otv\"os experiment and a search for possible variation of the gravitational acceleration by a suitably designed pendulum experiment. This discussion is meant to illustrate some of Dicke's characteristic methods, including recollections of how he assembled the Gravity Research Group. 

\begin{figure}[t]
\begin{center}
\includegraphics[angle=0,width=5.in]{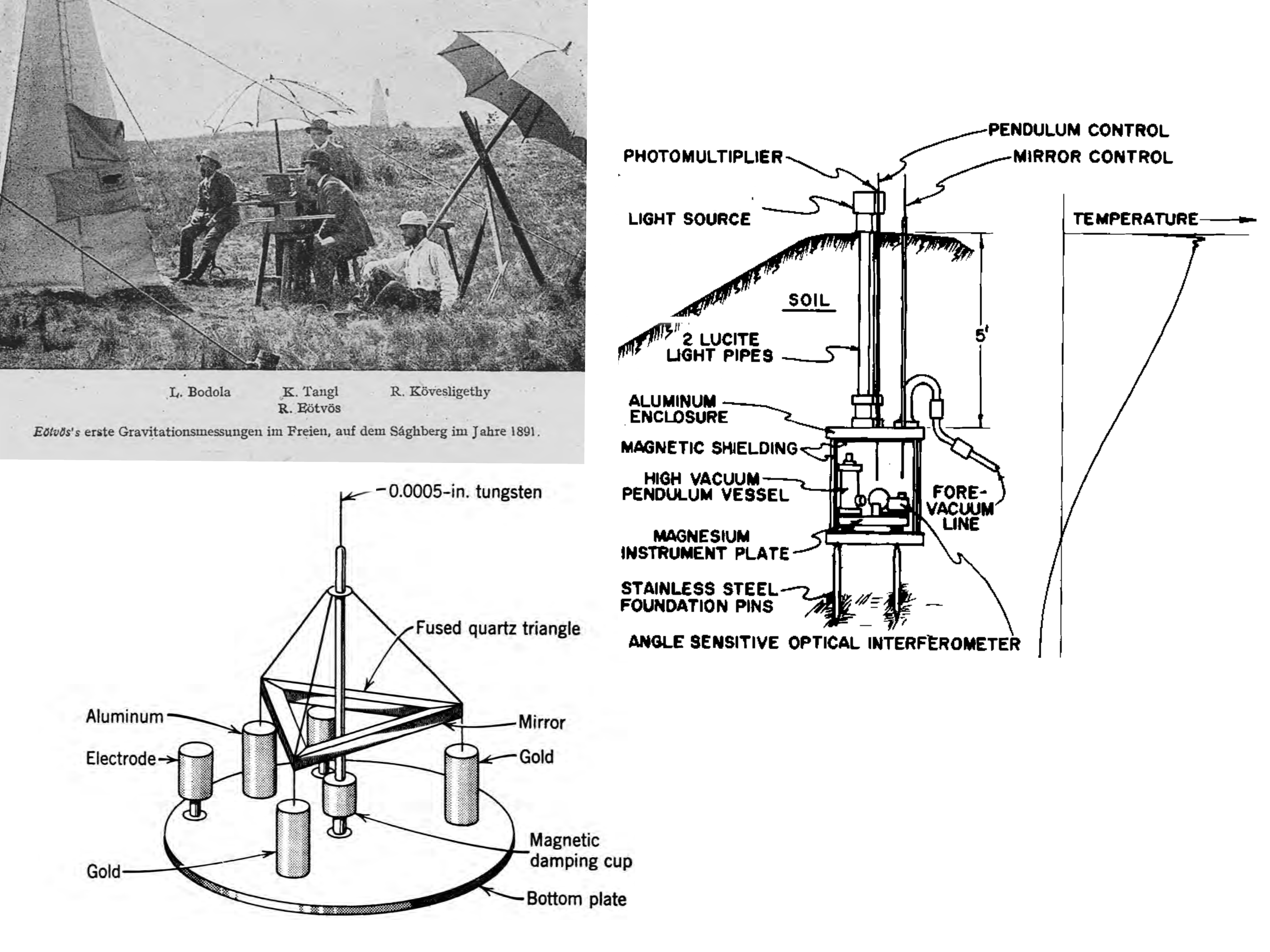} 
\caption{\label{Fig:Eotvos}  The photograph in the upper left shows E\"otv\"os and colleagues, likely measuring gravitational field gradients. On the right is an illustration of features common to the two repetitions of the E\"otv\"os experiment in Dicke's group: the balance placed in a vacuum, the torque detected remotely, and the instrument buried. On the lower left is a sketch of the torsion balance used in the static version (in an early design).}
\end{center}
\end{figure}
\subsection{The Princeton static and dynamic E\"otv\"os experiments}\label{sec:Eovtos}
The E\"otv\"os experiment tests a starting idea of general relativity, that the gravitational acceleration of a test particle is independent of the nature of the particle. Dicke's (1957a) comments about the considerable advances in technology from what was available to E\"otv\"os are illustrated in Figure~\ref{Fig:Eotvos}.  The photograph on the upper left shows E\"otv\"os and colleagues, likely measuring gravitational field gradients. He later turned this methodology to the comparison of gravitational accelerations of a variety of materials (E\"otv\"os, Pek{\'a}r, and Fekete 1922, E\"otv\"os  posthumously). To avoid disturbing the instrument E\"otv\"os had to let the balance come to rest in isolation, then approach it and promptly use the telescope for a visual observation of the orientation of the balance before it could respond to the gravitational field gradient produced by the mass of his body. The line drawing on the right, which shows the setup of the Princeton dynamic version, illustrates three of the improvements that Dicke mentioned in response to Bergmann's question: the balance was placed in a vacuum, to reduce dissipation and the attendant Brownian noise; the behavior of the balance was measured and recorded remotely, removing disturbances by the observer; and the system was buried to reduce the disturbing effects of temperature variations and the wind and the noise. Dicke also chose to compare the gravitational accelerations of test masses toward the Sun. The advantage was that a difference of accelerations toward the Sun would be manifest as an effect on the balance with a 24-hour period. That removed the need to rotate the apparatus, apart from checks of reproducibility. But this strategy of course required careful attention to diurnal disturbances. 

\medskip
\begin{figure}[ht]
 \begin{center}
\includegraphics[angle=0,width=5.in]{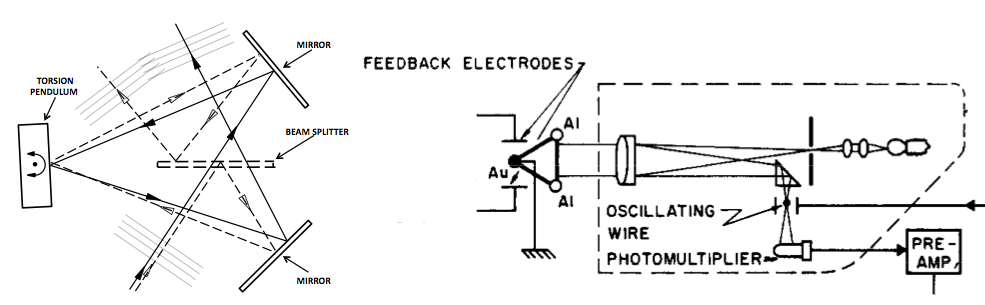} 
\caption{\label{Fig:EotvosDetectors} The angle-sensitive detectors in the Princeton E\"otv\"os experiments in the dynamic version, on the left (Liebes 1963, 2016), and the static version, on the right (Roll, Krotkov, and Dicke 1964).}
 \end{center}
\end{figure} 
The static version of the Princeton E\"otv\"os experiment employed a feedback loop to hold a torsion balance in place; the dynamic version tracked the oscillation frequency of a torsion pendulum. Dicke mentioned the static approach at the March 1957 Chapel Hill conference. In the fall of 1957 Sidney Liebes joined the Gravity Research Group and Dicke suggested to him the dynamic version and the optical interferometer method of precision timing of the pendulum. He left it to Liebes to design, construct, and operate the experiment. 

In the dynamic version the torsion balance was a 2.5-cm long tubular bar of fused silica with an aluminum test mass inserted in one end and a platinum mass inserted in the other. The equilibrium position of the bar was oriented east-west. The bar was suspended by a thin fused-silica fiber (which Liebes drew by the traditional crossbow technique: fix one end of a silica bar, attach the other end to an arrow without a point,  melt the center of the silica bar, shoot the arrow across the room, and then try to find the fiber). The amplitude of oscillation approached $90^\circ$, the period was approximately 8 minutes, and the decay constant in the high vacuum was several months, which is about $10^4$ oscillations. The apparatus was buried five feet beneath the bleachers of the Princeton football stadium, to suppress diurnal temperature variations. The upper end of the torsion fiber was attached to a gimbal mechanism that enabled eddy current damping of pendulum vibrations induced by football games and other disturbances.  

The left-hand part of Figure~\ref{Fig:EotvosDetectors} is a top view of the angle-sensitive optical interferometer Dicke proposed to Liebes. One element is an optical flat on one  side of the pendulum. The beam-splitter (semi-reflective mirror) caused the nearly monochromatic light from a low-pressure sodium lamp to follow the opposing paths through the interferometer marked by the solid and dashed lines. The emerging beams combined to form an interference pattern, the output of which was integrated by a photomultiplier. The bar is shown slightly displaced from the interferometric null-fringe position. In passage through optical null the fringes race apart and the photomultiplier detects a momentary characteristic pulse of light. The pulse timings were measured with a stabilized high-frequency oscillator and frequency counter system. A violation of the Weak Equivalence Principle would be manifest as a 24-hour variation of the gravitational torque on the pendulum by the Sun, which would produce a diurnal variation in the times of optical null as the torsion pendulum crossed in one direction, and an opposite shift in times of  optical null when crossing in the other direction. This experiment was reported at a conference (Liebes 1963), but not published.

The static version was buried in another then more remote part of the campus. The sketch of the torsion balance on the lower left side of Figure~\ref{Fig:Eotvos} shows the fused silica triangle that held the test masses, gold and aluminum, in an arrangement that suppresses tidal torques. This triangle is drawn in heavy lines in the right-hand sketch in Figure~\ref{Fig:EotvosDetectors}, showing the two feedback electrodes that straddled the gold test mass (but straddled the aluminum mass in the earlier design in Fig.~\ref{Fig:Eotvos}). Roll, Krotkov, and Dicke (1964) described this part of the experiment as follows:
\begin{quotation}
\noindent At the heart of the experiment is the instrumentation for measuring very small
rotations of the torsion balance \ldots The light is focused through a $25\,\mu$ slit, 
reflected from the aluminized flat on the
quartz frame of the torsion balance, deflected off the telescope axis by a small
prism, and the image of the slit focused on a $25\,\mu$-diam tungsten wire. By locating
this wire in the field of a small magnet and connecting it in a balanced bridge
oscillator circuit, it was made to oscillate at its mechanical resonance frequency
of about 3000 cps and with an amplitude of 25 to $50\,\mu$ \ldots When the diffraction pattern of the $25\,\mu$ slit produced by the 40~mm diameter telescope lens is centered exactly on the equilibrium position of the oscillating wire \ldots the photomultiplier will detect only the even harmonics of the
3000 cps fundamental frequency. As the torsion balance rotates slightly and
shifts the diffraction pattern off center, the fundamental frequency will begin to
appear in the photomultiplier output. The phase of the fundamental ($0^\circ$ or $180^\circ$
relative to the oscillator signal driving the wire) indicates the direction of rotation
of the pendulum, and its amplitude is proportional to the magnitude of the
rotation for sufficiently small angular displacements \ldots The full width at half maximum of this
curve (the ``line width'' which must be split by the detection apparatus) is about $30\,\mu$ or $3 \times 10^{-5}$ rad \ldots these processes are all performed by a lock-in amplifier.
\end{quotation}
The $25\,\mu$ tungsten wire was informally known as the wiggle-wire. The experimental result is discussed in Section~\ref{Sec:masses}. Peter Roll (2016) recalled that, in writing the paper on this experiment (Roll, Krotkov, and Dicke 1964),
\begin{quotation}
\noindent it was important for the final paper on the E\"otv\"os-Dicke experiment to document it in enough detail so that subsequent generations would not have to do as much legwork to understand exactly what was done and how.  Dicke's contributions to the paper were the rationale for doing it in the first place and the basic design of the apparatus. Bob Krotkov got the first versions of the equipment designed in detail, built, and working. I came along at the end, contributing some of details of the Au-Al-Al apparatus and procedures, the final analyses of data and error sources, and analysis of E\"otv\"os and Renner results. Bob Dicke was the master of both the theory behind the experiments and the experimental and apparatus design. His introduction to the 1964 paper explains the years he spent looking for a scalar gravitational field.
\end{quotation}

\begin{figure}[ht]
  \begin{center}
   \includegraphics[width=2.5in]{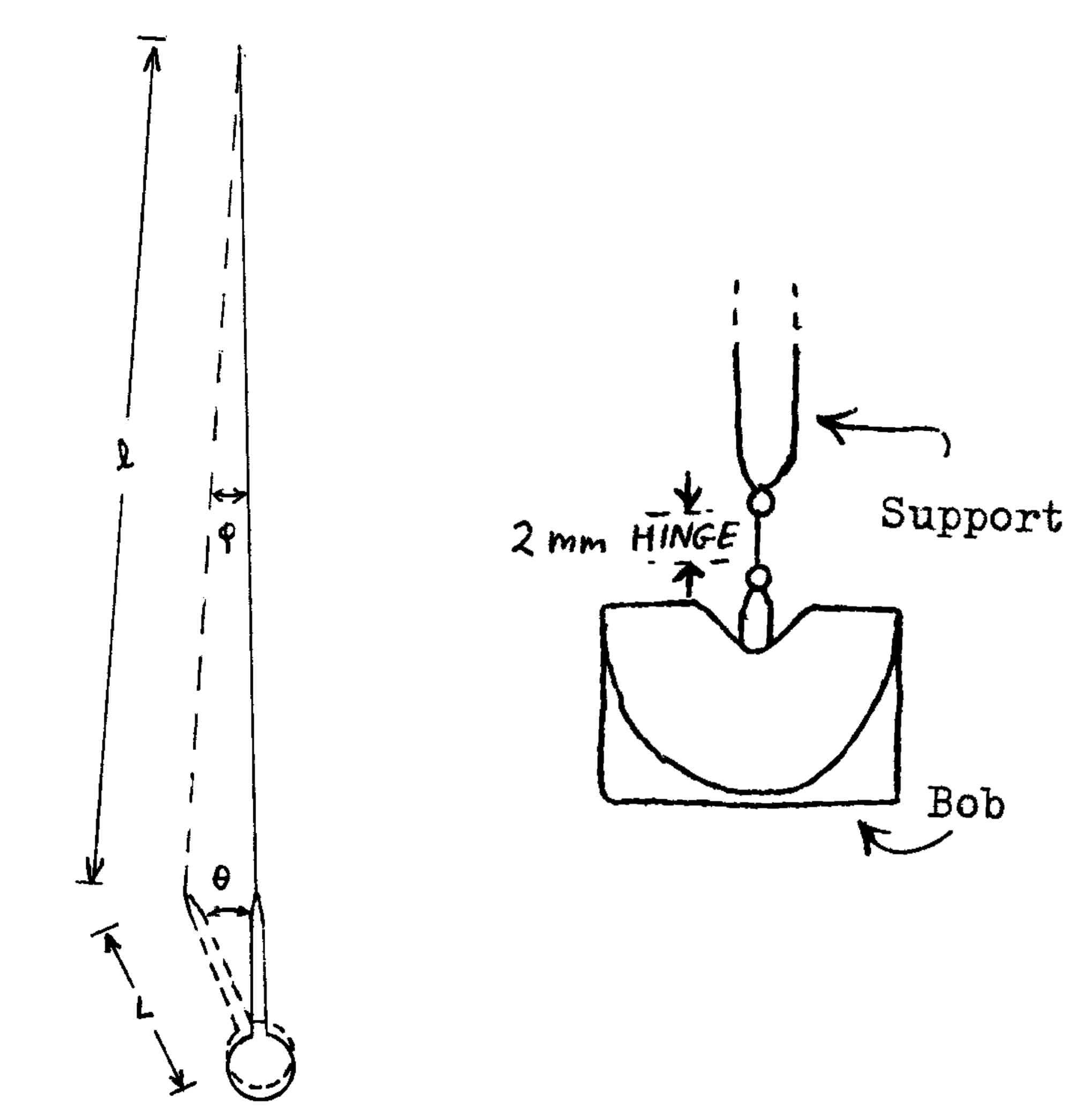}
    \caption{\label{Fig:pendulum} Schematics of the Hoffmann (left) and Curott (right) pendulum designs.}
   \end{center}
\end{figure} 
\subsection{Searching for variation of the gravitational acceleration}
The pendulum experiment Dicke mentioned at Chapel Hill was meant to test whether the gravitational acceleration at a fixed spot on Earth may vary as the Earth moves around the Sun, or as the universe expands. The ideas motivating this test are reviewed in Section~\ref{sec:Mach}. Features that Dicke mentioned at Chapel Hill are illustrated in Figure~\ref{Fig:pendulum}, in two different pendulum designs, the left-hand sketch from Bill Hoffmann's (1962) dissertation and the right-hand sketch from David Curott's (1965) dissertation. The rapid oscillations (22\,Hz in Hoffmann's experiment, 5\,Hz for Curott) made these pendulums less sensitive to external disturbances and much more sensitive to a slow variation of $g$, because more oscillations were observed in a given time. Each was drawn from a single piece of silica, with no knife edges that may wear, and very small energy loss in the vacuum. The pendulums were electrostatically driven. The dissipation times were about 6\,hours for Hoffmann, 20\,hours for Curott, or about $4\times 10^5$ oscillations in each experiment. The period measurement during free decay used a source of light that passed through a slit, was reflected off an optical flat on the bottom of the pendulum, and focused on the vertex of a prism. As the pendulum swung the light passed from primary detection by a photomultiplier on one side of the prism to detection by a photomultiplier on the other side. A frequency standard with good short-term stability was phase-locked to the pendulum to follow the average pendulum crossing time.

Dicke chose the designs for the two versions of the E\"otv\"os experiment, static and dynamic, and the two versions of the pendulum experiment. We may suppose he meant to explore possibilities for optimum methods. Was the time right for these exploratory experiments? Might they have been done much earlier? Hoffmann's (2016) response was 
\begin{quotation}
\noindent A small fused silica inverted pendulum could have been fabricated and operated in a vacuum before the 1950's.  But the accuracy of the measurement of the pendulum frequency depended on recent technological advances.  These included sensitive phototubes used to detect the light beam reflected from the pendulum; low noise, stable, battery-operated vacuum tubes developed for undersea communication cables and used with the phototubes; lock-in amplifiers pioneered by Bob Dicke and used for precision phase locking and monitoring weak signals; fast counters and digital printers; a commercial ultra-stable crystal oscillator and a commercial atomic clock (General Radio Atomichron) for precise timing; and a programable digital computer (IBM 650) for Fourier analysis of the measurements to identify an annual variation.  The experiment was built at the time transistors were coming into use and benefited by use of this new technology.

\ldots

\noindent The IBM 650 was purchased by Princeton and installed in the Gauss house, a Victorian home on Nassau Street, around 1957-58. Prior to that Princeton had limited use of the Institute for Advanced Study MANIAC computer.  During 1960 I was given the keys to the building for a night and spent the night there alone with with punched cards containing my data and a machine language Fourier transform program written by Bob Krotkov.  The result is in my thesis.  This would not have been possible two years earlier.
\end{quotation}
Curott (2015) expressed the same opinion, and added that his experiment
\begin{quotation}
\noindent couldn't have been done much earlier since it depended upon current (1960's) ultra-vacuum technology and electronic timing techniques. The emerging computer availability on campus also played an important role since daily corrections had to be made for minute by minute positions of moon and sun (tidal corrections). The data analysis would have been daunting, if not impossible, before electronic computers.
\end{quotation}

\subsection{Dicke's synchronous detection}
Dicke was the leading exponent of the method of synchronous lock-in amplifier detection employed in the Princeton E\"otv\"os and pendulum experiments discussed above. He used it earlier in the Dicke (1946) microwave radiometer for suppression of the effect of receiver noise by switching between the  source to be measured and a stable reference source, with detection of the output synchronized to the switching frequency. Lock-in amplifiers were also used in the Gravity Group measurement of the solar gravitational redshift (Sec.~\ref{sec:gravredshift}); the development of static gravimeters (Sec.~\ref{sec:gravitationalwaves}); the measurement of the shape of the Sun (Sec~\ref{sec:testingBD}); and the  Princeton search for the CMB (Sec.~\ref{Sec:CMB}). The first commercial lock-in amplifiers were produced in 1962 by the then privately held company, Princeton Applied Research, founded by Dicke and colleagues largely at Princeton University and the Princeton Plasma Physics Laboratory. Curott (1965) acknowledged use of a ``Prototype of a model commercially available from Princeton Applied Research Corp.'' In connection with the measurements of the shape of the Sun (Kuhn, Libbrecht, and Dicke 1988 and earlier references therein), Kuhn (2016) recalled that
\begin{quotation}
\noindent Dicke's solar oblateness measurements both with Goldenberg and Hill, and later with Ken and I had, at their heart, at least one analog lock-in amplifier. Even in the 80's when we had early microcomputers to do synchronous demodulation, Bob had a very clever scheme to measure the position of the solar image in reference to the occulting disk using a `JB8' lock-in amplifier. It made the Mt. Wilson oblateness measurements possible.\footnote{The initials JB indicate Jim Braults' early design.}
 \end{quotation}
Weiss (2016) emphasized 
\begin{quotation}
\noindent the critical idea in the Dicke E\"otv\"os
experiment in using quiet (low noise) sensors in a feedback loop to damp and position a
mechanical instrument. The idea of holding an instrument at a fixed position and then
reading it out by recording the force to hold it in that fixed position is critical to many
precision mechanical experiments that have followed the Dicke E\"otv\"os experiment. The
feedback system is used to keep the response of the system linear and if done cleverly can
be used to suppress normal modes of structures that through non-linearities in the system
cause noise in the mode that carries the physical information of the measurement being
made. LIGO uses many thousands of such feedback systems.
\end{quotation}
The reference is to the first successful detector of gravitational waves, the Laser Interferometer Gravitational-Wave Observatory (LIGO et al. 2016). Weiss (2016) added that 
\begin{quotation}
\noindent the technique of modulating a physical effect to make it measurable above the
$1/f$ noise used in the wiggle-wire telescope of the angle detector in the Dicke E\"otv\"os
experiment (Sec.~\ref{sec:Eovtos}) and in the Brault technique for finding the center of a spectral line (Sec.~\ref{sec:gravredshift}) was the other important technique developed by Dicke for precision
experiments. The idea did not originate with Dicke but it was developed to an art by him
and is now part of the stable of tricks used in virtually all precision measurements. The
technique has a name---suppressed carrier modulation detection---and uses the
lock-in amplifier.
\end{quotation}
The judgement of experimental colleagues at Princeton University was that, with its successors, the lock-in amplifier ``probably has contributed as much to experimental PhD theses as any device of the past generation'' (Happer, Peebles, and Wilkinson 1999). 

\subsection{Assembling the Gravity Research Group}
How did Dicke assemble the Gravity Research Group group? Recollections of the five main contributors to the pendulum and E\"otv\"os experiments discussed in this section offer a fair sample.
\begin{quotation}
\noindent David Curott: When I entered the graduate program I intended to go into Controlled Fusion Research. I assumed there was no reason to question General Relativity. Without my knowing about Dicke's Group, in my first summer I was assigned a summer research assistantship in Dicke's Gravity Group, and that opened my eyes to the need for gravity research and the opportunities offered by new technologies. Dicke and his group excited my imagination and I stayed in the Group.\medskip

\noindent Bill Hoffmann: I entered graduate school at Princeton in the fall of 1954, after graduating from a small college (Bowdoin), intending to be a theoretical physicist.  I soon found that this was not for me, but I was not interested in mainstream experimental work in atomic, nuclear, or high energy physics.  When I first met Bob Dicke on his return from Harvard sabbatical in 1955 he was all fired up about gravity experiments. His enthusiasm and ideas about gravitation experimentation captivated me.  I knew that this was the person I wanted to work with.  I was attracted by the challenges to be inventive and the stimulation from others in his group. \medskip

\noindent Robert Krotkov: I joined Dicke's group because he invited me. I don't remember just what research I did at first --- but I do remember where we first talked. It was in Palmer lab. As one came in the front door there was a hallway to the right, a hallway to the left, and straight  ahead stairs leading up to a landing. It was on that landing that we happened to meet each other. How one remembers the really important things! Eugene Wigner directed my PhD. With Wigner, I would carry out some calculations, come to him, and he would tell me what to do next. With Dicke, I got a big picture and saw where what we were doing would fit in.\medskip

\noindent Sidney Liebes: I've had a lifelong fascination with relativity, first kindled, in my youth, by reading George Gamow's {\it Mr. Tomkins in Wonderland}.  Nearing completion of my PhD experiment at Stanford, testing a prediction of quantum electrodynamics, I  asked George Pake whether he knew anyone doing experimental gravity physics.  George's response: ``Bob Dicke at Princeton.'' That prompted my phone call to Dicke, which resulted in an offer to join the faculty as an instructor and work in the Gravity Research Group. Dicke created an environment where I felt totally uninhibited in how I spent my time, never prompted, directed or monitored, beyond periodic Group meeting update contributions.  I believe that environment to have been a critical factor in my independent rediscovery of gravitational lensing. 
\medskip

\noindent Peter Roll: My first introduction to general relativity and gravitation was a summer reading assignment of Ernst Mach's book on Mechanics, and several ``wasted'' hours in the Yale Physics Library digesting a small textbook on tensor calculus and GR, when I should have been working on physics problems and papers. In the spring of 1960 I was completing my PhD dissertation in experimental nuclear physics at Yale and looking for a job. Bob Beringer, a senior faculty member at Yale and a colleague of Bob Dicke at the MIT Radiation Lab during the war, had just learned that Dicke was looking for a newly-minted PhD to join his group at Princeton, finishing the E\"otv\"os experiment that Bob Krotkov had started. Beringer encouraged me to look into the position and told me a bit about Dicke's background and accomplishments. After a telephone call and a trip to Princeton, my mind was made up --- I wasn't going to find another position or another place that was anywhere near as interesting and challenging.
 \end{quotation}
 
I might add that in the early 1960s, while I was attending Gravity Research Group meetings, I still had in mind writing a dissertation in theoretical elementary particle physics. But Bob had suggested that I look into constraints on possible evolution of the fine-structure constant from the degree of consistency of radioactive decay ages based on different long-lived atomic nuclei. That got me interested in the rhenium-osmium decay,
\beq
^{187}\hbox{Re}\rightarrow ^{187}\hbox{Os} + e^- + \bar\nu,
\eeq
because I had noticed that the quite small decay energy made the decay rate quite sensitive to a change of the fine-structure constant. Since the decay energy was not well known I started looking into how I might better measure it. But Dicke had come to know me well enough to instruct me, first that I ought to stick with theory, and second that I am better suited to the kind of theory we had been doing in his group.

To exemplify the diverse directions of research in the group I refer to the list of reports of work in the Gravity Research Group at the January 1963 meeting of the American Physical Society:\footnote{Bulletin of the American Physical Society, Volume 8, pp 27 -- 29} Dicke on {\it Mach's Principle and Laboratory Physics}; Brault on {\it Gravitational Red Shift of Solar Lines}; Liebes on {\it Test of the Principle of Equivalence}; Turner and Hill on {\it New Experimental Limit on Velocity-Dependent Interactions of Clocks and Distant Matter}; Peebles on {\it Experimental Restrictions on Generally Covariant Gravity Theories}; Faller on {\it Absolute Determination of the Gravitational Acceleration}; Hoffmann on {\it Pendulum Gravimeter for Monitoring the Gravitational Acceleration as  a Function of Time}; and Roll and Dicke on {\it Equivalence of Inertial and Passive Gravitational Mass}. By the time of this meeting other groups were making important contributions to experimental gravity physics, as reviewed in the next section, but none rivaled this searching range of investigations. 

Worth recording also is Dicke's style of operation with his group. He tended to explain in some detail the motivation for a proposed project, outline possible methods, sometimes in detail, and then stand back to let us get to work. I recall David Wilkinson remarking that Dicke followed with great interest the construction of the Roll and Wilkinson microwave radiometer, built at Dicke's suggestion to look for radiation left from a hot Big Bang. But Wilkinson recalled that Dicke ventured very few suggestions about how they were doing it. I also recall Dicke encouraging me to keep thinking of ideas about possible implications of the Roll-Wilkinson experiment, whether a detection or upper  limit.  But I think the only idea he offered was the possible connection of the baryon Jeans mass in the hot Big Bang model to globular star clusters (Peebles and Dicke 1968). 

\section{A review of experimental gravity physics through the naissance}\label{sec:tests}
This is a review of highlights of experiments in gravity physics from 1915 through the naissance to its nominal completion in about 1968, as experimental research in this subject became part of normal science. I mean to include what astronomers term observations, and also probes of gravity physics derived from measurements made for other purposes. I mention PhD theses completed in the course of research aimed at probing gravity, because students made considerable contributions, particularly so in Dicke's Gravity Research Group. (It's worth recalling that Dicke's group often met on Friday evenings; we complained but attended because the discussions were too interesting to miss.) This review of  how the subject grew is largely assembled from the published literature, but I refer to personal recollections from some of the actors, many of them Dicke's students and colleagues, as well as my own experience since arriving at Princeton as a graduate student in 1958 and joining the group soon after that. It should be understood that the sparse references to what happened after 1968 are meant only to illustrate what grew out of the naissance. Also, as a theorist, I am aware that a better-informed examination of experimental methods, and how they evolved during the naissance, would be a valuable addition to these summary comments.

\begin{figure}
\begin{center}
\includegraphics[angle=0,width=4.in]{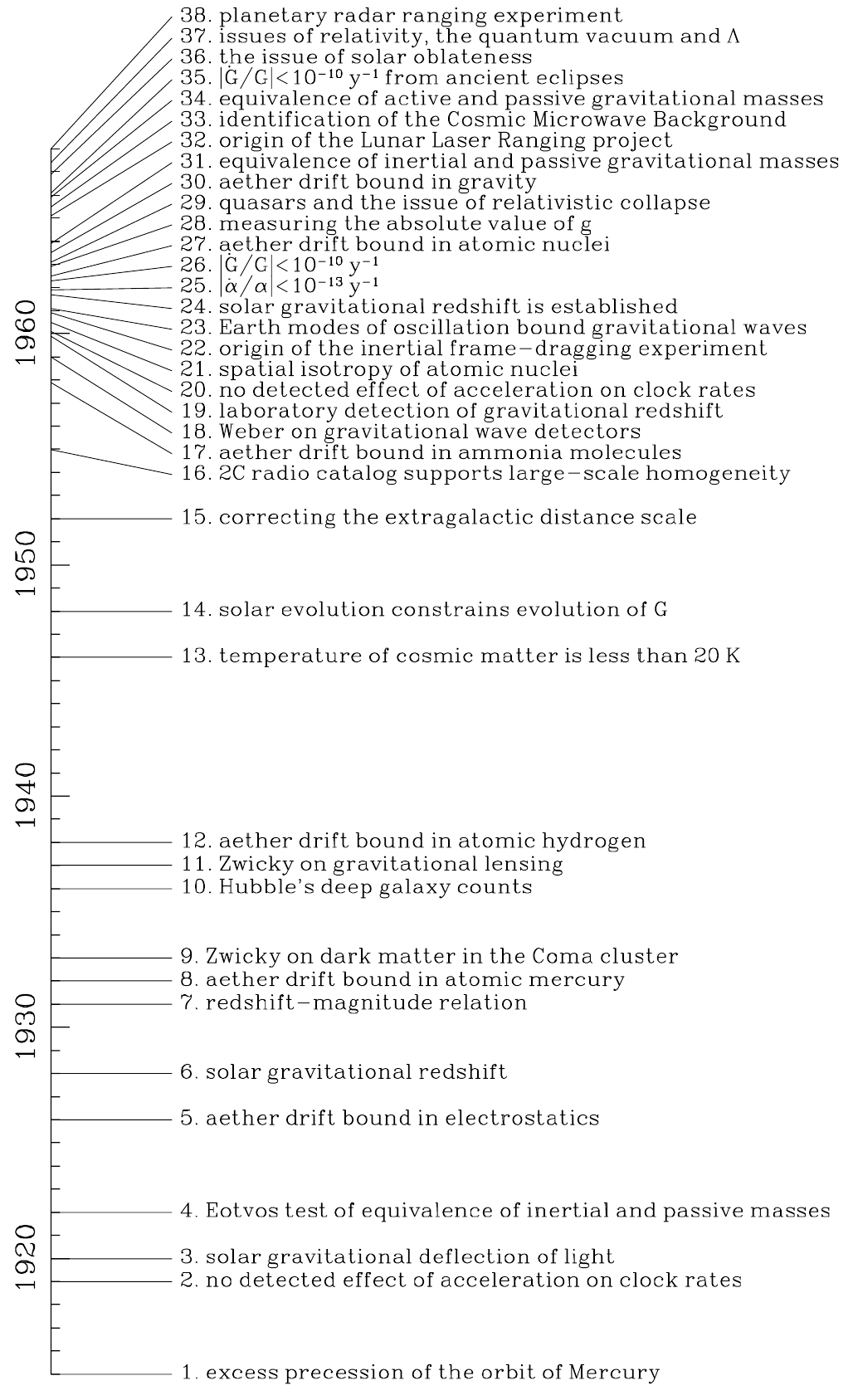} 
\caption{\label{Fig:timeline} A timeline of notable advances in experimental gravity physics, from the completion of general relativity theory in 1915 to the transition to a normal empirical science in the late 1960s. The exceptions to the rule that experiments are marked at completion are explained in the text.}
\end{center}
\end{figure}
\subsection{A timeline for experimental gravity physics}
The timeline in Figure~\ref{Fig:timeline} summarizes experimental developments in gravity physics from 1915 to the nominal completion of the naissance. The choice of developments to be marked required some creative accounting. The orbit of Mercury was the only demanding test of general relativity for a long time, so it ought to be marked; I put it at Einstein's 1915 relativistic interpretation of the excess precession (line 1 in the figure). I stretch the decade of the naissance to admit the Shapiro, Pettengill, Ash, et al. (1968) measurement of the relativistic time delay of planetary radar pulses that pass near the Sun (line 38 in the figure). I mark a few developments that were initiated during the naissance and later became important tests. Thus line~22 marks Schiff's (1960) argument for a gyroscope experiment to test the relativistic dragging of inertial frames (Sec.~\ref{sec:LenseThirrng}), which led to Gravity Probe~B (Everitt et al. 2015). The 1965 argument by the Dicke Group for a Lunar Laser Ranging Experiment (Alley, Bender, Dicke, et al. 1965; line 32 in the figure) is marked because it was an early step toward an extraordinarily demanding program of tests of gravity physics (Sec.~\ref{sec:LLR}). Zel'dovich's (1968, line 37) discussion of the quantum field zero-point contribution to the vacuum energy density is marked because, though little recognized then, it was and remains a major challenge for gravity and quantum physics. 

Zwicky (1933) showed that, under conventional physics, the stability of the Coma cluster of galaxies requires  ``dunkle Materie'' considerably in excess of what is seen in the stars (line~9 in Fig.~\ref{Fig:timeline}). The problem was noted by others; an example is Babcock's (1939) demonstration that the gaseous nebulae on the outskirts of the galaxy M31 are moving relative to the galaxy much more rapidly than would be expected from the observed mass of the stars. This and other evidence for dark matter was neglected during the 1960s, but it was so important to the aftermath (as briefly reviewed in Sec.~\ref{Sec:GR-CMB}) that it is marked in the figure, at Zwicky's paper. 

Line~11 marks another of Zwicky's (1937) contributions, his discussion of how a galaxy could act as a gravitational lens. It is entered because it was the first informed recognition of this possibly observable effect, though Zwicky's paper was little noticed until a reference to it appeared in broader analyses of gravitational lensing effects by Klimov (1963),  Refsdal (1964), who cited Einstein's (1936) paper, and Liebes (1964), who cited Einstein (1936) and Zwicky (1937), while Zwicky in turn cited Einstein (1936). This is discussed in Section~\ref{sec:lensing}. 

Jocelyn Bell Burnell found the pulsar phenomenon in 1967. This produced the first observational evidence for the existence of neutron stars, as pulsars (Gold 1968 and references therein). Pulsars are not marked in the timeline because their potential for demanding tests of gravity physics was seen only with the discovery of the Hulse and Taylor (1975) binary pulsar.

The paper on relativistic collapse by Oppenheimer and Snyder (1939) is important, but not marked because this theory had little to do with the evolving empirical situation in the 1960s. The evidence for the presence of supermassive black holes (or compact objects that act like them) in galactic nuclei was developing at this time. An early step was Burbidge's (1959) estimate of the  energy in plasma in the radio sources in the most luminous radio galaxies, about $10^{60}$ ergs. This is equivalent to the annihilation of about $10^6$\,M$_\odot$, which seemed excessive for energy production in stars. Schmidt's (1963) discovery of quasars added another energy problem, the source of the large optical luminosities of these compact objects. This forced community attention to the issues of gravitational collapse and black hole physics, as seen in the title of the 1963 Texas Symposium on Relativistic Astrophysics: {\it Quasistellar Sources and Gravitational Collapse} (Robinson, Schild, and Schucking 1965). Lynden-Bell (1969) reviewed the proposal that the energy source for quasars and luminous radio galaxies is relativistic collapse to a massive black hole, and he argued that  black holes remnant from quasar activity may be present in the nuclei of normal galaxies. This is now the paradigm. The discovery of quasars is entered in the timeline (line~29 in Fig.~\ref{Fig:timeline}) to mark its effect on thinking about gravitational collapse. 

The evidence for stellar mass black holes traces from the discovery of  Cygnus~X-1 as an X-ray source (Bowyer, Byram, Chubb, and Friedman 1965), and the evidence from optical identification that it is a binary system with a main sequence primary star and a companion X-ray source with mass at least 3\,M$_\odot$ (Webster and  Murdin 1972; Bolton 1972). That plausibly meant that the companion is too massive to be anything but a black hole. But the companion mass depends on an estimate of the mass of the primary, which depends on its distance, which was uncertain. The evidence now is that Cygnus~X-1 contains a stellar remnant black hole with mass $15\pm 1$\,M$_\odot$ (Orosz, McClintock, Aufdenberg, et al. 2011). Although the stellar mass black hole phenomenon is now well established, and Cygnus\,X-1 was discovered in 1965, the evidence for stellar mass black holes developed late enough not to be marked in the timeline. 

Incremental experimental advances in gravity physics are not marked. Thus the first measurement of the gravitational deflection of light passing close to the Sun (Dyson, Eddington, and Davidson 1920), is shown (line~3), but not the several repetitions, though as Trumpler (1956) emphasized they were very important in making the case that the solar deflection likely had been detected and is within about 10\% of the predicted value. Trumpler reported that two white dwarf stars have measured redshifts consistent with predicted gravitational redshifts. The theory and observation for one of the white dwarfs are within 15\% of present measurements, and they are consistent with general relativity. But the measured and predicted redshifts of the other white dwarf, although consistent with each other, both are one quarter of present results. Section~\ref{sec:gravredshift} reviews this situation. For the timeline it seems to be more appropriate to mark St. John's (1928) case for detection of the gravitational redshift of light from the Sun (line~6 in Fig.~\ref{Fig:timeline}). Although turbulent motions caused the measured redshift to vary across the face of the Sun, and to vary with atomic weight, the measurements generally were within about 30\% of the prediction. I  also mark Brault's (1962) solar redshift measurement, in line~24,  because his strategy of solar line choice, and his much improved ability to measure the line shape and line center, successfully suppressed the systematic variation across the face of the Sun, which considerably improved the case for detection of the gravitational effect. The Pound and Rebka  (1960) laboratory detection of the gravitational redshift is marked too (in line~19), because although its formal uncertainty was about twice Brault's the hazards of systematic errors in the two measurements were quite different. The consistency makes a very good case that the gravitational redshift effect really was detected by 1962, and determined to be within about 10\% of the relativistic prediction.

The following review of gravity experiments offers still more explanations of this timeline. Will's (1993) analysis of experimental  developments presented a timeline running from 1960 to 1980, in a useful complement to Figure~\ref{Fig:timeline}. Another's selection of entries for Figure~\ref{Fig:timeline} certainly could differ, but we may be sure it would show the strikingly abrupt change in density of entries in the late 1950s, at the naissance, which is the point of this figure. 

\begin{table}[ht]
\centering
\begin{tabular}{ r  l l}
\multicolumn{3}{c}{Table 1: Aether Drift Constraints}\\
\noalign{\medskip}
\toprule
\toprule
  timeline$^{\rm a}$ & experiment & bound$^{\rm b}$ \\
\noalign{\smallskip}
\toprule
---  \quad\  & isotropy of the velocity of light (1887) & $\epsilon\la 0.01$ \\
  5. \quad\  & static electromagnetic field (1926) &  $\epsilon\la 0.01$ \\
  8.  \quad\ & 5461\AA\ atomic mercury line (1932) & $\epsilon\la 0.03$ \\
 12. \quad\   & molecular hydrogen ion H$\beta$ line (1938) & $\epsilon\la 0.1$ \\
 17.  \quad\  & 24 GHz ammonia molecule inversion transition (1958) \qquad & $\epsilon\la 10^{-3.5}$\\
 21.  \quad\  & Zeeman splitting in the Li$^7$ atomic nucleus  (1960)  & $\epsilon\la 10^{-16}$\\ 
 27.  \quad\ &  14.4\,keV nuclear gamma-ray line (1962) & $\epsilon\la 10^{-4.5}$\\
 30.  \quad\ &  annual variation of gravitational acceleration (1962) & $\epsilon\la 0.1$ \\
\noalign{\smallskip} 
\bottomrule
\noalign{\smallskip}
\multicolumn{3}{l}{\footnotesize $^{\rm a}$Line number in Fig.~\ref{Fig:timeline}.}\\
\multicolumn{3}{l} {\footnotesize $^{\rm b}$Bound on the coupling parameter in the models in eqs. (\ref{eq:veff}) to (\ref{eq:KT})}\\ 
\multicolumn{3}{l} {\footnotesize\  for the aether drift velocity in eq. (\ref{eq:Michelson}).}\\
\end{tabular}
\end{table}
\subsection{Tests for preferred motion}\label{sec:preferredmotion}
The experiments reviewed in this subsection test whether the time kept by a local clock, such as transitions in a molecule, or atom, or atomic nucleus, or the behavior of electromagnetic fields, might be affected by position or motion, in a violation of the Strong Equivalence Principle. The violation would of course have to be more subtle than the Galilean transformation that was ruled out by the demonstration that the velocity of light is close to isotropic.  Some of these experiments were motivated by thoughts of an aether. Others might be attributed to the Machian thoughts discussed in Section~\ref{sec:Mach}, that the disposition and motion of all of the matter in the universe might define a preferred position or motion manifest in local physics. 

The results of measurements summarized in Table~1 are expressed as bounds on a parameter $\epsilon$ that is meant to be a measure of coupling to an effective aether. For definiteness I take the velocity $v$ of our motion through the aether to be comparable to our velocity relative to the thermal 3K Cosmic Microwave Background (the CMB discussed in Sec.~\ref{Sec:CMB}),
\beq
v \simeq 300\hbox{ km s}^{-1}.\label{eq:Michelson}
\eeq
The definition of  $\epsilon$ has to be {\it ad hoc}, of course, because our motion through an effective aether need not be comparable to our motion through the CMB, and, more important, because we do not have a viable theory of $\epsilon$. 

The first entry in the table, from before the start of the timeline, is the Michelson-Morley bound on the anisotropy of the velocity of light,\footnote{Michelson 1903, in his book {\it Light waves and their uses}, reviewed the situation, with the memorable conclusion that ``The theory may still be said to be in an unsatisfactory condition.''} which in a Galilean transformation bounds the effective motion relative to the aether to $v_e\la 4$~km~s$^{-1}$. The coupling parameter here is defined by 
\beq
v_e=\epsilon v,\label{eq:veff}
\eeq
with $v$ in equation~(\ref{eq:Michelson}). This implies the Michelson-Morley bound, $\epsilon\la 0.01$, entered in the last column of the first line of the table. 

Jaseja, Javan, Murray, and Townes (1964) compared frequencies of two neon-helium masers (in which helium pumps population inversion for neon emission lines at wavelength $\sim 10^4$\AA) placed in orthogonal directions on a platform that can rotate. The stimulated emission frequency is defined by the mirror separation, so the Galilean argument for the Michelson-Morley experiment applies here: the relative frequency difference is expected to be $\delta\nu/\nu = {1\over 2}(v_e/c)^2\cos\theta$, by the Galilean argument. The bound, $\delta\nu/\nu\la 10^{-11}$, on the effect of rotation of the system, so as to point it in different directions relative to an aether drift, indicates $v_e\la 1$~km~s$^{-1}$, a factor of three better than Michelson-Morley. The older method did so well with a much more poorly defined frequency because the same wave, or photon, sampled the path difference. 

The second entry in the table is Chase's (1926) revisit of a pre-relativity expression for the magnetic field $\vec B$ induced by uniform motion of a charge distribution at velocity $\vec v$ relative to the aether. A Galilean transformation of a static charge distribution and electric field produces magnetic field
\beq
\vec B = \epsilon\,\vec v\times \vec E/c. \label{eq:Chase26}
\eeq
With $\epsilon=1$ these electric and magnetic fields are a solution to Maxwell's equations.\footnote{Consider a static charge distribution $\rho$ with electric field that satisfies $\nabla\cdot\vec E=4\pi\rho$ and $\nabla\times\vec E=0$. To represent uniform motion through the aether at velocity $\vec v$ let the charge density be $\rho =\rho(\vec r-\vec v t)$, with current density $\vec j = \rho\vec v$ and $\vec E =\vec E(\vec r-\vec v t)$. Then equation~(\ref{eq:Chase26}) with $\epsilon=1$ satisfies Maxwell's equations.}  We  know now that this solution does not represent a static situation, of course, but the constant $\epsilon$ in equation~(\ref{eq:Chase26}) allows a measure of the possible coupling of the local electromagnetic field to Chase's effective  ``stationary ether"  determined by the rest of the matter in the universe. Equation~(\ref{eq:Chase26}) would indicate that a charged capacitor moving at velocity $\vec v$ relative to the aether induces a magnetic field whose energy depends on the angle between $\vec E$ and $\vec v$, producing a torque to minimize the energy. Electromagnetic energies are large enough that Chase's torsion balance yielded an impressively tight constraint: the velocity of the experimental apparatus relative to the local stationary ether has to be $v\la 4$~km~s$^{-1}$ if $\epsilon=1$. Motion relative the stationary ether defined by the CMB (eq.~[\ref{eq:Michelson}]) gives the bound on $\epsilon$ in the last column of the table. The first column is the line number in the timeline in Figure~\ref{Fig:timeline}. 

Kennedy and Thorndike (1932) sought to complete  the argument for special relativity from the Michelson-Morley experiment by testing whether the time kept by a source of light  for an optical interferometer might vary with the motion of the source relative to some preferred frame (while the interferometer arm lengths might be unchanged,  or perhaps change in some other way). Their interferometer with unequal arm lengths could then show a variation of the interference pattern as the motion of the light source changed. The frequency of the light source as a function of velocity $\vec w$ of the mercury atoms relative to an effective aether may be modeled as
\beq
\nu = \nu_o(1\pm\epsilon\, w^2/c^2), \qquad \vec w = \vec u + \vec v, \label{eq:model}
\eeq
where $\vec u$ is the circumferential velocity of the interferometer due to the rotation of the Earth,  $\vec v$ is the velocity of the Earth relative to the aether, and the parameter $\epsilon$ replaces the factor $1/2$  in the Kennedy-Thorndike model (their eq.~[5]). The diurnal fractional frequency shift Kennedy and Thorndike looked for is then 
\beq
{\rm\Delta \nu\over\nu} = {2\epsilon\,u\,v\over c^2}\cos\,\phi, \label{eq:KT}
\eeq
where $\phi$ is the angle between the two vectors (with the sign in eq.~[\ref{eq:model}] absorbed in $\phi$), and the terms that are nearly isotropic for purpose of the diurnal measurement are ignored. It is worth pausing to consider the numbers entering their result. The light was the spectral line of atomic mercury  at wavelength  $\lambda = 5461\,$\AA. The difference of path lengths in the two arms was $\rm\Delta s\sim 30$\,cm, which translates to $n = \rm\Delta s/\lambda\simeq 6\times 10^5$ wavelengths. The measured bound on the diurnal fringe shift was $\delta n < 6\times 10^{-5}$, which gives $\delta n/n\la 10^{-10}$. We are considering the model $\delta n/n=\delta\nu/\nu$. The circumferential (Earth rotation) speed is $u\sim 0.3$\,km\,s$^{-1}$. These numbers in equation~(\ref{eq:KT}), more carefully used, give the Kennedy-Thorndike bound on the velocity of the Earth relative to the aether, $v<(24\pm 19)/\epsilon$~km~s$^{-1}$. With the adopted Earth velocity (eq.~[\ref{eq:Michelson}]) this translates to the bound on $\epsilon$ in the third line in the table. It depends on the impressively tight bound on the fringe shift $\delta n/n$.

 Ives and Stilwell (1938) tested the second-order transverse Doppler shift in a beam of hydrogen consisting largely of the molecular hydrogen ion H$_2^+$ and the heavier ion with three protons bound by two electrons. They measured the longitudinal and transverse shifts in the analog of the H$\beta$ line in the molecules at $\lambda\sim 5000$\,\AA. The measured transverse Doppler shifts in beams moving in the North, South, East or West directions, were found to differ by no more than about $\rm\Delta\lambda\sim 0.003$\,\AA. This bounds the fractional shifts in times kept by the ions to $\rm\Delta \nu /\nu\la 10^{-6}$, for ions moving at beam velocities $u\sim 10^3$\,km\,s$^{-1}$. The bound on $\epsilon$ from equation~(\ref{eq:KT}) is entered in the fourth row in the table. Recent precision tests of the relativistic Doppler effect by variants of the Ives and Stilwell experiment considerably improved this bound. Botermann, Bing,  Geppert, et al. (2014) reported that measurements of  frequency shifts of hyperfine structure lines of Li$^+$ ions moving at speed $w/c=0.3$ agree with the relativistic prediction to a few parts in $10^9$. Since they do not report problems with reproduceability as the Earth rotates and moves around the Sun we may take it that their bound in equation~(\ref{eq:model}) bounds the coupling parameter to  $\epsilon\la 10^{-8}$ for the hyperfine structure lines. 

Cedarholm, Bland, Havens, and Townes (1958) used ammonia beam masers for an aether drift test based on M\o ller's (1956, 1957) considerations of how atomic clocks might test relativity. The experiment compared the frequencies of two masers with ammonia beams that moved in opposite (antiparallel) directions at beam velocity $u\simeq 0.6$\,km\,s$^{-1}$. The fractional change of frequency when the system was rotated by $180^\circ$ was bounded to $\rm\Delta \nu/\nu\la 10^{-12}$, in trials repeated for a year (Cedarholm and Townes 1959). M\o ller (1957) and Cedarholm et al. (1958) argued for the expression in equation~(\ref{eq:KT}), without the parameter $\epsilon$, from a consideration of the Doppler effect on radiation reflected by the walls of a resonant cavity moving through an effective aether. But it seems best to follow Turner and Hill (1964) in going directly to the model in equation~(\ref{eq:KT}), including $\epsilon$, as a fitting function. Here the bound on $\rm\Delta \nu/\nu$ translates to the bound on $\epsilon$ entered in the fifth row of the table.

Cocconi and Salpeter (1958) wrote that, ``If Mach's Principle holds, we might then expect that the slight asymmetries in the distribution of matter at large would result in slight deviations from at least some of the laws of mechanics and gravitation which are commonly assumed to be exact,'' and could produce a ``diurnal variation in the period of a quartz crystal (or a pendulum) clock.''  They did not refer to similar arguments by Sciama (1953) and Dicke (1957b), or the pendulum experiment Dicke (1957a) mentioned at Chapel Hill, but it is easy to imagine Cocconi and Salpeter were thinking along the lines of one of the many other trails of thought tracing back to Mach's arguments. They went on to consider the possibility that the electron inertial mass is slightly different for motions transverse and parallel to a preferred direction set by the large-scale distribution of matter, and pointed out that an inertial mass anisotropy would cause the Zeeman splittings between atomic levels with neighboring magnetic quantum numbers, $m$, to differ, depending on the orientation of the magnetic field relative to some preferred direction, because the patterns of electron motions relative to the magnetic field differ for different values of $m$. Cocconi and Salpeter (1960) suggested that the test could be even more sensitive if the Zeeman splittings were probed by the M\"ossbauer (1958) effect. Hughes, Robinson, and Beltran-Lopez (1960), Drever (1960, 1961), and Virgilio Beltran-Lopez (1962, PhD Yale) turned to a still more sensitive test by nuclear magnetic resonance measurements of Zeeman splittings of the four levels of the $I = 3/2$ spin of the atomic nucleus of Li$^7$. In a simple model of a P$_{3/2}$ neutron in the potential well of the six inner nucleons, the constraint on anisotropy of the nucleon inertial mass was found to be $\delta m/m < 10^{-20}$. This impressed Dicke; it is the subject of my first paper with him (Peebles and Dicke 1962a). In the model for $\delta m/m$ expressed in the form of  equation~(\ref{eq:model}), with the nucleon velocity approaching a few percent of the velocity of light, the bound on coupling to an effective aether is in the fifth row in Table~1. 

Two groups independently tested for preferred motion using M\"ossbauer's (1958) discovery and explanation of the very narrow nuclear $\gamma$-ray absorption line spectrum produced when the recoil momenta of the nucleus emitter and absorber are taken up by the crystal lattices rather than by the atomic nuclei. In the report of their experiment, Champeney, Isaak, and Khan (1963) referred to Ruderfer (1960), who proposed placing the $\gamma$-ray source at the center of a turntable and the M\"ossbauer absorber near the edge. The origin of the other  experiment, at Princeton, may be traced back to a November 1959 letter from Dicke to Robert Pound at Harvard. Pound (2000) recalled that Dicke wrote
\begin{quotation}
\noindent I note from your recent note in the Physical Review Letters that we have been
inadvertently treading on each otherÕs research. For the past couple of months
Ken Turner has been working full time on the very problem you discuss.
\end{quotation}
The problem Dicke mentioned in this letter was a laboratory detection of the gravitational redshift by means of the M\"ossbauer effect. Pound recalled that Dicke had considered using a silver isotope that Pound and colleagues, with more experience in condensed matter, felt was not likely to produce a usefully  strong absorption line, and that Dicke sent his graduate student, Turner, to Pound to learn the technology. Dicke (1963 p. 187) recalled that 
\begin{quotation}
\noindent It started out about September 1959 as an attempt to measure the gravitational red shift using the M\"ossbauer effect which had just been discovered, but it soon became apparent that there were two other groups working on this problem, and to avoid a horse race it was dropped about November in favor of the one to be described.
\end{quotation}
The experiment ``to be described'' was  by  Kenneth Turner (1962, PhD Princeton). Champeney et al. and Turner both used a Co$^{57}$ $\gamma$-ray source and Fe$^{57}$ absorber. Turner placed the source near the rim of a standard centrifuge wheel and the absorber and detector near the axis. The conclusions from the two experiments are that our effective velocity in Earth's equatorial plane relative to the effective aether is limited to $\epsilon v=160\pm 280$~cm~s$^{-1}$ (Champeney et al. 1963) and $\epsilon v=220\pm 840$~cm~s$^{-1}$  (Turner and Hill 1964). These results in the model in equation~(\ref{eq:veff}) are summarized in the seventh row in the table.

Dicke's interest in the idea that gravity may be affected by motion relative to a preferred frame led to the experiment he mentioned at the 1957 Chapel Hill Conference (and is discussed in Sec.~\ref{sec:GravityGroup}), to check whether the gravitational acceleration $g$ at a fixed position on Earth might vary as the Earth moves around the Sun or as the universe expands. This was the subject of the doctoral dissertation experiments by William Hoffmann (1962, PhD Princeton) and David Curott (1965, PhD Princeton).  Hoffmann concluded ``that the amplitude of any annual variation of the gravitational constant, is less than 4 parts in $10^8$,'' which is comparable to what may be inferred from other $g$ measurements  and ``has considerable promise for accurate g measurements.'' Curott reported ``a frequency increase of $1.7\pm .4$ parts in $10^9$ per day,'' but this tentative indication of a detection did not pass later tests. The fitting function in equation~(\ref{eq:KT}), expressed as the fractional variation of $G$ as the Earth moves around the Sun with speed $u=30$\,km~s$^{-1}$, with a conservative constraint from these two experiments, $\delta G/G\la 10^{-7.5}$, limits the parameter $\epsilon$ in equation~(\ref{eq:KT}) to the value in the eighth row in the table.

The intended point of Table~1, which is meant to be a fair sample of what experimentalists were doing  and the span of dates of the experiments, is the following. The search for tests of special relativity, or the idea of some sort of effective aether, or the Machian considerations in Section~\ref{sec:Mach}, or the elegance of some other holistic world picture that would relate local to global physics, has been persistently interesting enough to enough people to motivate many experimental explorations. All this work has not revealed any departure from the Strong Equivalence Principle. This consideration continues in Section~\ref{sec:fractal}, following discussions of other probes of the Strong Principle. 

\subsection{Time kept by accelerating clocks}\label{sec:acceleration}
The time kept by an accelerated molecule, atom, or atomic nucleus may be affected by mechanical stresses that distort local wave functions and electromagnetic fields, but within broadly accepted ideas these mechanical effects may be computed, or estimated, using standard atomic or condensed matter physics, even when one is considering the possibility that the physical parameters in the computation may depend on position or on velocity relative to some aether. But tests for an intrinsic effect of acceleration on timekeeping must be considered. Rutherford and Compton (1919) briefly reported the test entered in line~2 of the timeline in Figure~\ref{Fig:timeline}: they found no effect on the rate of decay of radioactive material fixed to the edge of a spinning disk. Ageno and Amaldi (1966) presented an edifying review of improvements of this experiment, including their own version.  The timeline marks (in line~20) the great advance in sensitivity afforded by the M\"ossbauer (1958) effect in the centrifuge experiment by Hay, Schiffer, Cranshaw, and Egelstaff (1960). The setup was similar to the anisotropy tests (Turner 1962; Champeney et al. 1963), except that Hay et al. tested the mean transverse Doppler effect. The measured fractional second-order Doppler shift, $\delta\nu/\nu\sim 10^{-13}$, was found to agree reasonably well with the relativistic $v^2/2c^2$ prediction. This means the fractional shift in the intrinsic atomic nucleus clock rate due to its acceleration must be well below a few parts in $10^{13}$ at the largest experimental acceleration, $6\times 10^7$\,cm\,s$^{-2}$. The characteristic relativistic acceleration, $c^2/r\sim 10^{33}$\,cm\,s$^{-2}$, defined by the radius $r$ of the nucleus is much larger, however.   

\subsection{Gravitational redshift, relativistic timing, and tired light}\label{sec:gravredshift}
Trumpler (1956) mentioned the test of gravitational redshift in two white dwarf stars (in multiple systems, so that the radial velocities of the companions yield the correction for the motions of the systems). He reported that the observed and predicted redshifts of Sirius\,B are $v_{\rm obs}=19$\,km~s$^{-1}$ and $v_{\rm pred}=20$\,km~s$^{-1}$. This was without attribution, but Adams (1925) and Moore (1928) both measured $v_{\rm obs}=21$\,km~s$^{-1}$,  close to Trumpler's number. These two measurements were meant to be independent: Adams used the 100-inch reflector at Mount Wilson, Moore the 36-inch refractor at Mount Hamilton, which might be expected to be differently affected by light scattered from the main sequence companion star Sirius A. Both referred to Eddington for the prediction, $v_{\rm pred}=20$\,km~s$^{-1}$, which apparently was satisfactorily close to the two  measurements. But Greenstein, Oke, and Shipman (1971) argued that the Adams and  Moore 
\begin{quotation}
\noindent spectra were badly  contaminated  by  Sirius A light, and the  results  depended  on measurements  of  metallic  lines, such  as the Mg  II line $\lambda$4481, which  are  now  known  not  to  occur  in  white  dwarfs.  Consequently,  these redshifts are of historical interest only. 
\end{quotation}
Greenstein, Oke, and Shipman, in observations when Sirius A had moved further away from the white dwarf Sirius B, reducing the problem with scattered light, found $v_{\rm obs}=89\pm 16$\,km~s$^{-1}$ and $v_{\rm pred}=83\pm 3$\,km~s$^{-1}$.  (The measurements for Sirius B have not changed much since then: Barstow, Bond, Holberg, et al.\ 2005.) We see that in the 1950s the two measurements of the redshift of Sirius B  were consistent, but both were wrong by a factor of four. They apparently confirmed the relativistic prediction, but it too was wrong, by a like factor. Hetherington (1980) and Greenstein, Oke, and Shipman (1985) debated the meaning of this interesting situation. 

The second white dwarf star Trumpler mentioned is 40\,Eridani\,B, for which Popper (1954) found $v_{\rm obs}=21\pm 4$\,km~s$^{-1}$ and $v_{\rm pred}=17\pm3$\,km~s$^{-1}$. Greenstein and Trimble (1972), making use of the developing art of image intensifiers, found $v_{\rm obs}=23\pm 5$\,km~s$^{-1}$ and $v_{\rm pred}=20\pm9$\,km~s$^{-1}$ for this white dwarf, consistent with Popper.  These numbers agree with the relativistic prediction, and have not changed much since then. 

Trumpler (1956) remarked that ``For more than 30 years I have been working on a program of measuring the radial velocities (Doppler shifts) of stars in galactic star clusters.'' This well-experienced observer did not pause to consider whether the observations of the radial velocities of the two white dwarf stars might be questionable. This is not a criticism of Trumpler, but rather a serious cautionary example of the hazards of empirical science. This discussion continues in Section~\ref{sec:establishment}.

The other early test of the gravitational redshift, St. John's (1928) measurements of the redshift of light from the Sun, was vexed by turbulence and outflows manifest as distortions of absorption line shapes, variations of the measured line shifts from center to limb of the Sun, and systematic variations of the line shifts with the binding energy of the ion, which correlates with the depth of formation of the line in the Solar atmosphere. But the values of the line shifts were roughly in accord with relativity, usually to about 30\%, arguably good enough to be entered in Figure~\ref{Fig:timeline} (line~6), but not a very convincing detection. 

James Brault (1962, PhD Princeton), at Dicke's suggestion, improved the situation. Brault measured the shift of the solar sodium D$_1$ absorption line at 5896 \AA, which is strong enough to allow a tight measurement of the line shape. And the sodium ionization potential is small enough that the line is thought to largely originate above the turbulence in the photosphere but below the chromosphere, in a region where non-gravitational perturbations might be expected to be minimal. Brault used a wavelength modulation technique that he showed stably defined the line center, as follows. The position of the output slit of the spectrometer oscillated at frequency $\omega$, so that the narrow band of wavelengths admitted to the photodetector varied with time as 
\beq
\lambda(t) = \lambda_o + \delta\lambda\,\sin\omega t. \label{eq:BraultScan}
\eeq
The first term on the right-hand side, $\lambda_o$, was adjusted until the photodetector output showed no component at the slit oscillation frequency $\omega$; only the harmonics were detected. The value of $\lambda_o$ at this point defined a measure of the line center. This strategy will be recognized as  similar to the phase-sensitive lock-in amplifier technique used in the Roll, Krotkov, and Dicke (1964) E\"ot\"vos experiment reviewed in Section~\ref{sec:GravityGroup}. Brault probed the line shape by measuring how $\lambda_o$ depended on the scan amplitude $\delta\lambda$ in equation~(\ref{eq:BraultScan}). Brault demonstrated that, in his chosen line and range of scan amplitudes, $\lambda_o$ is insensitive to $\delta\lambda$. This is the wanted signature of a satisfactorily symmetric line shape. He also showed that the line center defined this way is insensitive to position on the Sun, scanning from center to limb. These two results make a reasonable case that the measurements were not seriously affected by nongravitational disturbances. Brault's conclusion was that ``The ratio of the observed red shift to the theoretical value is found to be $1.05 \pm 0.05$.'' 
\begin{figure}[t]
\begin{center}
\includegraphics[angle=180,width=3.in]{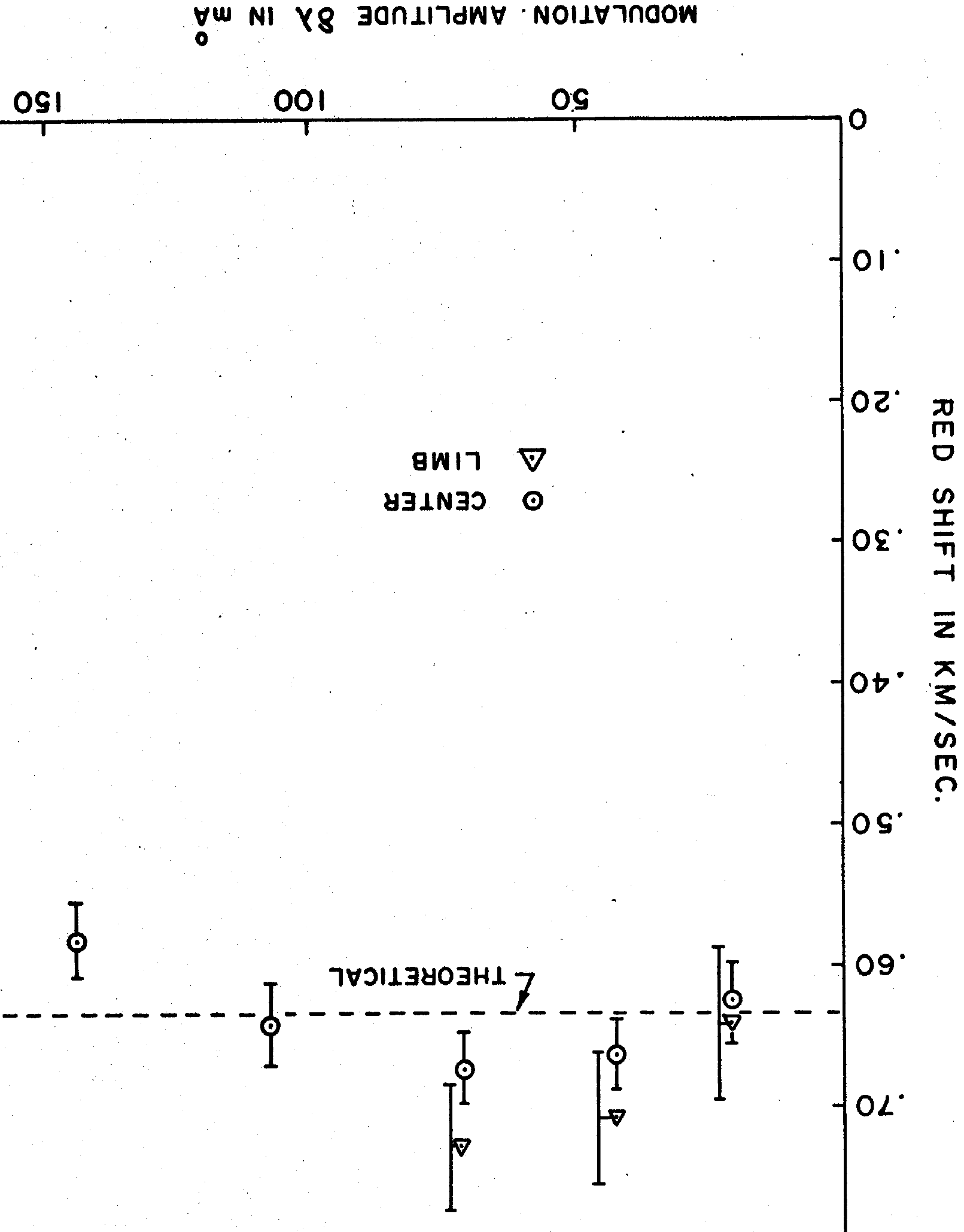} 
\caption{\label{Brault} Brault's (1962) test of the solar gravitational redshift.}
\end{center}
\end{figure}

This beautiful experiment was fully published only in Brault's thesis. Dicke (1963, pp. 189 -- 191) outlined the experiment, and I take the liberty of showing in Figure~\ref{Brault} Brault's (1962) summary figure. The horizontal axis is the amplitude $\delta\lambda$ of the spectrum scan. The vertical axis shows the measured line shifts at solar center and limb; other figures in the thesis showed samples across the full face of the Sun. The figure shows the insensitivity to scan amplitude and position on the Sun that make the case for a reliable test of the relativistic  prediction marked by the dashed line. 

Pound and Rebka (1960) used M\"ossbauer's (1958) effect to obtain a 10\% laboratory detection of the gravitational redshift of $\gamma$-rays falling 23\,m through helium in a tower at Harvard. The precision was comparable to Brault's, but the situation was very different, which is important in the search for systematic errors. Also, the laboratory experiment was under much better control and capable of improvement. Pound and Snider (1964) brought the laboratory precision to $1\%$. 

An even more demanding test of the gravitational redshift and relativistic time-keeping grew out of another great advance in technology, atomic clocks. M\o ller's (1957) early recognition of their promise is discussed in Section~\ref{sec:GRinthe50s}. At the Chapel Hill Conference Dicke (1957a) mentioned work at Princeton on rubidium atomic clocks, in collaboration with Carroll Alley (1962, PhD Princeton). After considerably more development (Mark Goldenberg 1960, PhD Harvard), Vessot et al. (1980) reported a test of gravitational timekeeping by the Gravity Probe A rocket flight of an atomic hydrogen maser to 10,000~km altitude. The timing measurements  agreed with the prediction of general relativity theory  to about a part in $10^4$. 

Irwin Shapiro's (1964) ``fourth test of general relativity,'' which probes another aspect of relativistic timing, used planetary radar astronomy (as reviewed in Pettengill and Shapiro 1965) to check the relativistic prediction of an increase in return times of radar pulses reflected by Mercury or Venus when the line of sight passes close to the Sun. Shapiro (2015) recalled that
\begin{quotation}
\noindent My ``entry'' into testing GR was not influenced as far as I could tell by Bob's broad and well broadcast - in scientific meetings and the like - approach to (re)start this experimental field.  I didn't attend any of these meetings, nor did I read any of the proceedings.  I had the idea to enter it from the prospects for radar astronomy and my knowledge of the Mercury test; my first thought, in the late 1950s, was to check on, and improve, the measurements of the relativistic advances of the perihelia of the inner planets.  From there I went on to think of more than a half dozen different tests, almost all not original with me, save for the radar/radio approach.  All but two of them were eventually carried out by my colleagues, students, and me, and also by others. 
\end{quotation}
Shapiro, Pettengill, Ash, et al. (1968) termed their radar timing measurements ``preliminary,'' but they presented a clear detection of the relativistic prediction of the time delay. This experiment was an elegant addition to the tests of  general relativity. It is the last entry in the timeline in Figure~\ref{Fig:timeline}, line~38; it is taken to mark the end of the naissance of experimental gravity physics. More recent measurements of microwave signals transmitted from the ground to the Cassini spacecraft, retransmitted by the spacecraft, and detected on the ground, as the line of sight to the spacecraft passed near the Sun,  established the relativistic effect of the Sun on the timing of radiation to a few parts in $10^5$ (Bertotti, Iess, and Tortora  2003).

Yet another aspect of relativistic timing is the shift to the red in the spectra of distant galaxies of stars. In the 1950s and earlier it was reasonable to ask, with Zwicky (1929), whether the starlight might have been shifted to the red by some physical process operating along the line of sight, rather than by the expansion of the universe. This question helped inspire Kennedy and Thorndike (1931) to check one conceivable physical effect on the frequency of propagating light: they measured the effect on the frequency of the 5641\,\AA\ mercury line after moving through 50,000 volts potential difference, ``because it has required only a modification of apparatus devised for another purpose,'' with a null result. 

Tolman (1930, eq.~30) pointed out that Zwicky's ``tired light'' model for the cosmological redshift is tested by measuring how the surface brightnesses of galaxies vary with the redshift. Under standard local physics, and assuming the light propagates freely through a spacetime described by a metric tensor, Liouville's theorem tells us that the integrated surface brightness (energy received per unit time, area, and solid angle) varies with the ratio of the observed wavelength $\lambda_{\rm obs}$ of a spectral feature to the laboratory wavelength $\lambda_{\rm em}$ at the source as $i\propto (\lambda_{\rm em}/\lambda_{\rm obs})^4$. One of the four powers of $\lambda_{\rm em}/\lambda_{\rm obs}$ may be attributed to the loss of energy of each photon as its frequency decreases, one to time dilation of the rate of detection of photons, and two powers to aberration of the solid angle of a bundle of radiation. In a simple tired light model in a static universe only the first factor would operate: $i\propto (\lambda_{\rm em}/\lambda_{\rm obs})$. The test is important but the precision is limited by the variable properties of galaxies (Geller and Peebles 1972; Sandage 2010). A much tighter test follows from Tolman's  demonstration that the four factors serve to preserve the form of a thermal radiation spectrum as a homogeneous universe expands and freely propagating radiation cools. The spectrum of the  Cosmic Microwave Background radiation discussed in Section~\ref{Sec:CMB} is quite close to thermal, as shown in Figure~\ref{Fig:CMB}. Since the universe is observed to be optically thin to radiation at these wavelengths,  the tired light model predicts that the CMB spectrum cannot remain thermal as the radiation is redshifted, contrary to the measurements. The tired light model clearly is wrong. The measured spectrum is consistent with freely propagating radiation in a very close to homogeneous expanding universe with standard local physics. 

\begin{figure}[h]
\begin{center}
\includegraphics[angle=0,width=5.in]{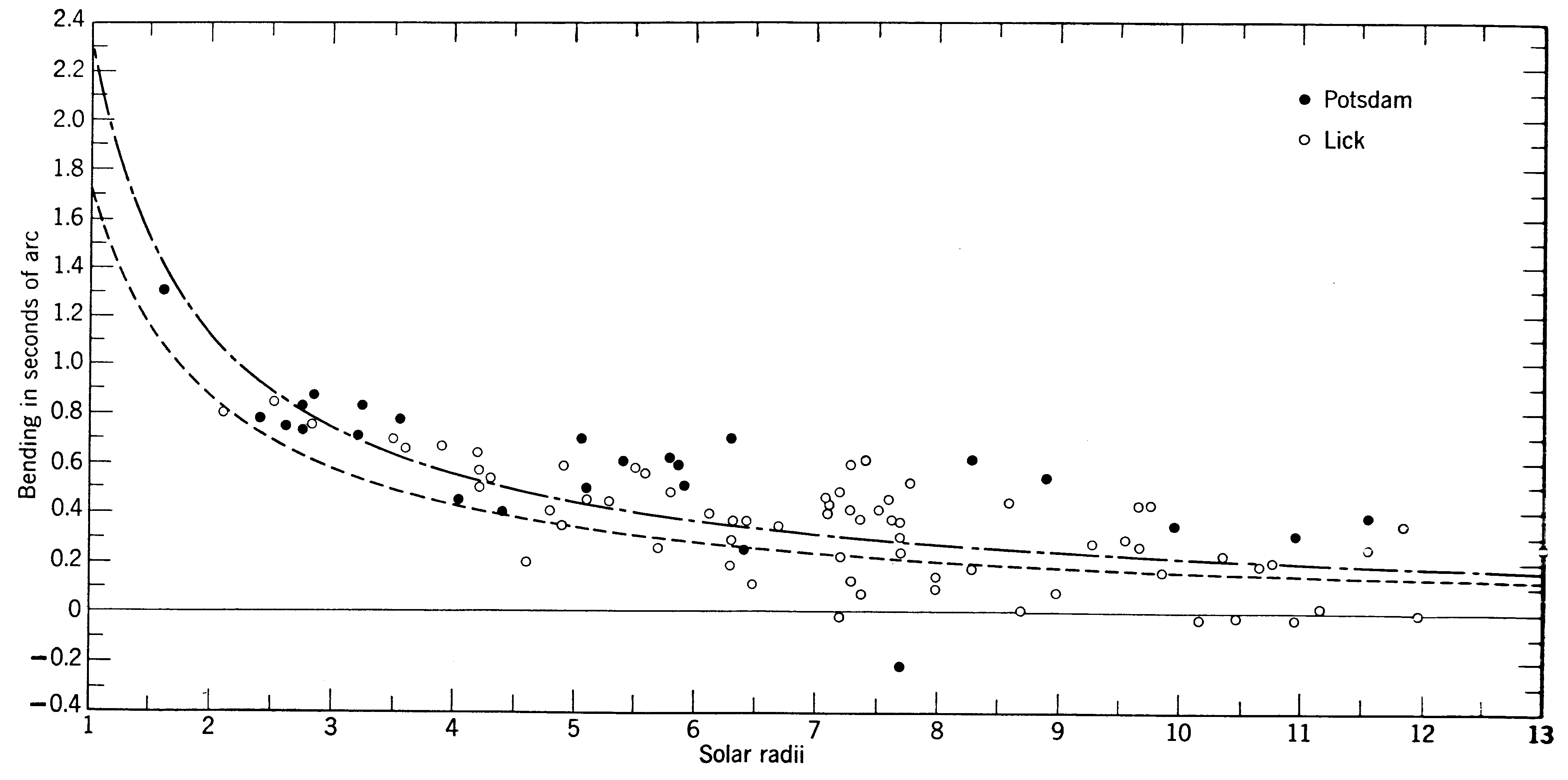} 
\caption{\label{Fig:gravdeflection}  Radial star displacements measured during the 1922 (open circles) and 1929 (filled circles) solar eclipses, assembled and reviewed by Bertotti, Brill and Krotkov (1962). The dashed curve is the relativistic prediction, the dot-dashed curve a fit to the Potsdam data.}
\end{center}
\end{figure}
\subsection{Gravitational lensing and deflection of light}\label{sec:lensing}
Bertotti, Brill and Krotkov (1962) presented a careful review of the state of the tests of general relativity. Figure~\ref{Fig:gravdeflection} shows their summary of the test of the gravitational deflection of light by the Sun for two sets of observations in favorable star fields. In their assessment, the most important uncertainty was the conversion of distances of stars from the Sun on the plates to angular distances from the Sun in the sky, in the eclipse plates and in the comparison dark sky exposures. The conversions differed because of temperature differences and uncertainties in the positions of plates in the telescope. Bertotti et al. concluded that the observed pattern of radial shifts of angular distances during the eclipse ``does not contradict the $1/r$ law predicted by general relativity, but neither does it give much support to such a dependence on distance.'' Indeed, one sees in Figure~\ref{Fig:gravdeflection} that the case for the $1/r$ law rests on the one star closest to the Sun in one of the eclipses.

At the 1955 Bern Conference Trumpler (1956) listed ten measurements of the solar gravitational deflection of light, all with probable errors of about $10\%$, and concluded that ``If one considers the various instruments and methods used and the many observers involved, the conclusion seems justified that the observations on the whole confirm the theory.'' This of course assumes the $1/r$ law. We may conclude that, at the level of these somewhat tepid assessments of the weight of the evidence, general relativity in 1955 had passed two critical tests, from the orbit of the planet Mercury and the observations of deflection of starlight by the Sun. As discussed in Section~\ref{sec:gravredshift}, the third classical test, gravitational redshift, was more doubtful.

Shapiro (1967) pointed out that radio interferometer observations of radio-loud quasars could detect the predicted gravitational deflection. The first results (Seielstad, Sramek, and Weiler 1970; Muhleman, Ekers, and Fomalont 1970) were about as precise as the optical observations but capable of considerable improvement. Indeed, the solar gravitational deflection is now detected in directions over a large part of the celestial sphere (Shapiro, Davis, Lebach, and Gregory 2004). Will (2015) reviewed the history of this test.  

Gravitational deflection causes a sufficiently compact mass to act as a gravitational lens. Renn, Sauer,  and Stachel (1997) reviewed the early history of papers on this effect.  Notable among them was Einstein's (1936) publication of his earlier thoughts about the double images and  increase of apparent magnitude (intensity enhancement when lensing increases the solid angle of the image while preserving the surface brightness) to be expected when one star passes nearly in front of another. Renn et al. reported that Einstein was not optimistic about the possibility of an observation of these lensing effects, but the intensity enhancement produced by lensing by stars and planets is now well observed, and termed microlensing. Zwicky (1937)  assessed the prospects for detecting Einstein's (1936) effects when the gravitational lens is a galaxy, with emphasis on the observationally important effect of intensity enhancement. Zwicky's vision of gravitational lensing by the mass observed to be concentrated around galaxies is now a valuable probe of the mass distribution on larger scales, in the dark matter outside the concentration of stars in galaxies. Zwicky's informed discussion is marked as line~11 in Figure~\ref{Fig:timeline}. 

Klimov (1963),  Liebes (1964), and  Refsdal (1964) independently presented more detailed analyses of the prospects of observing effects of gravitational lensing. Klimov discussed lensing of a galaxy by the mass concentration in a foreground galaxy close to the line of sight, and  took note of the Einstein ring produced at close alignment if the lens is close enough to axially  symmetric.  Liebes and Refsdal mainly discussed lensing by ``point-like'' mass concentrations such as stars, and they emphasized the phenomenon of intensity enhancement that has proved to be so observationally important. Liebes considered the lensing signature of planets around stars, and the possibility of observing lensing of more remote stars in our galaxy by intervening visible stars or dark or faint objects that might contribute to the mass of the Milky Way. This is the line of ideas taken up by the MACHO search for massive dark  objects (e.g.  Alcock, Allsman,  Alves, et al.  2000). Liebes cut a plastic lens that simulated the properties of a gravitational lens (Liebes 1969). A disk source viewed through the lens exhibited dual and ring images; a target placed in the lensed beam of a bright source exhibited the intensity amplification. Refsdal anticipated lensing measures of Hubble's constant $H_o$ and lensing measures of galaxy masses. 

Walsh, Carswell, and Weymann  (1979) were the first to identify an observation of lensing: a double image of a quasar.  Wheeler pointed out in remarks at the 1959 Royaumont Conference (Lichnerowicz. and Tonnelat 1962, pp. 269 -- 271) that lensing produces an odd number of images,  but the odd image in the Walsh et al. observation is demagnified, perhaps obscured by dust in the lensing galaxy, and certainly hard to observe. Gravitational lensing now probes the mass distributions in clusters of galaxies (Hoekstra, Bartelmann, Dahle, et al. 2013) and the mean mass distribution around galaxies (Bahcall and Kulier 2014).

\subsection{Gravitational waves}\label{sec:gravitationalwaves}
Joseph Weber's interest in gravitational waves traces back at least to the analysis by Weber and Wheeler (1957) of the {\it Reality of the Cylindrical Gravitational Waves of Einstein and Rosen} (in a physical system that is translationally invariant in one direction, but not axisymmetric). Although the paper by  Einstein and Rosen (1937) presented an argument for the reality of gravitational waves that carry energy, in analogy to electromagnetic waves, the idea was still controversial. One sees this in Einstein's note at the end of the paper (after a skeptical reception of an earlier version by {\it The Physical Review}), that he had corrected the conclusion ``after the departure of Rosen.'' In his contribution to the proceedings of the 1955 Bern Conference Rosen concluded by endorsing the ``conjecture  \ldots that a physical system cannot radiate gravitational energy.'' The point that gravitational waves certainly carry energy, which can be deposited as work by a gravitational wave acting on a viscous body, was made by Weber and Wheeler (1957), by Feynman (in DeWitt 1957, p. 143) at the Chapel Hill Conference,  and by Bondi (1962), at the Royaumont Conference. Bondi also argued that gravitational waves can be produced by a nongravitational explosion, but  that, for a purely gravitational binary star system, ``I am somewhat doubtful whether such a system will radiate at all.'' (I suspect that this was because Bondi did not consider the effect of radiation reaction on the equation of motion of the point-like stars.) Despite this quite confusing theoretical situation, Weber (1962) presented to the Royaumont conference his practical examination of how to build a gravitational wave detector. This was one of the earliest steps toward the development of experimental gravity physics (and accordingly is entered in line~18 in Fig.~\ref{Fig:timeline}). 

Nancy Roman, who was the first Chief of the Astronomy, Solar Physics, and Geophysics Programs at the NASA Office of Space Sciences, saw the growing possibility and interest in better probes of relativity afforded by space science, and organized the July 1961 NASA {\it Conference on Experimental Tests of Theories of Relativity}. The proceedings (Roman 1961) record Weber's report of progress in building a gravitational wave detector (Weber 1960), and his discussion of the measurements of modes of acoustic oscillation of the Earth, including the interesting quadrupole modes that would be excited by the long wavelength gravitational waves of general relativity (Forward,  Zipoy, Weber, et al. 1961; marked as line~23 in Fig.~\ref{Fig:timeline}). Weber discussed the possibility of placing gravimeters on the Moon, which he expected would be a quieter place to look for the excitation of quadrupole modes of acoustic oscillation by gravitational waves. A decade later the Apollo~17 astronauts placed a Lacoste-type spring gravimeter on the Moon (Giganti, Larson, Richard, and Weber 1973). An unfortunate 2\% miscalculation of the masses needed to trim the balance to the gravitational acceleration at the position where the gravimeter was placed on the Moon prevented operation at design sensitivity. Interest in this approach continues: Lopes and Silk (2014) analyzed the possibility of detecting gravitational wave excitation of quadrupole acoustic oscillations of the Sun. 

Weber (1969, 1970)  reported evidence that his bar detectors had found gravitational waves, based on coincident detection of events: unusual departures from the mean noise fluctuations in bar detectors separated by 1,000\,km, at dimensionless strain estimated to be $h\sim 10^{-16}$. This attracted considerable interest from theorists and experimentalists. The complaint that the indicated gravitational wave strain corresponds to an unreasonably large energy density certainly was worth noting, but the central issue was whether the events were real, perhaps signaling the effect of gravitational waves in some better theory, or perhaps signatures of some other new phenomenon. Tyson and Giffard (1978) reviewed several independent experiments that did not confirm Weber's event rates. But the neutrinos detected from supernova 1987A showed an interesting correlation with Weber Bar events in detectors in Maryland and Rome (Aglietta, Badino, Bologna, et al. 1989). It is difficult to find a community consensus of what this might mean. In a review of Weber Bar detectors, Aguiar (2011) wrote: ``Did the bars detect gravitons from SN1987A or some other particles that excited the bars by thermoelastic processes? In any case, we hope that another supernova will solve this problem.''

In the scalar-tensor gravity theory there could be observable effects of temporal or spatial variations of the scalar field value that determines the local strength of gravity. Morgan, Stoner, and Dicke (1961) searched for an annual periodicity  of earthquakes that might have been triggered by an annual variation in the scalar field value as the Earth moves around the Sun. Their conclusion was that ``The occurrence of this periodicity would be understandable if the gravitational constant were to vary as the earth-sun distance changes or as Earth's velocity relative to  a  preferred coordinate frame changes; however, the observed periodicity cannot be interpreted as conclusive support for such a hypothesis.''  A scalar wave could excite the Earth's 20-minute breathing mode (the high-Q nearly spherically symmetric mode with no nodes in the radial function). Jason Morgan (1964, PhD Princeton) looked for geophysical, lunar and planetary indications of this effect, without significant detection. 

Dicke's interest in the possibility of a laboratory detection of the low-frequency modes of oscillation of the Earth that might be driven by long wavelength tensor or scalar gravitational waves (as discussed by Forward, Zipoy,  Weber, et al. 1961) led to projects in his group to build or modify LaCoste-type (spring) gravimeters, with sensitivity increased by making use of the precision detection of the test mass deflection afforded by the phase-sensitive lock-in amplifier techniques discussed in Section~\ref{sec:GravityGroup} (Robert Moore 1966, PhD Princeton; Weiss and Block 1965;  Block  and  Moore 1966). This work contributed to the creation of  networks for low-frequency seismology that produced detailed detections of Earth's low-frequency modes of oscillation. Jonathan Berger (2016) recalled that, when he was a graduate student, 
\begin{quotation}
\noindent Barry Block came to IGPP (Walter Munk's institution) at UCSD in 1965 (or 1966) followed shortly thereafter by Bob Moore.  Bob brought with him 2 (I think) modified LaCoste gravimeters which he had developed for the thesis work at Princeton and afterwords at U. Maryland.  At the same time, Freeman Gilbert and George Backus (also at IGPP) were developing the theory and mathematical methods for inverting normal mode data to resolve details of Earth structure.  These instruments soon produced some spectacular observations of the Earth's normal modes from relatively frequent earthquakes that whetted appetites for more such data.
\end{quotation}
This grew into the project, International Deployment of Accelerometers, with Berger as director, that is now part of the Global Seismographic Network. It yields probes of the internal structure of the Earth and measures of earthquakes, storm surges, tsunamis, and underground explosions.\footnote{As discussed in the web sites for Project IDA, at \url{http://ida.ucsd.edu/}, and  the Global Seismographic Network, at \url{https://www.iris.edu/hq/programs/gsn}.} 

Let us pause to consider how the analysis by Forward,  Zipoy, Weber, et al. (1961) of the possible effect of low frequency gravitational waves on Earth's modes of oscillation helped interest the experimental gravity community in gravimeters, which aided the development of global seismology, which detected low frequency Earth oscillations, which Boughn, Vanhook, and O'Neill (1990) turned into further exploration of the bounds on the energy density in long wavelength gravitational waves. And let us consider also that the search for detection of gravitational waves may be dated to have begun with Weber's paper in the proceedings of the 1959 Royaumont Conference. A quarter of a century later precision timing showed that the Hulse-Taylor binary pulsar system is losing energy at the rate expected from radiation of gravitational waves (Taylor and  Weisberg 1982). A quarter of a century after that LIGO detected gravitational waves (LIGO Scientific Collaboration and Virgo Collaboration 2016), completing Weber's vision. 

My impression is that the very qualities that Weber needed to pioneer this difficult science made it exceedingly difficult for him to deal with critical reactions to his early results. Weber in the early 1960s impressed me for his great determination, his indifference to experts who were not sure these waves even exist, his love of the chase for a wonderful phenomenon, and his energetic accounts in seminars at Princeton on how he was building and instrumenting his detectors. To be noted also is his checks of significance of signals by coincidences in detectors first separated by a few kilometers, then 1000\,km. The near coincidence of events in the well-separated LIGO interferometers was a key element in the first convincing gravitational wave detection (LIGO et al. 2016).

\subsection{Masses: active, passive, inertial, and annihilation}\label{Sec:masses}
The equations
\beq
F = m_ia,\qquad  F = m_ig = {GM_am_p\over r^2}, \qquad E = m_ec^2, \label{eq:kindsofmass}
\eeq
define four masses to be assigned to an object. In the first equation $F$ is the force, mechanical or gravitational, on an object with inertial mass $m_i$ moving with acceleration $a$. In the second equation $m_p$ is the passive gravitational mass of the object that is moving with acceleration $g$ due to the gravitational force of attraction by a second body with active gravitational mass $M_a$ at distance $r$. In the last equation $m_e$ is the mass defined by the annihilation energy $E$. These definitions follow Bondi (1957), who was largely concerned with the possibility of negative mass. (At the time some wondered whether  antimatter falls up, as in Schiff 1958.) To be discussed here is the empirical situation in nonrelativistic physics, where in the standard model the four masses are equal. (The situation is more complicated in relativistic situations. In general relativity an ideal fluid with mass density $\rho$ and pressure $p$, with $c=1$, has active gravitational mass density $\rho + 3p$, inertial and passive gravitational mass densities $\rho + p$, and energy density $\rho$.)

The E\"otv\"os experiment discussed in Section~\ref{sec:GravityGroup} tests whether the ratios $m_i/m_p$ of inertial to passive gravitational masses, and hence gravitational accelerations, may depend on the natures of compact nearly massless test particles. E\"otv\"os et al. (1922) bounded differences of gravitational accelerations among a considerable variety of materials to about 1 part in $10^{8}$ (line~4 in Fig.~\ref{Fig:timeline}). The static experiment in Dicke's group, with the synchronous detection scheme illustrated in Figure~\ref{Fig:EotvosDetectors} at the heart of the instrument, found that the gravitational accelerations of gold and aluminum toward the Sun differ by no more than 3 parts in $10^{11}$ (Roll, Krotkov, and Dicke 1964; line~31 in Fig.~\ref{Fig:timeline}).  Braginski{\v i} and Panov (1971) found that the fractional difference of gravitational accelerations of aluminum and platinum is less than about 1 part in $10^{12}$.  The present bound reaches parts in $10^{13}$ on fractional differences of gravitational accelerations toward the Earth and Sun (Adelberger, Gundlach, Heckel, et al. 2009). 

A difference in the ratio $m_a/m_p$ of active to passive gravitational masses for different materials is in principle detectable as a difference in the measured values of Newton's constant $G$ in Cavendish balance experiments using different materials, but precision measurements of  $G$ are difficult.  In Kreuzer's  (1968) survey of the literature measurements of  $G$, using materials ranging from glass to mercury, scatter by a few parts in $10^3$, indicating a similar bound on fractional differences of $m_a/m_p$ among the elements. To improve the constraint Lloyd Kreuzer (1966, PhD Princeton; Kreuzer 1968; line~34 in Fig.~\ref{Fig:timeline}), floated a solid body in a fluid at near neutral buoyancy, meaning the passive masses of the body and the fluid it displaced were nearly the same. If the active masses of the body and the displaced fluid were sensibly different it would be detected as a change in the gravitational attraction of a Cavendish-type balance as the body was moved through the fluid toward and away from the balance. The inevitable departure from exact neutral buoyancy was measured by the force needed to support the body in the fluid as the fluid mass density was adjusted, with the balance response extrapolated to zero support. Kreuzer concluded that the ratios of active to passive gravitational masses of bromine (which dominated the mass of the fluid) and fluorine (which dominated in the solid) differ by less than about 5 parts in $10^5$. (The reanalysis of Kreuzer's data by Morrison and  Hill 1973 confirms Kreuzer's conclusion.) 

If active and passive gravitational masses were not equivalent then, as Kreuzer (1966) discussed, a dumbbell consisting of two different massive bodies held at fixed separation by a rod would accelerate in the absence of any external force. This is a situation we are conditioned to reject but must consider, on the principle that unexamined assumptions in science ought to be challenged. Following Kreuzer's experiment, Bartlett and Van Buren (1986) pointed out that, if the composition of the Moon were not spherically symmetric, as suggested by a difference between the centers of figure and mass of the Moon, then the departure from equivalence of active and passive mass would contribute to the acceleration of the Moon. Lunar Laser Ranging measurements place a tight bound on this effect (Sec.~\ref{sec:LLR}). 

As a graduate student, at Dicke's invitation, I pored over measurements of binding energies of atomic nuclei and nuclear spectroscopy mass measurements, to test the equivalence of mass and binding energy, but without publishing. Much more recently Rainville, Thompson, Myers, et al. (2005) and Jentschel, Krempel, and  Mutti (2009) examined budgets of energy and mass in radiative nuclear transitions. The $\gamma$-ray wavelength was measured by crystal Bragg spectroscopy, and converted to annihilation mass by standard values of $\hbar$ and $c$. The inertial mass difference was measured by  ``cyclotron frequencies (inversely proportional to the mass) of ions of the initial and final isotopes confined over a period of weeks in a Penning trap'' (from Rainville et al. 2005). The conclusion from these impressively precise measurements is that the inertial and annihilation energy masses agree to a few parts in $10^7$. 

\subsection{Gravitational frame-dragging}\label{sec:LenseThirrng}
Michelson and Gale's (1925) paper on {\it The Effect of the Earth's Rotation on the Velocity of Light}   is notable for the heroic size of the Sagnac interferometer, $2010\times1113$ feet (640 by 320 meters). Earth's rotation was detected, but that is a long way from detecting the relativistic prediction of frame-dragging (eq.~[\ref{eq:LenseThirrng}]). Pugh (1959), and Schiff (1960) in collaboration with experimentalists at Stanford, realized that advances in artificial satellite technology offered a realistic possibility of detection of frame-dragging. The demanding condition for detection is that the drag on the satellite by light pressure and the last traces of the atmosphere be canceled well enough to suppress torques on the gyroscopes. This was discussed at length at the NASA Conference (Roman 1961). This   prompted marking the origin of Gravity Probe~B in Figure~\ref{Fig:timeline} (line~22) at 1961. The successful completion a half century later, in a satellite in polar orbit, detected the predicted geodetic precession in the plane of the orbit, and it detected the inertial frame-dragging normal to the orbit to be expected from the rotation of the Earth. The frame-dragging was measured to be  $37.2\pm 7.2$ milli arc seconds per year, consistent with the relativistic prediction (Everitt 2015 and the following 20 papers on Gravity Probe B).  Ciufolini and Pavlis (2004) found a roughly matching result by analysis of the precision tracking of two high-density LAGEOS Earth satellites, which do not have provision for cancellation of atmospheric drag, but are dense enough to allow a reasonably secure correction.
 
 \subsection{Absolute measurements of gravitational acceleration and the
  gravitational constant}\label{sec:abs-g}
Precision measurements of the absolute value of the gravitational acceleration $g$ have practical applications, as in monitoring ground water level changes, and some that are purely curiosity-driven, as in probes of the Machian ideas reviewed in Section~\ref{sec:Mach}. In the 1950s measurements of the absolute value of $g$ (expressed in established standards of length and time, with surveys that transferred the value of $g$ from one place to another) used carefully designed pendulums or freely falling objects timed by photoelectric detection of intersections of light beams. In his PhD thesis James Faller (1963, PhD Princeton) described origins of his interferometer measurement:
\begin{quotation}
\noindent About six years ago, it was suggested that an interferometric method might also provide a better approach to the problem of an absolute ``g'' determination$^{12}$. In particular, the falling object might be one plate  of an interferometer. The work, probably due to the advent of inertial navigation, was never brought to a conclusion as the particular interest at that time was concerned with the possibility of navigating submarines gravitationally. It is, however, the suggestion of employing an interferometer in order to make an absolute ``g'' determination that has been taken up and made use of in the experiment described here. 
\end{quotation}
Faller's reference 12 in this quote reads ``J. G. King and J. R. Zacharias of M.I.T.'' King\footnote{Interview of John G. King by George O. Zimmerman on 2009 November 18, Niels Bohr Library \& Archives, American Institute of Physics, College Park, MD USA} recalled that
\begin{quotation}
\noindent We wanted to take two mirrors and make a Fabry-Perot interferometer out of them, and this is pre-laser, so you do it with collimated light, filtered. And now the upper plate is held up by electrostatic field, so you shut the field off and it drops so suddenly that it can't wiggle sideways, and it falls down, and the interference fringes go swittt, and you measure that, and that gives you g. Interestingly, this was going on I think at Hycon Eastern, a small local company.
\end{quotation}
In his thesis Faller emphasized the great problem with this plan: the interference pattern in a Fabry-P\'erot interferometer is exceedingly sensitive to the relative orientation of the plates, and if one of the plates is falling its orientation is exceedingly difficult to control. The solution in Faller's thesis and the many later versions of his experiment was to replace the plates by corner reflectors.

\begin{figure}[t]
\begin{center}
\includegraphics[angle=0,width=4.5in]{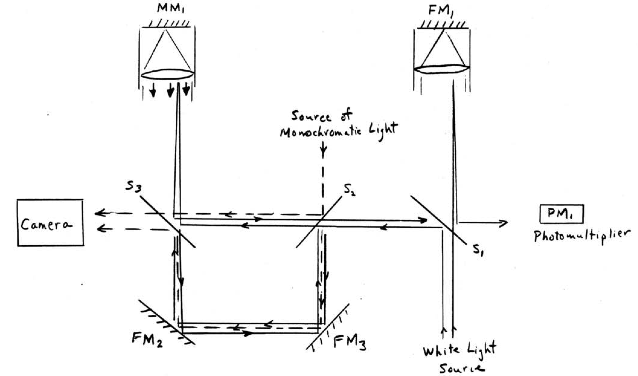} 
\caption{\label{Fig:Faller}  A preliminary sketch of Faller's thesis measurement of the absolute value of the gravitational acceleration, from Jim Wittke's Princeton Graduate Alumni Newsletter, 1962}
\end{center}
\end{figure}
The arrangement in Faller's experiment is sketched in Figure~\ref{Fig:Faller} (in the experiment marked as line~28 in Figure~\ref{Fig:timeline}). Light from the source at the lower right partially passed through the half-silvered mirror S$_1$, returned from the fixed corner reflector FM1 on the upper right, and was reflected by S$_1$ to the photomultiplier. Part of the light from the source was reflected by S$_1$ and S$_2$, returned by the freely falling corner reflector MM1 on the upper left, and arrived at the photomultiplier to be combined with the light from the other path. The fixed mirrors FM2 and FM3 served as calibration.  The bandwidth of the light source was broad enough that the position of the freely falling corner reflector at a fringe, relatively bright or dark, was defined by the condition that at the photomultiplier the phase as a function of wavelength at fixed falling corner reflector position was at an extremum. The precision this allowed in Faller's thesis measurement of $g$ was comparable to what was achieved by other methods, but could be improved. Faller and Hammond (1967) showed that the newly available stabilized lasers with sharply defined wavelength allowed sharply defined fringes and a much better measurement. The present standard (advertised in \url{http://www.microglacoste.com/fg5x.php}), with applications such as long-term monitoring of water table levels and continental uplift following the last Ice Age, in Faller's (2015) words ``pretty much is the legacy of what, with Bob's help, I started in Palmer Lab.''

One would also like to have an accurate absolute measurement of Newton's gravitational constant, $G$, the better to weigh the Earth and the rest of the solar system, as well as to establish a tight system of fundamental physical parameters. An indication of the difficulty is that Heyl (1930) measured $G=6.670\times 10^{-8}$~cm$^3$~g$^{-1}$~sec$^{-2}$, with a scatter among gold, platinum, and glass weights at about a part in $10^4$. That is to be compared to the present CODATA standard,  $G=6.6741\times 10^{-8}$~cm$^3$~g$^{-1}$~sec$^{-2}$, with fractional uncertainty reduced from Heyl by about a factor of two. Faller (2014a) reviewed this situation.	  

\subsection{Evolution of Newton's gravitational constant}\label{Sec:Gdot}
Section~\ref{sec:Mach} reviewed Dirac's Large Numbers Hypothesis (LNH), that the strength of the gravitational interaction (as measured by the value of  $G$ in standard units) may be very weak because it has been decreasing for a very long time. The idea has been probed in several ways. 

\subsubsection{Meteorites, eclipses, and geophysics}
Teller (1948) pointed out that a larger $G$ in the past would imply a hotter Sun, which if too hot would violate evidence of early life on Earth. Teller made an important point (which is marked as line~14 in Fig.~\ref{Fig:timeline}).  But the Hubble age then was underestimated by a factor of seven, meaning the constraint on the evolution of $G$ is significantly less tight than Teller estimated. Dicke (1962b) added the consideration that Earth's climate is seriously affected by the greenhouse effect of water vapor and the opposite effect of clouds. He argued that the net effect is quite uncertain, but that this consideration further weakens Teller's constraint. Dicke (1962b) also noted that stellar evolution models with standard gravity physics predict that the luminosity of the Sun has been increasing: The conversion from hydrogen to helium that keeps the Sun shining has been increasing the mean molecular weight in the core, which increases the temperature required to support it, which increases the luminosity. Dicke's colleague at Princeton University, Martin Schwarzschild, estimated that the present luminosity of the Sun, relative to the luminosity minimum at formation 4.5~Gyr ago, is $L_{\rm present}/L_{\rm initial}=1.6$. (This estimate is in Schwarzschild 1958, eq.~[23.8]. A more recent estimate in Ribas 2010 is $L_{\rm present}/L_{\rm initial}=1.3$.) A larger $G$ in the past might have helped keep Earth warm enough for early life. But, as Dicke argued, the relation between the history of the surface temperature of the Earth and the luminosity of the Sun was quite uncertain. The problem still with us.

A quantitative bound on the past luminosity of the Sun (in Peebles and Dicke 1962b; the title, {\it The temperature of meteorites and Dirac's cosmology and Mach's Principle}, makes manifest Dicke's motivation) considered that if  $G$ were larger in the past it would mean that parent meteorite bodies would have been warmer, the diffusive loss of radiogenic argon faster, and the apparent  potassium-argon radioactive decay ages of meteorites smaller than their true ages. We were following Goles, Fish, and Anders (1960), who used this consideration to constrain the environments of parent meteorite bodies under the assumption of conventional physics. We concluded that the strength of the gravitational interaction could not have been decreasing more rapidly than
\beq
-{1\over G}{dG\over dt} \la 1 \times 10^{-10}\hbox{ yr}^{-1}. \label{eq:meteorites}
\eeq
And we assured the reader that ``This limit does not seem to rule out any of the cosmologies in which the strength of the gravitational interaction is variable.'' That included the estimate in equation~(\ref{eq:Gt}) based on the LNH.  This result is marked at line~26 in Figure~\ref{Fig:timeline}.

Ideas about how the effect of a decreasing value of $G$ might be manifest in geophysical phenomena were mentioned in Section~\ref{sec:Mach}. In particular, a secular change of $G$ would affect the Earth's moment of inertia, its  distribution of angular momentum in the solid Earth and in atmospheric and oceanic currents, and its orbit around the Sun, all of which could have a detectable effect on timings of solar eclipses, which have been recorded back to the time of Babylon some 3000~yr ago. Curott (1966) examined the eclipse evidence, and Dicke (1966)  analyzed the many geophysical considerations that may help determine the effect of a secular evolution of $G$ on the evolution of the angular velocity of the Earth's surface and hence the predicted eclipse timings. Dicke concluded that the evidence bounds the evolution of $G$ at about the level of equation~(\ref{eq:meteorites}). This was by an approach that is so different that it merits a separate entry as line~35 in Figure~\ref{Fig:timeline}.

\subsubsection{Lunar Laser Ranging}\label{sec:LLR}
At the Chapel Hill Conference, Dicke (1957a) expressed his interest in better measurements of the motion of the Moon. Though not explained in the proceedings, he was at the time seeking to test whether the strength of gravity has been decreasing as the universe expands (Sec.~\ref{sec:LNH} eq.~[\ref{eq:Gt}]). This is the stated goal in the papers on tracking artificial satellites or the Moon by Hoffmann, Krotkov, and Dicke (1960) and Dicke, Hoffmann, and Krotkov (1961) (with, in the first of these papers, acknowledgement of contributions by all 13 members of the Gravity Research Group at that time, some much more important for the project than others). These two papers discussed  the prospects for precision optical tracking of the angular position of a satellite by a ``searchlight illuminating a corner reflector,'' or else ``sunlight illuminating a sphere,'' or a ``flashing light on the satellite.'' The proposal had become more specific in the paper by Alley et al. (1965) with the title, and the argument for, {\it Optical Radar Using a Corner Reflector on the Moon}. The author list included  Dicke, with Alley, Bender, and Faller, who were his former graduate students, Wilkinson, who was  then an assistant professor in Dicke's group, and two who were not in the group: Henry Plotkin at NASA Goddard Space Flight Center and Peter Franken at the University of Michigan, Ann Arbor. The paper mentioned the possibility of a ``few interesting measurements,'' including ``an accurate check on lunar orbit theory,''  ``Simultaneous ranging by several stations, with known geographical positions, could be used to measure the size and shape of the earth,'' a ``connection could be established between the American and European geodesic networks,'' and ``disturbances or periods which are not explained by the perturbations contained in the lunar theory \ldots if seen, might be identified with gravitational wave effects.''

Recollections of how this line of thought grew into the Lunar Laser Ranging Experiment are in the proceedings of Session 2 of the 2014 $19^{\rm th}$ International Workshop on Laser Ranging,  Annapolis. In these proceedings Plotkin (2014) recalled that in 1960 he moved to the Goddard Spaceflight Center, where his ``job was supposed to be developing optical systems for photographing satellites and stars in order to calibrate Microwave and Radio tracking systems,'' and that he stopped at Princeton along the way to discuss the task with Dicke, Carroll Alley, and others. Dicke ``suggested that when I get to NASA I consider putting cube corners on the moon or launching cube corners into space orbits and photographing reflections from powerful searchlight illumination.''  In an undated document\footnote{\url{http://ntrs.nasa.gov/archive/nasa/casi.ntrs.nasa.gov/19680025204.pdf}} Plotkin reported that by 1964 ``NASA launched the first of the satellites with arrays of fused-quartz cube corner retroreflectors to act as cooperative targets for laser radar stations" which ``achieved ranging precision of about 1 meter.'' This is the methodology for the Lunar Laser Ranging Experiment (and for the precision tracking of the high mass density LAGEOS satellites used to map Earth's gravitational field, and, as a byproduct, detect relativistic inertial frame dragging; Sec.~\ref{sec:LenseThirrng}). Also in the Annapolis proceedings are Faller's (2014b) recollections of the corner reflectors he used in his absolute measurement of the acceleration of gravity (Sec.~\ref{sec:abs-g}), his work on the design of the arrays of corner reflectors to be placed on the Moon, and the serious challenge he and colleagues faced in finding the first photons to be received by reflection from the corner reflectors, made even more difficult because the corner reflector arrays were not placed at the planned position on the Moon. Faller (2014b) also presented Dicke's  previously unpublished essay on {\it Lunar Laser Ranging Reminiscences}. Alley (1972) published his recollections of the project. Bender (2015) recalled that 
\begin{quotation}
\noindent in Sept., 1968 when we first heard that the Apollo 11 astronauts would only be allowed to spend 2 hours on the lunar surface, compared with the 4 hours originally planned, Bob Dicke being the one who asked if a reflector package could be considered as a contingency experiment for the Apollo 11 landing  may well have carried more weight with the NASA people than if he hadn't been involved.
\end{quotation}
The NASA Apollo 11 astronauts placed this first package of corner reflectors on the Moon in July 1969. The USSR Lunokhod~1 unmanned lunar rover left the second package on the Moon after launch in November 1970. 

The origin of the Lunar Laser Ranging Experiment is marked at line~32 in the timeline in Figure~\ref{Fig:timeline}, at  publication of the paper by Alley et al. (1965) that argued for this program. Faller, Winer, Carrion et al. (1969) reported that measurements of the distance to the corner reflectors ``can now be made with an accuracy approaching 15 cm.''  Distance measurements to the lunar corner reflectors now approach a precision of $1$~mm, and they tightly constrain possible departures from general relativity theory, including the Strong Equivalence Principle mentioned in Section~\ref{sec:WeakStrongEP}. The inferred relative gravitational accelerations in the Sun-Moon-Earth system bound the fractional differences of the ratios of inertial to passive gravitational masses of the Earth and Moon to no more than about a part in $10^{13}$ (Williams, Turyshev, and Boggs 2012). It is particularly relevant for the thread of this history that the evolution of the strength of gravity is bounded to 
\beq
{1\over G}\left|{dG\over dt}\right| \la 10^{-12}\hbox{ yr}^{-1}. \label{eq:LLR-Gdot}
\eeq
Dicke's dream of tests of gravity physics from precision measurements of the motion of the Moon, which he first mentioned in print in 1957, has been abundantly realized. But Hoffmann, Krotkov, and Dicke (1960) had proposed tracking satellites to look for evolution of $G$ ``that might be expected to amount to about one part in $10^{10}$ per year.'' This is two orders of magnitude larger than the bound.

\subsubsection{Other constraints}
The strength of the gravitational interaction in the early universe is constrained by the pattern of variation of the Cosmic Microwave Background radiation (the CMB) across the sky (Sec.~\ref{Sec:CMB}). The pattern is largely determined by the evolving dynamical interaction of the spatial distributions of matter and radiation through to redshift $z\sim 1000$, when in the standard model the primeval plasma combined to largely neutral atomic hydrogen, eliminating the non-gravitational interaction between matter and radiation. Bai, Salvado, and Stefanek (2015) found that the standard six-parameter $\rm\Lambda$CDM cosmology (defined in Sec.~\ref{Sec:GR-CMB}), with $G$ as a seventh free parameter, fitted to the Planck CMB satellite data (as in Planck Collaboration  2015a), with other cosmological tests, constrains the value of $G$ (in standard units for the rest of physics) at redshift $z\sim 1000$ to be within about 8\% of the laboratory value. The precision is modest but the reach is impressive.

If, in still earlier stages of expansion of the universe, the strength of gravity had been different from now, making the rate of expansion and cooling different from the prediction in general relativity, it would have affected the formation of the light elements at redshift $z\sim 10^9$ (Sec.~\ref{Sec:CMB}). The effect of a change of expansion rate was illustrated in Figure~2 in Peebles (1966). Dicke (1968) explored this effect in the scalar-tensor gravity theory. But the light element abundance measurements have been found to be in line with the constant strength of gravity in the $\rm\Lambda$CDM cosmology.

Also to be noted is the impressively tight and direct bound on the present rate of change of the strength of gravity from timing measurements of the orbit of a binary pulsar (Zhu, Stairs, Demorest, et al.\ 2015), at precision comparable to equation~(\ref{eq:LLR-Gdot}). This measurement and the many other constraints reviewed here on the possible evolution of the strength of gravity conflict with an elegant idea, Dicke's  readings of Mach's Principle and Dirac's Large Numbers (Sec.~\ref{sec:Mach}). The discussion of how empirical evidence can, on occasion, conflict with elegant ideas, continues in Sections~\ref{sec:testingBD} and~\ref{sec:fractal}. 

\subsection{Evolution of the weak and electromagnetic interactions}\label{sec:alphadot}
Dicke (1957a,b) proposed extending Dirac's LNH, that the strength of gravity is very small because it has been decreasing for a very long time, to a decreasing strength of the weak interaction, but more slowly  because the weak interaction is not as weak as gravity. This would mean $\beta$-decay rates are decreasing. Dicke (1959a) tested for this by comparing radioactive decay ages of meteorites from $\alpha$- and $\beta$-transitions, with inconclusive results. 

The electromagnetic interaction is even less weak, but Dicke (1959b) pointed to the long-standing idea that the Planck length $\sqrt{\hbar G/c^3}$ may play some role in determining an effective momentum cutoff for quantum field theory (e.g. Landau 1955; Deser 1957; Arnowitt, Deser, and Misner 1960). If $G$ were decreasing, and the Planck length decreasing with it, maybe the fine-structure constant $\alpha = e^2/\hbar c$ that measures the strength of the electromagnetic interaction is decreasing, perhaps as the absolute value of the logarithm of the Planck length (Landau 1955). Dicke set me the dissertation topic of an evolving electromagnetic interaction (James Peebles 1961, PhD Princeton; Peebles and Dicke 1962c). A dynamical value of $\alpha$ implies a long-range fifth force of interaction that could violate the constraint from the E\"otv\"os experiment. The thesis finessed that by introducing two metric tensors as well as the scalar field to replace $\alpha$ (since  developed in much more detail in Lightman and Lee 1973). I found empirical limits on the rate of evolution of $\alpha$ from estimates of how a changing fine-structure constant would affect relative rates of radioactive decay by $\alpha$-particle emission, nuclear fission, positron emission, and electron capture and emission. Estimates of the consistency of published radioactive decay ages of meteorites and terrestrial samples led to the conclusion that ``the data could not be used to eliminate a change in $\alpha$ of 0.1\% in the past 4.4 billion years.'' (This is entered in line~25 in Fig.~\ref{Fig:timeline}.) 

Recent bounds are better. Examples are 
\beqa
&& {1\over\alpha}{\left|d\alpha\over dt\right|} \la 10^{-16} \hbox{ yr}^{-1}\hbox{ at } z=0,\nonumber\\
&& {1\over t}{|\rm\Delta\alpha|\over\alpha} \la10^{-16}\hbox{ yr}^{-1}\hbox{ at } z=0.14,\label{eq:alphadot}\\
&& {1\over t}{|\rm\Delta\alpha|\over\alpha} \la10^{-14}\hbox{ yr}^{-1}\hbox{ at } 0.2\la z\la 1,\nonumber\\
&& {1\over t}{|\rm\Delta\alpha|\over\alpha} \la10^{-12.5}\hbox{ yr}^{-1}\hbox{ at } z=1200.\nonumber
\eeqa
In the second through fourth lines $\rm\Delta\alpha/\alpha$ is the allowed fractional change in the value of the fine-structure constant and $t$ is the time to the present from the time at the redshift $z$ of observation. The first line is a ``preliminary constraint'' from comparisons of optical atomic clock rates of aluminum and mercury ions (Rosenband, Hume, and Schmidt 2008). The second line is from bounds on shifts of the  resonant energies for slow neutron capture by fission products in the Oklo natural nuclear reactor. The geophysical considerations are complicated, but the many reanalyses (Shlyakhter 1976: Damour and Dyson 1996) argue for reliability of the bound. The third line is an example of constraints from line spacings in quasar spectra (Albareti, Comparat, Guti{\'e}rrez, et al.\ 2015). The last line is the constraint from the pattern of variation of the Cosmic Microwave Background radiation across the sky, fitted to the $\rm\Lambda$CDM cosmological model (Planck Collaboration 2015a; Sec.~\ref{Sec:GR-CMB}).  For another review of  the history of ideas about the value and possible evolution of the fine-structure constant see Kragh (2003).

The length of the list in equation~(\ref{eq:alphadot}) is another illustration of the persistent fascination with the idea that local physics may be influenced by the global nature of the universe in which local physics operates. Motivations for this idea have evolved since the Machian arguments that so fascinated Dicke and others, but  interest in the holistic concept continues.

\subsection{Tests of the scalar-tensor theory}\label{sec:testingBD}
Brans and Dicke (1961) found that the most demanding constraint on the scalar-tensor gravity theory was the measured precession of the orbit of the planet Mercury. They concluded that this measurement required that the parameter $w$ in the theory (eq.~[\ref{eq:BD}]) satisfy $w>6$. This of course assumes a correct accounting of the masses that determine the orbit. But Dicke (1964) reminded the reader that, before  relativity, the excess precession in Newtonian gravity theory was imagined to be caused by mass not taken into account in the standard computations, and that 
\begin{quotation}
\noindent One old suggestion seems not to have  had  the  attention that  it  deserved. If the  Sun  were  very  slightly  oblate, the   implied   distortion   of   the   sun's   gravitational   field would  result  in  a  rotation  of  the perihelia  of  the  planets. To  produce  a  discrepancy  in  Mercury's  orbit as  great  as 8 per  cent  of  the  Einstein  value  would  require  an  excessively small visual oblateness,  only  5 parts  in  $10^5$ amounting to  a difference  between  the solar  equatorial  and  polar radii of only 0.05" arc.
\end{quotation} 
One might imagine that the Sun is slightly oblate because the solar interior is rotating more rapidly than the surface, maybe a result of spinup as matter was drawn into the growing Sun, in line with the far shorter rotation periods of the gas giant planets Jupiter and Saturn. The rotation of the solar surface might have been slowed by the transfer of angular momentum to the solar wind, as Dicke inspired me to analyze in some detail.

Dicke (1964) reported that he, Henry Hill, and Mark Goldenberg had ``built a  special instrument   to   observe   photoelectrically the   Sun's    oblateness,   and    preliminary    measurements were  made  during  a  few  weeks  toward  the  close  of  the summer   of   1963.'' Dicke and Goldenberg (1967) later reported that the measured fractional difference of equatorial and polar radii is $5.0\pm 0.7$ parts in~$10^5$, about what Dicke (1964) felt would allow an interesting value of the scalar-tensor parameter,  $w$, in equation~(\ref{eq:BD}).\footnote{\label{footnote:wager}The situation in the late 1960s was summarized by Dicke (1969). He reported that the contribution to the precession of the obit of Mercury by the Dicke-Goldenberg measurement of the  oblateness of the solar mass distribution is $3.4\pm 0.5$ seconds of arc per century. This contribution subtracted from the measured precession, $43.10\pm 0.44$, is the precession to be attributed to the departure from Newtonian gravity physics. And this precession divided by the relativistic prediction, 43.03, is $0.920\pm 0.015$. This would be a serious challenge for general relativity. In the scalar-tensor theory with $w=5$ this ratio is $(4+3w)/(6+3w) = 0.905$, or one standard deviation below the measurement. At about this time the first radio interferometer measurements  (Seielstad, Sramek, and Weiler 1970; Muhleman, Ekers, and Fomalont 1970) showed that the relativistic deflection of light by the sun is $1.02\pm 0.12$ times the relativistic prediction. In the scalar-tensor theory with $w=5$ the deflection is $(3+2w)/(4+2w)=0.93$ times the relativistic prediction. This again is about one standard deviation below the measurement. And one standard deviation is a quite acceptable degree of consistency. The gravitational redshift is the same in general relativity and the scalar-tensor theory. That is, in 1969 Dicke's evidence was that the  scalar-tensor theory with $w=5$ is  consistent with the classical tests, provided the solar oblateness measurement was correct. The scalar-tensor solar gravitational deflection of light, 0.93 times the relativistic prediction at $w=5$, is the gravitational deflection in the Wheeler-Dicke wager shown in Figure~\ref{Fig:WDbet}.} The  result was not widely welcomed, in part because it challenged general relativity theory, and certainly in part because identifying and measuring a level surface on the Sun is seriously challenging. Subsequent measurements by Hill, Clayton, Patz, et al.\ (1974), and Hill and Stebbins (1975), used techniques that Hill, then in the Gravity Research Group, and his Princeton graduate student Carl Zanoni, originally meant to be used for measurement of the gravitational deflection of starlight passing near the Sun without the aid of a Solar eclipse. This required design for the strong rejection of diffracted sunlight (Carl Zanoni 1967, PhD Princeton; Zanoni and Hill 1965), which benefitted the design of the Hill et al. oblateness measurements. The conclusion was that the solar oblateness is too small for a serious effect on the orbit of Merury. This proved to be consistent with the probe of the internal rotation of the Sun afforded by the splitting of frequencies of modes of solar oscillation with different azimuthal numbers $m$. This revealed that the solar interior is rotating nearly as a solid body at close to the mean of the rotation rates seen at different latitudes on the solar surface (Thompson, Christensen-Dalsgaard, Miesch, and Toomre 2003). Rozelot and Damiani (2011) reviewed the history and status of this subject. We may conclude that the mass distribution in the Sun is not likely to have disturbed the test of gravity theory from the orbit of Mercury. The Dicke and Goldenberg (1967) measurement is marked at line~36 in the timeline (Fig.~\ref{Fig:timeline}), as a step toward closing the case for this important test of gravity physics. 

The remarkably tight bound on the scalar-tensor gravity theory from the Cassini satellite radar timing experiment is $w>10^{4.5}$  (Bertotti, Iess, and Tortora  2003). In simple solutions for the expansion of the universe in the scalar-tensor theory this  bound implies that the rate of change of the strength of gravity is limited to $-\dot G/G\la 10^{-14}\hbox{ yr}^{-1}$, tighter even than the more direct bounds from the Lunar Laser Ranging Experiment (eq.~[\ref{eq:LLR-Gdot}]) and pulsar timing (Zhu, Stairs, Demorest, et al.\ 2015). Mchugh (2016) reviewed the history of tests that have progressively tightened the bound on $w$. The original vision of the scalar-tensor theory certainly must be scaled back to at most a small perturbation to general relativity. But the scalar-tensor theory still fascinates, figuring in the search for a deeper gravity physics, as one sees by its frequent mention in the review, {\it Beyond the cosmological standard model} (Joyce, Jain, Khoury, and Trodden 2015). And an excellent experimental program similarly can lead in unexpected directions. The solar distortion telescope Dicke first designed in the early 1960s was turned to detection of variation of  the solar surface temperature as a function of solar latitude during five years of the solar cycle, a serious contribution to the continuing attempts to interpret the observed variation of the solar constant with the sunspot cycle (Kuhn, Libbrecht, and Dicke 1988).  Kuhn, Bush, Emilio, and Scholl (2012) reviewed the history and present status of precision space-based measurements of the shape of the Sun. 

\subsection{Cosmology and the great extension of tests of gravity physics}\label{Sec:cosmology}
General relativity theory inspired modern cosmology, and it was recognized in the mid-1950s  that cosmological tests might in turn test relativity. Thus Walter Baade spoke at the 1955 Berne Conference, {\it Jubilee of Relativity Theory}, on progress in the measurement of a fundamental datum for cosmology, the extragalactic distance scale. This was an important step toward what has grown to be the rich science of cosmology.

\subsubsection{The expanding universe}\label{sec:expandinguniverse}
Hubble (1929) found the first reasonably clear evidence of the linear relation between galaxy redshifts and distances,
\beq
v=H_or,\label{eq:HubbleLaw}
\eeq
which Lema\^\i tre (1927) had shown is to be expected if the universe is expanding in a homogeneous and isotropic way.\footnote{Lema\^\i tre derived this linear relation from a solution to Einstein's field equation, but it follows more generally from standard local physics in a near homogeneous spacetime described by a  metric tensor.} The cosmological redshift, $z$, is defined by the ratio of wavelengths of features in the observed spectrum and the laboratory wavelengths at emission, as $1 + z = \lambda_{\rm obs}/\lambda_{\rm em}$. At small $z$ the redshift may be considered to be a Doppler shift, where the recession velocity is $v= cz$. The distance $r$ may be inferred from the inverse square law using estimates of the intrinsic luminosities of observed objects, expressed as absolute magnitudes, using the observed energy flux densities, expressed as apparent magnitudes. A commonly used term thus is the redshift-magnitude relation (though one also uses the relation between intrinsic linear sizes and observed angular sizes). This redshift-magnitude relation also describes observations of objects at greater distances, where the redshift $z$ is large, and one looks for the relativistic effects of spacetime curvature. 

Hubble and Humason (1931) considerably improved the case for equation~(\ref{eq:HubbleLaw}), reaching  redshift $z\sim 0.07$ for giant galaxies, and Hubble (1936) showed still better evidence in a larger sample that reached $z\simeq 0.15$. These results were very influential in the development of cosmology.  The Hubble and Humanson paper on measurements of the redshift-magnitude relation accordingly is marked at line~7 in Figure~\ref{Fig:timeline}.  

Another important early advance in cosmology was Hubble's (1936) deep counts of galaxies as a function of limiting apparent magnitude (line~10 in Fig.~\ref{Fig:timeline}). The counts were not inconsistent with Einstein's Cosmological Principle, that the distribution of the galaxies is homogeneous in the large-scale average. The counts reached an impressively large redshift (estimated at $z\sim 0.4$ in Peebles 1971). Systematic errors in Hubble's counts were problematic enough that they could not rule out Mandelbrot's (1975) argument discussed in Section~\ref{sec:fractal} for a fractal galaxy distribution with fractal dimension $D$ well below three. But one could conclude that, if there were an observable edge to the universe of galaxies, the most distant ones would be flying away at near relativistic speeds. 

It was understood early on (e.g., Tolman 1934 \S 185) that the relativistic Friedman-Lema\^\i tre cosmological models predict that at high redshift there may be a departure from the redshift-magnitude relation implied by equation~(\ref{eq:HubbleLaw}), depending on the model parameters (including the mean mass density and the mean curvature of space sections at constant world time). Interest in detecting a departure from equation~(\ref{eq:HubbleLaw}) increased with the introduction of the Steady State cosmology, which makes a definite prediction for the form of the redshift-magnitude relation (Bondi and Gold 1948; Hoyle 1948). Humason, Mayall, and Sandage (1956)  reported progress in measurements of this relation, which had reached redshift $z\sim 0.2$, a modest advance over Hubble (1936) two decades earlier. Sandage (1961) presented a detailed analysis of the prospects for further advances in the application of this and other cosmological tests. We might take as prophetic Sandage's remark that (in his italics) ``{\it If observations show $q_o$ to be $-1$, we cannot decide between a steady-state universe and a Lemaitre-Eddington universe.}'' The parameter $q_o$ is a measure of the departure from equation~(\ref{eq:HubbleLaw}), and the value $q_o=-1$ is predicted by the Steady State cosmology and  by the general relativity Friedman-Lema\^\i tre model if the expansion rate is dominated by Einstein's cosmological constant, $\rm\Lambda$.  As it happens, well-checked measurements of the redshift-magnitude relation for supernovae (of type Ia, the explosions of white dwarf stars) are close to $q_o=-1$ (Riess, Filippenko, Challis, et al. 1998; Perlmutter, Aldering, Goldhaber, et al.\ 1999). This is the degeneracy Sandage noted. The degeneracy is broken, and the classical Steady State cosmology convincingly ruled out, by other cosmological tests. The evidence briefly reviewed in Section~\ref{Sec:GR-CMB} is that we live in an evolving universe now dominated by Einstein's $\rm\Lambda$.

Establishing the value of Hubble's constant, $H_o$, in equation~(\ref{eq:HubbleLaw}) by astronomical methods requires determination of the intrinsic luminosities of  extragalactic objects, a difficult task. Hubble and Humason (1931) slightly increased Hubble's (1929) estimate to $H_o\simeq 560$ km~s$^{-1}$~Mpc$^{-1}$, a result that was generally accepted until the 1950s. For example, in the monograph {\it Relativity, Thermodynamics, and Cosmology}, Tolman's (1934, \S 177) comment about the measured value of $H_o$ was that ``It is believed that the uncertainty in the final result is definitely less than 20 per cent.'' But this meant that the characteristic expansion time is $1/H_o\sim 2\times 10^9$~yr, which was uncomfortably short, roughly half the largest radioactive decay ages of terrestrial minerals and estimates of stellar evolution ages. It was not widely discussed then, but Lema{\^i}tre's 1931 model, with the values of $\rm\Lambda$, the cosmic mean mass density, and space curvature chosen so the universe passed through a time of slow expansion, could reconcile the large estimate of $H_o$ with a long expansion time It did require a very special adjustment of  parameters, however.

Bondi and Gold (1948) pointed out that the Steady State cosmology allows galaxies of arbitrarily great age, albeit with great scarcity, thus allowing a large age for our particular galaxy. But the Steady State cosmology predicts that the mean age of the galaxies is  $1/(3H_o)\sim 6\times 10^8$~yr (for the estimate of $H_o$ accepted then). Gamow (1954) remarked that this would mean that ``we should find ourselves surrounded by a bunch of mere youngsters, as the galactic ages go.'' He remarked that this seemed inconsistent with the observation that our Milky Way galaxy and neighboring galaxies have similar mean stellar spectra, indicating similar stellar evolution ages.

Walter Baade improved the situation with his announcement at the 1952 Rome Meeting of the International Astronomical Union that he had found a correction to the extragalactic distance scale: he found that cepheid variable stars of type I (younger, with larger heavy element abundances) are considerably more luminous than previous estimates. This increased estimates of extragalactic distances, and reduced Hubble's constant to about $H_o=180$ km~s$^{-1}$~Mpc$^{-1}$.\footnote{Baade's presentation to the IAU is reported without details in the Transactions of the IAU, volume~8, p. 397, in a summary of the talks presented. We can be sure Baade also reported his distance scale correction at the  1955 Bern Conference, as discussed in Sec.~\ref{sec:GRinthe50s}, though he did not contribute a paper to the proceedings.}  Sandage (1958) improved the situation still more by his correction for misidentification of gaseous nebulae as luminous stars. This brought his estimate to $H_o\simeq 75$ km~s$^{-1}$~Mpc$^{-1}$, ``with a possible uncertainty of a factor of 2.'' For further commentary on these and related developments see Trimble (1996). The Planck Collaboration (2015b) found $H_o= 68\pm 1$ km~s$^{-1}$~Mpc$^{-1}$ indirectly derived from precision CMB anisotropy measurements. This is close to Sandage's central value, and still large enough to require the ``disreputable $\rm\Lambda$,'' as Robertson (1956) put it at the Bern Conference.

The revision of the distance scale was influential enough to be marked in the timeline (line~15, at Baade's announcement of his correction at the 1952 IAU conference). Its importance  is illustrated by Bondi's comment added to the second edition of his book, {\it Cosmology} (Bondi 1960, p. 165): ``It is not easy to appreciate now the extent to which for more than fifteen years all work in cosmology was affected and indeed oppressed by the short value of $T$ ($1.8\times 10^9$~years) so confidently claimed to have been established observationally.'' The remark is fair, but one might ask why the theorists did not challenge the observers. 

At the Bern Conference  Klein (1956) proposed an alternative to the large-scale homogeneity assumed  in the Friedman-Lema\^\i tre and Steady State models. Perhaps the galaxies are drifting apart into empty space after an explosion of a local concentration of matter. Klein may not have been aware that  Hubble's (1936) deep galaxy counts require near relativistic recession velocities, hence a relativistic explosion. But still the idea was well worth considering, because Einstein's picture of a homogeneous universe was not apparent in the surveys of the galaxy distribution available at that time. Oort (1958) emphasized this in his commentary on the observational situation in cosmology, in the proceedings of the Solvay Conference on {\it La structure et l' \'evolution de l'univers} (Stoops 1958). Oort commenced his paper with the statement that ``One of the most striking aspects of the universe is its inhomogeneity.'' He went on to review the observations of clustering of galaxies on the largest scales that could be surveyed then. But Oort was willing to estimate the cosmic mean mass density, because the assumption of homogeneity in the average over still larger scales was not inconsistent with Hubble's (1936) deep galaxy counts. 

Measures of the large-scale distribution of extragalactic objects were starting to improve in the late 1950s.  At the Chapel Hill Conference, Lilley (1957) discussed the Second Cambridge Catalogue of Radio Sources  (Shakeshaft, Ryle, Baldwin, et al. 1955). The main point of interest then was the count of sources as a function of flux density, for which the Steady State cosmology makes a definite prediction. This test was spoiled by side-lobe confusion of source identifications. But the map of angular positions of the radio sources was not so seriously afflicted, and the strikingly uniform distribution of sources across the sky was not what one would expect in a fractal universe. (This is further discussed in Sec.~\ref{sec:fractal}.) The radio source map, with Hubble's deep counts and the observed linearity of the redshift-distance relation, was among the first observational indications that Einstein's argument from Mach's Principle for the Cosmological Principle might be right (and for this reason the 2C catalog is entered in line~16 in Fig.~\ref{Fig:timeline}). More  evidence for the Cosmological Principle came with the discovery that space is filled with the near uniform sea of microwave radiation discussed next.

\subsubsection{The sea of cosmic microwave radiation}\label{Sec:CMB}
In the late 1940s George Gamow, with his student Ralph Alpher and their colleague Robert Herman, developed a theory of element formation in the early stages of expansion of the universe in an initially hot dense Friedman-Lema\^\i tre cosmological model (Ralph Alpher 1948, PhD George Washington University; Gamow 1948; Alpher, Follin, and Herman 1953). In this theory space is filled with a sea of thermal radiation, now known as the Cosmic Microwave Background, or CMB. The properties of this radiation have been read now in considerable detail that yields demanding tests of general relativity theory. The story of how the thermal radiation in this hot Big Bang model was predicted, and the community came to recognize its existence, is complicated. My analysis of how Gamow and colleagues arrived at their cosmology with its thermal radiation is in Peebles (2014). The book, {\it Finding the Big Bang} (Peebles, Page, and Partridge 2009), recalls how the radiation was discovered, interpreted, and its properties explored. Peebles~(2012) adds personal recollections. A few aspects of this CMB story are entered here as part of the history of how experimental gravity physics grew. 

The CMB research program began in the summer of 1964 when Dicke gathered Peter Roll, David Wilkinson, and me, junior faculty and a postdoc in his Gravity Research Group, to discuss a proposal for what the universe might have been doing before it was expanding. Dicke suggested that the universe may have been collapsing, maybe in a cycle of expansion and contraction. The bounce would have been seriously  irreversible, producing entropy largely in the form of a sea of thermal radiation. This idea of irreversibility of a bouncing universe traces back at least to Tolman (1934p. 443), who pointed out that if the bounce conserved entropy then in a cyclic universe each cycle would last longer. Dicke proposed a specific physical model for entropy production in the bounce. And it is very typical of Dicke that his idea suggested an experiment: a search for the radiation. 

Ideas about early universe physics have changed, but Dicke's thought about entropy production is worth recording here. Stars in the previous cycle would have been converting hydrogen into heavier elements, releasing nuclear binding energy of several million electron volts for each nucleon that became bound in a heavier atomic nucleus. This binding energy would have been radiated as some $10^6$ starlight photons per nucleon, because the starlight photons have energy on the order of an electron volt. If the contraction before the bounce to the next expansion were deep enough, then during the contraction these starlight photons would have been blueshifted to energies above several million electron volts, enough to  photodissociate the elements heavier than hydrogen that were formed in stars during the previous expansion and collapse. That would yield fresh hydrogen for the generations of stars in the next expansion. Just a few of the blueshifted starlight photons would have been needed to release each nucleon from a heavier atomic nucleus, and the remaining $10^6$ photons per nucleon would relax to a sea of thermal radiation, in a seriously irreversible process.\footnote{The estimate in Dicke and Peebles (1979) of the number of cycles needed to produce the present entropy per nucleon can be improved. For a more direct approach, let the energy density in the Cosmic Infrared Background (the CIB) produced by stars and active galactic nuclei during the cycle of expansion and contraction be $f$ times the energy in the CMB. Suppose the bounce produces no entropy apart from the thermalization of the CIB, and suppose nucleons and radiation are conserved in the bounce. Then an easy exercise shows that the entropy per nucleon in the CMB   after the bounce is $(1+f)^{3/4}$ times the entropy per nucleon before the bounce. The observed energy density in the CIB is about $f\sim 0.1$ times the energy in the CMB (Hauser, Arendt,  Kelsall,  et al. 1998). If the CIB is not going to receive much more energy in the rest of the present cycle, then in this model the entropy per nucleon in the next cycle will be $1.1^{3/4}$ times the entropy per nucleon in this cycle, roughy a 10\% increase. Not addressed, of course, is what might have caused the bounce, and how the quite clumpy distribution of matter, including black holes, just prior to the bounce could have become so very close to the near homogeneous condition we may expect to obtain at the start of the next cycle, as it did at the start of our cycle according to the established cosmology. The Steinhardt and Turok (2007) model offers a way to deal with these issues, but in a manner that eliminates the entropy Dicke was thinking about.}

Dicke suggested that Roll and Wilkinson build a microwave radiometer that might detect this thermal CMB radiation. This was timely for Roll, because the Roll, Krotkov, and Dicke (1964) E\"otv\"os experiment was completed, and it was timely for Wilkinson, because he had not yet settled on a long-term project after arrival from completion of his PhD at the University of Michigan. Dicke suggested that I look into possible theoretical implications. At the time this was to me just another of Dicke's many ideas, most of which I found interesting to explore. This one was very  interesting. It was speculative, to be sure, but that was not out of the ordinary in the Gravity Research Group. And Dicke's idea has proved to be wonderfully productive. 

While at the MIT Radiation Laboratory during the Second World War Dicke invented much of the microwave radiometer technology Roll and Wilkinson used. We had to remind him that Dicke, Beringer, Kyhl, and Vane (1946) had already used a Dicke radiometer to place a limit on the CMB temperature. They reporting a bound of 20~K on ``radiation from cosmic matter at the radiometer wave-lengths,'' 1.0~to 1.5~cm (line~13 in Fig.~\ref{Fig:timeline}). The radiometer Roll and Wilkinson built used Dicke's technique of suppression of receiver noise by switching between the horn antenna and a reference source of thermal radiation at known temperature, for which Roll and Wilkinson used liquid helium. 

A half decade before Dicke's suggestion to Roll and Wilkinson, an experimental microwave receiving system at the Bell Telephone Laboratories detected more noise than expected from accounting of noise sources in the instrument (DeGrasse, Hogg, Ohm, and Scovil 1959). This curious result was repeatable (as Hogg 2009 recalled), but not widely discussed. The noise originating in the receivers in the Bell experiments was so small that Dicke's switching technique was not needed, but Penzias and Wilson (1965) used it to remove any chance that the  ``excess antenna temperature'' could be attributed to some error in the noise accounting, again using a liquid helium reference source. (They could switch much more slowly than in the Roll and Wilkinson radiometer, because their system noise was so small.) Penzias and Wilson also carefully tested and eliminated possible terrestrial sources of the excess radiation. They had an important measurement. They learned of a possible interpretation when they learned of the Roll and Wilkinson search for a sea of radiation left from the early universe.

It was clear that this excess radiation, if extraterrestrial, must be close to isotropic, because the signal did not  vary appreciably as the Earth and antennas rotated. Partridge and Wilkinson (1967) improved this to a demonstration that the large-scale variation of the CMB intensity across the sky is less than about 3 parts in $10^3$. This was another early argument for Einstein's Cosmological Principle. Space was known to be close to transparent at these wavelengths, because radio sources were observed at high redshift, so the isotropy had to mean either that the universe is nearly uniformly filled with this radiation, or else that the  distribution of the radiation is significantly inhomogeneous but close to spherically symmetric, with us near the center of symmetry. The latter seemed less likely then, and it is now convincingly ruled out by the cosmological tests. 

In Gamow's (1948) hot Big Bang picture, and the Dicke, Peebles, Roll, and Wilkinson (1965) interpretation of the Bell Laboratories excess noise, the radiation intensity spectrum would be expected to be close to thermal. This is because the spectrum would have to have been very close to thermal in the early universe, when the high density and temperature would have produced statistical equilibrium, and the CMB thermal spectrum would be expected to have been preserved as stars and galaxies formed, because the CMB heat capacity is large compared to what is expected to be readily available in matter. In the Steady State cosmology one is free to postulate continual creation of radiation as well as matter, and one is free to postulate the spectrum of the radiation at creation. But the radiation spectrum would not have relaxed to thermal because, as has been noted, the universe is observed to be transparent at these wavelengths. And since the detected radiation would be the spectrum at creation convolved over redshift it would require an exceedingly contrived spectrum at creation to make the detected spectrum close to thermal. One also had to consider that the Bell excess noise might only be the sum of local sources of microwave radiation, making it consistent with the Steady State cosmology or a cold Big Bang model. But again, in this local source picture the spectrum would not likely be close to thermal. The spectrum measurements thus were critical to the interpretation of this radiation.  

The Penzias and Wilson (1965) and Roll and Wilkinson (1966) experiments operated at 7.4~cm and 3.2~cm wavelength, respectively. Their measurements at the two wavelengths were consistent with the same effective radiation temperature, which argued for a close to thermal spectrum in the long wavelength, Rayleigh-Jeans, part of the spectrum. An important early advance was the recognition that the observation of absorption of starlight by interstellar cyanogen (CN) molecules in the first excited level as well as the ground level offered another measure of the spectrum, at the wavelength 0.26\,cm of transition between the ground and excited levels. McKellar (1941) had translated the observed ratio of populations in the excited and ground levels to an effective temperature $T=2.3$\,K. George Field, Patrick Thaddeus, and Neville Woolf (whose recollections are in Peebles, Page, and Partridge 2009, pages 75-78, 78-85, and 74-75 respectively) recalled their prompt recognition that the Bell radiation, if close to thermal, could account for the observed CN excitation. Iosif Samuilovich Shklovsky (1966) in the Soviet Union independently made the same point. New measurements of the CN excitation temperature were found to be consistent with the Bell and Princeton radiometer measurements (Field, Herbig, and Hitchcock 1966; Thaddeus and Clauser 1966). That meant the CMB intensity at 0.26~cm and the measurements at 7.4~cm and 3.2~cm wavelength are reasonably consistent with a thermal spectrum, including the departure from the Rayleigh-Jeans power law about at the CN wavelength. The departure from the Rayleigh-Jeans power law was demonstrated also by a radiometer intensity measurement at 0.33~cm (Boynton, Stokes, and Wilkinson 1968). These results are marked at line~33 in the timeline in Figure~\ref{Fig:timeline}.

\begin{figure}[ht]
\begin{center}
\includegraphics[angle=0,width=4.in]{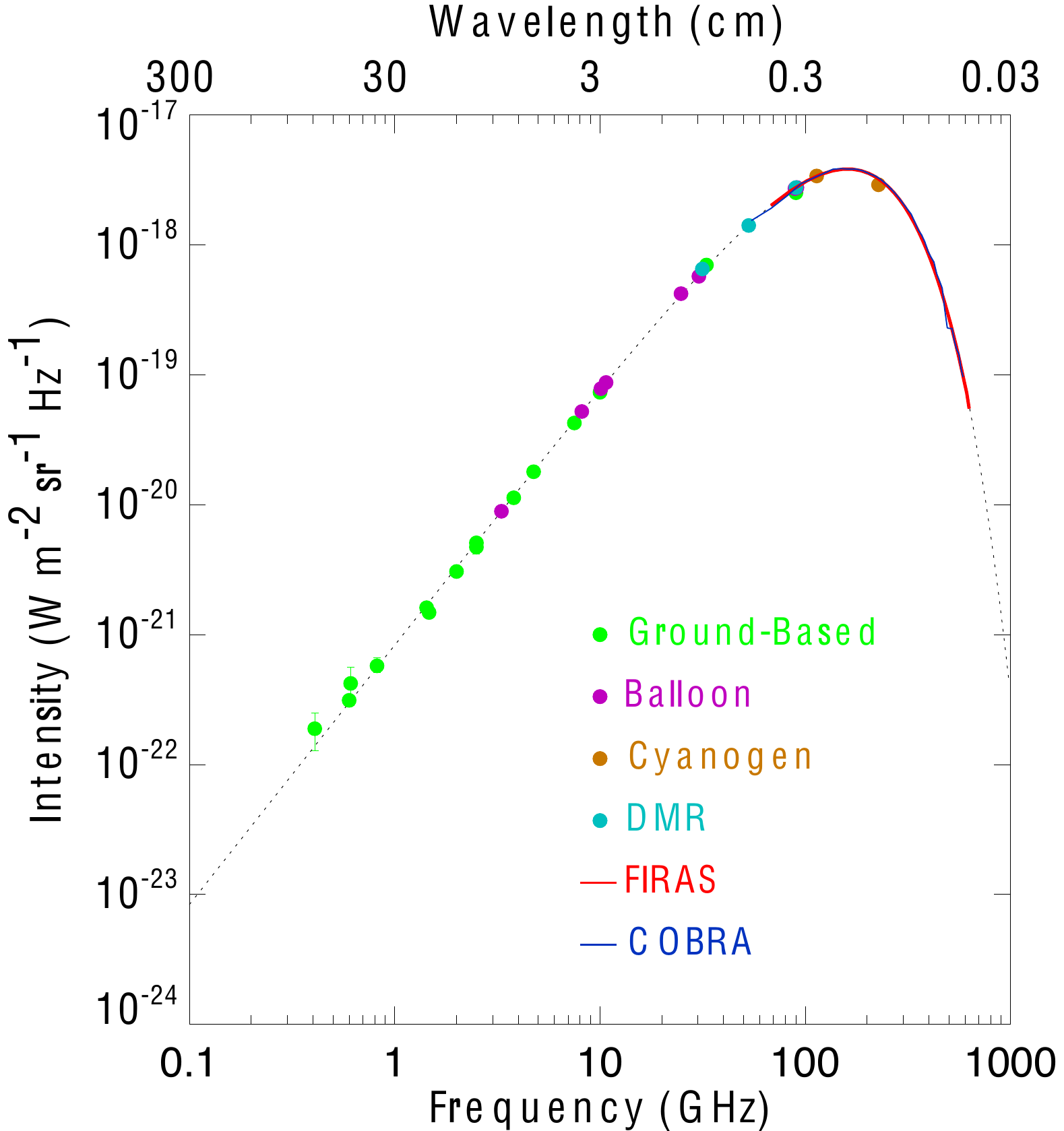} 
\caption{\label{Fig:CMB}\footnotesize The energy (intensity) spectrum of the Cosmic Microwave Background  radiation that nearly uniformly fills space (Kogut 2012). The dotted curve is the theoretical Planck blackbody spectrum; the solid curve near the peak shows measurements.}
\end{center}
\end{figure}
In the 1960s discussions  of how cosmology and gravity physics might be tested by CMB measurements were beclouded by indications that there is a considerable excess of energy relative to a thermal intensity spectrum at wavelengths near and shorter than the thermal peak. In the hot Big Bang picture this excess would require the postulate that explosions of some sort in the early universe added considerable radiation energy. Such explosions surely would have obscured any primeval patterns imprinted on the radiation by whatever imprinted the departures from exact homogeneity in the primeval mass  distribution that grew into galaxies. This uncomfortable situation was resolved a quarter of a century after identification of the CMB, by two independent groups (Mather, Cheng, Eplee, et al. 1990; Gush, Halpern, and Wishnow 1990). Both showed that the spectrum is very close to thermal over the radiation intensity peak. Kogut (2012) compiled these and the other spectrum measurements shown in Figure~\ref{Fig:CMB}. \footnote{It is to be noted that most of these data are from measurements of the differences between the CMB intensity and the intensity of radiation from black thermal sources at well-calibrated temperatures. Since the theoretical Planck blackbody spectrum is not as well tested as these comparison measurements, the accurate conclusion is that the CMB spectrum is very close to thermal.} 

The strikingly close agreement with a thermal spectrum has two important implications. First, we have noted that space in the universe as it is now is observed to be close to transparent. This means that the CMB would not have relaxed to the thermal spectrum shown in Figure~\ref{Fig:CMB} in the universe as it is now. Formation of the thermal spectrum requires that the universe expanded from a state dense and hot enough to have forced thermal relaxation. This serious evidence that our universe is evolving, drawn from such a simple figure (albeit one based on many far from simple measurements), is to be ranked with the memorable advances in the exploration of the world around us. The spectrum does not offer a serious constraint on gravity physics, however, because preservation of the thermal spectrum as the  universe expands follows if spacetime is well described by a close to homogeneous and isotropic line element with standard local physics. The thermal spectrum in Figure~\ref{Fig:CMB} argues for cosmic evolution, but it does not argue for general relativity theory.

Second, the demonstration in 1990 that the CMB spectrum is close to thermal meant that the CMB need not have been seriously disturbed from its primeval condition. This meant that there was a chance that measurements of the CMB could be mined for comparison to what is predicted by theories of  cosmic evolution in the early universe. That is, we might not have to deal with the complexities of nongravitational disturbances of the CMB by the processes of galaxy formation. Peebles and Yu (1970) introduced the radiative transfer computation that predicts the primeval oscillations in the power spectra of the distributions of the CMB and and of the matter, under the assumptions of general relativity theory and adiabatic initial conditions (meaning the primeval spatial distribution $n_\gamma(\vec x)$ of the CMB photons and the distribution $n(\vec x)$ of the matter particles have the same small fractional departures from exact homogeneity, $\delta n_\gamma/n_\gamma=\delta n/n$). The first pieces of evidence that these assumptions are on the right track were found some three decades after Peebles and Yu, from measurements of the angular distribution of the CMB and the spatial distribution of the galaxies. This history was reviewed in Peebles, Page, and Partridge (2009). 

\subsubsection{The Cosmic Microwave Background and general relativity}\label{Sec:GR-CMB}
To add to the picture of how tests of gravity physics grew out of research in the 1960s, including the discovery and early steps in exploration of the Cosmic Microwave Background, I offer this overview of the CMB-based cosmological tests that reached fruition well after the naissance. These tests treat gravity physics in linear perturbation from the relativistic Friedman-Lema\^\i tre solution. The linearity greatly simplifies theoretical predictions, but of course it means the tests are limited to this approximation. The linear treatment of gravity and the effect of departures from homogeneity on the rate of expansion of the universe remains a good approximation even at low redshifts, when the mass distribution has grown strongly nonlinear, because the gravitational potentials are still small, on the order of $(v/c)^2\sim 10^{-6}$, everywhere except close to neutron stars and massive compact objects, presumably black holes. On the scale of things these compact objects act as dark matter particles. 

The standard and accepted six-parameter $\rm\Lambda$CDM cosmology starts with the cosmologically flat Friedman-Lema\^\i tre solution to Einstein's field equation. This solution is perturbed by departures from homogeneity that are assumed to be initially adiabatic, Gaussian, and near scale-invariant. The stress-energy tensor in Einstein's equation is dominated by the CMB, nucleons, neutrinos, the hypothetical nearly collisionless initially cold nonbaryonic dark matter, or CDM, and Einstein's cosmological constant, $\rm\Lambda$. That is, the term $\rm\Lambda$CDM signifies a considerable number of assumptions in addition to the presence of $\rm\Lambda$ and CDM. 

Gravity in general relativity theory in linear approximation is represented by two potentials, sourced by the active gravitational mass density and the energy flux density. At redshifts well above $z_d\simeq 1200$ baryonic matter was fully ionized and the plasma and radiation acted as a fluid made slightly viscous by diffusion of the radiation. This matter-radiation fluid has mass density $\rho$, pressure $p$, active gravitational mass density $\rho + 3p$, and inertial and passive mass densities $\rho+p$, where $\rho$ is the energy density and $p$ is the pressure. Near redshift $z_d$ the CMB radiation is to be described by its photon distribution function in single-particle phase space, with the CMB evolution in phase space described by the Boltzmann collision equation for the photons propagating according to the equation of motion in the gravitational potentials and scattering off the free electrons. The dark matter is modeled as an initially cold ideal gas that behaves as Newton would expect, while the baryons are treated as a fluid that suffers radiation drag. The CMB and neutrino sources for the gravitational potentials are to be computed as integrals over the distribution functions in their single-particle phase spaces.

The two points of this perhaps unduly schematic accounting are that the measurements of the CMB probe a considerable variety of elements of general relativity theory, and that the measurements are interpreted in a model that depends on a considerable number of assumptions in addition to general relativity. It is important therefore  that there is a considerable variety of measurements. The variety is large enough that an adequate review is too long for this paper, so I refer instead to the Planck Collaboration (2015b) discussion of  the empirical situation from the CMB measurements. 

\begin{figure}[ht]
\begin{center}
\includegraphics[angle=0,width=3.75in]{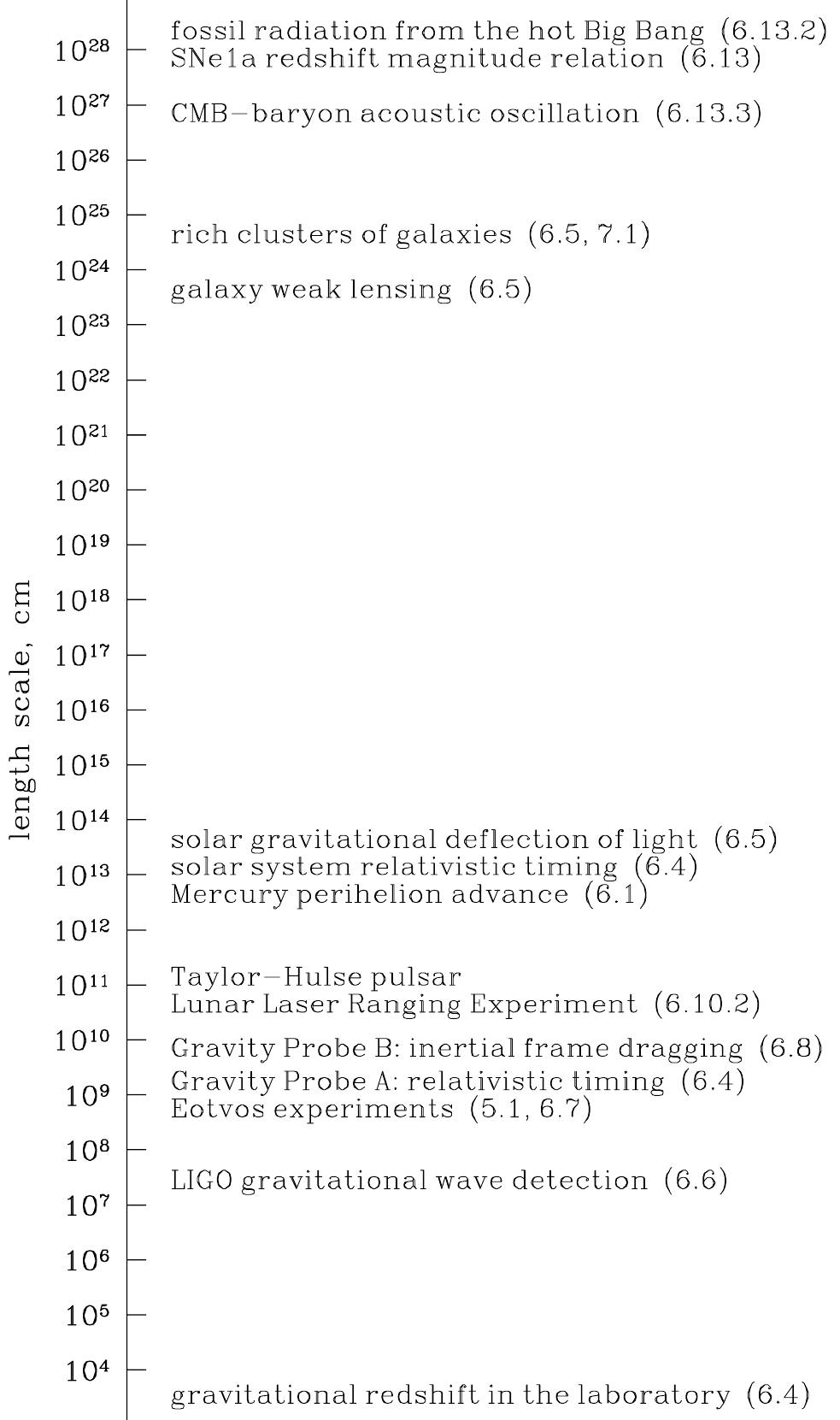} 
\caption{\label{Fig:lengthscale}  Length scales of probes of gravity physics.}
\end{center}
\end{figure}

\subsection{On the empirical case for general relativity theory}\label{theempiricalcase}
An assessment of the weight of empirical evidence for an idea or theory in natural science is of course informed by the degree of precision of the supporting measurements, but at least equally important is the variety of tests that probe the situation in independent ways. For consider that a measurement interpreted by the wrong theory may be precise but quite wrong. The situation for tests of gravity physics and general relativity is illustrated in Figure~\ref{Fig:lengthscale}, where the probes are ordered by characteristic length scale, from the laboratory to the edge of the observable universe. The section numbers in parentheses indicate discussions of lines of research that originated in the 1960s and earlier. The picture has been made more complete by adding the results from observations of binary pulsars (Hulse  and  Taylor 1975), as noted in Section~\ref{sec:testingBD}, and the LIGO et al. (2016) gravitational wave detection. The picture could be made even more complete by marking the tests of the gravitational interaction on scales down to 0.01\,cm (Adelberger, Gundlach, Heckel, et al. 2009). Still other probes of gravity physics are discussed by the Planck Collaboration (2015b) and references therein. The point of the figure is that gravity has been looked at from many sides now, in phenomena that are observed and analyzed by a variety of methods, by measurements that span a broad range of length scales. The case for general relativity theory as a  good approximation to what actually has been happening rests on the abundance and variety of these tests. In my opinion they make the case about as good as it gets in natural science. 

Might there be a still better approximation, perhaps one that eliminates the need for the  hypothetical dark matter? The case for its existence is indirect, from the consistency of the network of cosmological tests. A precedent might be noted: Before the Reines-Cowan detection there was compelling indirect evidence for the existence of neutrinos from measured decay energy spectra and transition rates (as reviewed in Blatt and Weisskopf 1952, though there still was uncertainty about the Fermi and Gamow-Teller interactions). There is the big practical difference that the expected detection rate for neutrinos was known, while the parameters for direct detection of the dark matter are far less well characterized. But in the philosophy of this paper the weight of the indirect evidence for neutrinos in the early 1950s was about as good as the weight of the present indirect evidence for the dark matter of the $\rm\Lambda$CDM theory (at the time of writing). We may be surprised, of course, and  general relativity with its dark matter and cosmological constant certainly might be disestablished. That seems quite unlikely, but an arguably likely, and welcome, development would be the establishment of a still  better theory. It may show us how these hypothetical components, $\rm\Lambda$ and CDM, fit into the rest of physics, perhaps within an extension of the standard model for particle physics, perhaps in a generalization of general relativity. Some such eventuality seems likely, simply because it has been the normal course of events in the history of physics.

\section{Lessons from the naissance of experimental gravity physics}\label{sec:lessons}
This history of the empirical exploration of gravity in the 1960s offers some lessons that are  largely of historical interest because there no longer is much left of the low-hanging fruit of the 1960s. But some lessons are of broader significance, and are conveniently examined in some detail in the context of experimental gravity physics because the subject was such a small science in the 1960s. 

\subsection{Nonempirical evidence and empirical surprises}\label{sec:fractal}
The story of how gravity physics grew during the naissance shows the power of concepts of logic and elegance advocated by influential scientists informed by previously successful advances in science. This may be termed the importance of standard and accepted belief systems. The story also shows the power of empirical evidence, which can surprise us, and on occasion force revisions of the belief systems. 

In 1915 there was only one seriously demanding test of general relativity theory, from the motion of the planet Mercury, which orbits the Sun at roughly $10^{13}$\, cm. The empirical situation was little improved in 1955. But  a half century after that Einstein's theory has proved to pass a demanding network of tests on scales ranging from the laboratory to the edge of the observable universe, at the Hubble length $H_o^{-1}\sim 10^{28}$\,cm. The latter is an extrapolation of fifteen orders of magnitude in length scale from the best evidence Einstein had, from Mercury's orbit, and the former, down to the laboratory, is a similarly enormous extrapolation. The successes of these extrapolations of a theory discovered a century ago is a deeply impressive example of the power of nonempirical evidence. But not all elegant ideas can be productive, of course. I offer three examples drawn from this history of the naissance of experimental gravity physics. 

The distribution and motion of the mass around us has been shown to affect local inertial motion, in the Lense-Thirring effect. It has seemed natural to some to imagine that other aspects of how local physics operates might also be related to what is happening in the rest of the universe. We have an excellent empirical case that our universe is expanding and cooling from a very hot dense early state, and it has seemed natural to look for evolution of how local physics operates, perhaps to be described by evolution of the values of its dimensionless parameters. Perhaps proper clock rates defined by molecules, atoms, or atomic nuclei, which may be small enough that tidal stresses may be ignored, and may be adequately shielded from electromagnetic fields, cosmic rays, and all that, may still be affected by position or motion or cosmic evolution. As discussed in Section~\ref{sec:Mach}, this holistic concept has inspired some to think of an effective aether; some to think of Mach's Principle; Jordan to introduce a scalar-tensor theory that would express Dirac's Large Numbers Hypothesis; and Dicke to think of this scalar-tensor theory as expressing both  the LNH and Mach's Principle. An holistic concept led Einstein to his prediction of the large-scale homogeneity of the universe. If Einstein had consulted astronomers he would have received no encouragement, but the observable universe proves to agree with his thinking. But although this holistic thinking, which Einstein so successfully applied, has continued to seem beautiful to some, it has not led to anything of substance since Einstein. The searching tests reviewed in Section~\ref{sec:tests} agree with the idea that local physics, including gravitation, is quite unaffected by what the rest of the universe is doing.

For a second example, consider the opinion of influential physicists in the 1950s that the cosmological constant $\rm\Lambda$ is inelegant if not absurd. The situation then was reviewed in Peebles and Ratra (2003, \S III). The power of this thinking is seen in the 1980s through the 1990s in the general preference for the relativistic Einstein-de Sitter cosmological model, with  $\rm\Lambda=0$ and negligible mean space curvature. The first clear evidence that the observed galaxy peculiar velocities are smaller than expected in the Einstein-de Sitter model, if the space distribution of matter is fairly traced by the observed distribution of the galaxies, was presented by Davis and Peebles (1983). It was countered by the hypothesis that the galaxy space distribution is significantly more clumpy than the mass. The consequences of this biasing picture for the evolution of cosmic structure were first  examined by Davis, Efstathiou, Frenk, and White (1985). Biasing agrees with the observed greater clustering of the rare most massive galaxies, but it is not naturally consistent with the observed quite similar distributions of the far more abundant normal and dwarf galaxies. This was first seen in the CfA galaxy redshift survey maps (Davis, Huchra, Latham, and Tonry 1982, Figs. 2a and 2d). By the arguments summarized in Peebles (1986), the biasing picture was a nonempirical postulate meant to preserve $\rm\Lambda=0$ and avoid curvature of space sections. Some (including de Vaucouleurs 1982; Peebles 1986; Brown and Peebles 1987; Peebles, Daly, and Juszkiewicz 1989; Maddox et al. 1990; and Bahcall and Cen 1992) took seriously the growing empirical evidence that, assuming general relativity theory, either $\rm\Lambda$ or space curvature differs from zero. But the title of a paper adding to this evidence, {\it The baryon content of galaxy clusters: a challenge to cosmological orthodoxy} (White, Navarro, Evrard, and Frenk 1993), illustrates the general reluctance to abandon the orthodox Einstein-de Sitter cosmology with $\rm\Lambda=0$.

The preference for negligible large-scale space curvature was largely inspired by the inflation picture of the early universe, which in some scenarios was meant to account for the observed large-scale homogeneity by exceedingly rapid expansion of the early universe that might have  stretched out and thus suppressed observable departures from homogeneity. The great expansion also would have stretched out and suppressed space curvature. This preference for suppressed space curvature proves to agree with stringent bounds obtained much later (Planck Collaboration 2015b, \S 6.2.4, and references therein). It is another impressive success for nonempirical evidence. But the empirical evidence is that we must learn to live with $\rm\Lambda$ (or something in the stress-energy tensor that acts like it), even though its numerical value is nonzero but absurdly small compared to natural estimates from quantum physics. 

For the third example, consider Mandelbrot's (1975) argument that the galaxy space distribution may be a fractal. (In a simple fractal distribution, particles are placed in clusters, the clusters placed in clusters of clusters, the clusters of clusters in clusters of clusters of clusters, and so on up.) The mathematics of fractals is elegant, and there are interesting applications. Thus Mandelbrot explained that the measured length of the coastline of Brittany depends on the length resolution used to make the measurement. And Mandelbrot took note of de Vaucouleurs' (1970) point, that the  galaxy distribution in surveys then available certainly resembled a fractal. This line of argument, if more widely pursued earlier, could have had the beneficial effect of driving observational tests. But as it happened, interest in the elegance of a fractal universe started to grow just as empirical evidence for Einstein's large-scale homogeneity was at last emerging (as discussed in Sec.~\ref{Sec:cosmology}, and reviewed by Jones, Mart{\'{\i}}nez, Saar, and Trimble 2004). 

Elegant ideas tend to be lasting. The Steady State cosmology is elegant, as is the fractal concept, and the two arguably find their places in the multiverse of eternal inflation. The role of the scalar field that Jordan and Dicke envisioned is now seen to be at most an exceedingly subdominant effect at low redshifts, but the possible role of scalar fields in gravity physics still is discussed. Examples are superstring theories (Uzan 2003), models for inflation, and the search for a deeper gravity physics (Joyce, Jain, Khoury, and Trodden 2015). Also to be considered is the notion that the cosmological constant, $\rm\Lambda$, in the present standard cosmology may be an approximation to the energy of a slowly evolving scalar dark energy field (Peebles and Ratra 1988, who referred back to Dicke's  thinking about the slow evolution of the strength of gravity). The very weak strength of the gravitational interaction is still discussed, now under the name of the gauge hierarchy issue.  The idea that the gravitational interaction has been growing weaker as the universe expands, which so interested Dirac, Jordan, and Dicke (Secs.~\ref{sec:Mach} and~\ref{Sec:Gdot}), still is explored (Secs.~\ref{sec:LLR} and~\ref{sec:testingBD}). Also still explored is the possible evolution of the fine-structure constant, as exemplified by equation~(\ref{eq:alphadot}), and evolution of the other dimensionless physical parameters, though now inspired largely by issues of completion of quantum physics (as reviewed by Uzan 2003).  

Another old issue of completeness is the considerable difference between the relativistic constraint on the vacuum energy density, to be represented by the value of Einstein's cosmological constant if the vacuum is invariant under Lorentz transformations, and simple estimates of the sum of energies of the fields of particle physics. Early thinking was reviewed in Peebles and Ratra (2003, Sec. B.3). In particular, Rugh and Zinkernagel (2002) recalled Wolfgang Pauli's recognition in the 1930s that the zero-point energies of the modes of the electromagnetic field presented a problem. In the English translation (Pauli 1980) of Pauli's (1933) {\it Die allgemeinen Prinzipien der Wellenmechanik}, Pauli wrote that, in the quantization of the electromagnetic field, 
\begin{quotation}
\noindent a zero-point energy of ${1\over 2} \hbar\omega_r$ per degree of freedom need not be introduced here, in contrast to the material oscillator. For \ldots this would lead to an infinitely large energy per unit volume because of the infinite number of degrees of freedom of the system \ldots Also, as is obvious from experience, it does not produce any gravitational field.  
\end{quotation}
The zero-point energy of each mode of a matter field, which is computed by the same quantum physics, must be real to account for measured binding energies, but these zero-point energies also add up to a absurdly large---negative for fermions---mean vacuum energy density. This curious situation was discussed in passing in Gravity Research Group meetings in the 1960s, but not to the point of publishing, because we did not know what to make of it. Zel'dovich (1968) published, and pointed out that the vacuum energy likely appears in the form of Einstein's cosmological constant, though with an apparently absurd value. Zel'dovich's paper was visible enough, and the issue for gravity physics and quantum physics is deep enough, to be entered in the timeline as line~37.

The Anthropic Principle, for which one might claim nonempirical evidence, offers the thought that we live in an element of a multiverse, or in a part of a really extended universe, in which the sum of zero-point energies, latent heats, and whatever else contributes to the vacuum energy density, happens to be small enough to allow life as we know it (Weinberg 1987). The earlier version of this line of thinking that we heard in the Gravity Research Group was that the age of the expanding universe has to be on the order of $10^{10}$ years, to allow time for a few generations of stars to produce the heavy elements of which we are made, but not so much time that most of the stars suitable for hosting life on a planet like Earth around a star like the Sun have exhausted their supply of nuclear fuel and died (Dicke 1961b). The resolution of the vacuum energy puzzle by the nonempirical evidence of the Anthropic Principle is considered elegant by some, ugly by others, a healthy situation in the exploration of a serious problem.

\subsection{Prepared and unprepared minds}
It is observable in this history of experimental gravity physics that ``chance favors the prepared mind'' (in an English translation of the thought attributed to Pasteur). Consider, for example, how quickly Pound and Rebka turned the announcement of M{\"o}ssbauer's effect into a laboratory detection of the gravitational redshift (Sec.~\ref{sec:GRinthe50s}). In his recollections Pound (2000) mentioned Singer's (1956) discussion of how an atomic clock in an artificial satellite might be used to detect the gravitational redshift. Pound recalled that he and Rebka recognized that the M{\"o}ssbauer effect presented an opportunity and ``a challenge; namely, to find a way to use it to measure relativistic phenomena, as I had wanted to do with atomic clocks; however, the clocks had not proved sufficiently stable.'' Dicke's letter to Pound (Sec~\ref{sec:preferredmotion}) showed that, when Pound and Rebka (1959) announced their intention to do this experiment, at least two other groups also were working on it. We see that minds were well prepared for this experiment. Wilson and Kaiser (2014) considered how Shapiro's thinking about planetary radar ranging experiments helped prepare his mind for the measurement of the relativistic effect on the time delay of planetary radar pulses that pass near the Sun (Sec.~\ref{sec:gravredshift}). Dicke was prepared and searching for probes of gravity physics. For example, as NASA was learning to fly rockets Dicke and his group were exploring how to use this new space technology for precision measurements of the orbits of satellites, perhaps by tracking angular positions of corner reflectors illuminated by searchlights, and then, when the technology allowed it, turning to pulsed or continuous wave lasers for precision ranging to corner reflectors on the Moon. The results of Lunar Laser Ranging, after considerably more work, had an intended consequence, the production of demanding tests of gravity physics. For Dicke, of course, an unintended consequence was a much tighter bound on evolution of the strength of gravity than he had expected.

We may consider also a situation in which minds were not prepared. The research by Gamow and colleagues in the late 1940s, on element formation in the early stages of expansion of a hot Big Bang cosmology, has grown into a demanding set of tests of gravity physics (as reviewed in Sec.~\ref{Sec:GR-CMB}). Gamow (1953) outlined this research in lectures at the 1953 Ann Arbor {\it Symposium on Astrophysics}. But I have not found any mention of the research by Gamow's group in the proceedings of the four international conferences on general relativity and gravitation that I have discussed in this history: Bern in 1955,  Chapel Hill in 1957,  Royaumont in 1959, and NASA in 1961. There is no mention of it in the proceedings of the next  conference in this series, the 1962  Warszawa and Jablonna {\it Conf\'erence internationale sur les th\'eories relativiste de la gravitation} (Infeld 1964), there is no mention in two IAU Symposia where it might have figured:  the {\it Paris Symposium on Radio Astronomy}, 1958 (Bracewell 1959), and {\it Problems of Extra-galactic Research}, Berkeley California, 1961 (McVittie 1962), and no mention in the {\it Solvay Conference on Cosmology}, Brussels 1958 (Stoops 1958).  Bondi's (1952, 1960) book, {\it Cosmology}, which was the best review of research in cosmology during the naissance of experimental gravity, gave references to papers on element formation by Gamow's group, in a list of papers for further reading, but there was no discussion of these ideas in the text. We may conclude that, until 1965, the leading figures in relativity and cosmology were not prepared for the sea of thermal microwave radiation, the CMB. They did not know about Gamow's (1948) ideas or else they did not consider them promising. 

Donald Osterbrock attended Gamow's Ann Arbor lectures, and remembered Gamow's idea of helium production in a hot Big Bang (as Osterbrock 2009 recalled). Geoffrey Burbidge also was at the Ann Arbor conference. Burbidge (1958) later commented on evidence that the helium abundance in our galaxy is larger than might be expected from production by stars. He did not mention Gamow's thinking, however. If Burbidge attended Gamow's lectures he had forgotten them, or considered them unpromising.  Osterbrock and Rogerson (1961) added to the evidence for a large helium abundance in the Milky Way, even in the apparently older stars that have lower abundances of heavier elements. They pointed out that ``the build-up of elements to helium can be understood without difficulty in the explosive formation picture." Their reference was to Gamow (1949). This is the first published announcement of a possible detection of a fossil from a hot early stage of expansion of the universe. It received no significant notice. Hoyle and Tayler (1964), apparently independently, also announced the possible significance for cosmology of the large cosmic helium abundance. We cannot know how this paper would have been received if another candidate remnant from a hot early universe, the sea of microwave radiation, had not been announced at close to the same time. This radiation was first detected, as anomalous noise in microwave telecommunications experiments, at the Bell Telephone laboratories (Sec.~\ref{Sec:CMB}). Engineers were aware of the anomaly in 1959 (DeGrasse, Hogg, Ohm, and Scovil 1959). Five years later Penzias and Wilson had made a good case that the anomalous noise could not be instrumental or terrestrial. But their minds were not prepared for cosmology until they learned of the search for fossil microwave radiation by Dicke's Gravity Research Group. 

What delayed general recognition of Gamow's hot early universe theory?  Weiss (2016) pointed to two factors that in his experience might have contributed to the delay. First, Gamow's plan had been to account for the origin of all the elements (apart from possibly subdominant contributions by nuclear burning in stars). By 1950 Enrico Fermi and Anthony Turkevich, at the University of Chicago, had concluded that element buildup in the hot early universe would very likely end at the isotopes of hydrogen and helium. The failure of Gamow's plan may have tended to obscure the still viable---and now established---idea of formation of isotopes of the lightest elements. The second factor was the absence of experimentalists in Gamow's group.\footnote{Fermi and Turkevich certainly understood experiments. But their contribution to Gamow's program was a theoretical analysis, which they did not bother to publish, apart from giving their results to Alpher and Gamow.} This is to be contrasted with the active interactions of theory and practice in Dicke's group. In 1964 Peter Roll and David Wilkinson had little prior experience in microwave technology. The prior feeling I had for cosmology was that the homogeneous  expanding universe solution to Einstein's equation seemed to be grossly oversimplified so as to offer workable exam problems; I thought it was to be compared to the acceleration of a frictionless elephant on an inclined plane. But we encouraged each other in our willingness to learn experiment and theory from Dicke's challenges.

Another factor that must be mentioned was Gamow's supreme lack of interest in details. In the 1950s he continued to write important papers; a notable example was the demonstration of the significant time-scale challenge to the Steady State cosmology (Gamow 1954, as discussed in Sec.~\ref{sec:expandinguniverse}). But Gamow (1953a, 1953b, 1956) turned from his perceptive 1948 physical argument for a hot Big Bang to a far less persuasive picture. His new picture assumed that the expansion rate predicted by the Friedman-Lem\^\i tre equation became dominated by space curvature  just when the mass densities in nonrelativistic matter and in the sea of thermal radiation passed through equality (with cosmological constant $\rm\Lambda = 0$). Knowing the present matter density and Hubble's constant (for which Gamow took $\rho_m(t_o)= 10^{-30}\hbox{ g cm}^{-3}$ and $H_o^{-1}=10^{17}$\,s or a little longer), the model predicts present temperature $T_o=6$ to~7\,K (Gamow 1953a, 1956).  This specific prediction, of a not very low temperature, could have encouraged an experimental program aimed at detecting Gamow's radiation.\footnote{Bernard Burke (2009, p. 182) concluded that the resources were available to detect the CMB at $T_o=2.725$\,K in the 1950s. But I do not know of any evidence that anyone actually considered looking for Gamow's 6 to 7\,K thermal radiation.} But a closer look could have discouraged the project, because it would have revealed that there was no basis for the starting assumption. The approach only places an upper bound of about 50\,K on the present temperature, and no lower bound.\footnote{In his popular book, {\it Creation of the Universe}, Gamow (1952) arrived at present temperature $T_o= 50$\,K (p. 52), which indeed follows from his adopted age of the universe, $t_o=10^{17}$\,s, under the assumption that the expansion rate was dominated by radiation up to the present. This is a physically reasonable upper bound on the present radiation temperature in a relativistic cosmology given the present age. Gamow may not have meant it to be a serious estimate of $T_o$, however. On page 78 he wrote that at the time of equality of mass densities in matter and radiation the radiation temperature was $T_{\rm eq}=300$\,K. With his standard value for the present matter density this scales to present temperature $T_o=7$\,K, the value in Gamow (1953b).} An interested experimentalist would have had to go back to Gamow (1948) to see a well-motivated physical argument for a present temperature of this order. Chernin (1994) argued for the elegance of Gamow's (1953a, 1953b, 1956) simplifying assumption that allowed a succinct and reasonably accurate account of cosmic evolution with present radiation temperature close to the Alpher and Herman (1948) estimate,\footnote{The situation in 1948 was confused (Peebles 2014). The Alpher and Herman CMB temperature estimate was based on their fit to the broad range of observed cosmic element abundances. This approach was soon found to fail. But Gamow's (1948) estimate of the conditions for light element formation, which are close to the now established cosmology, extrapolate to $T_o=8$\,K for Gamow's assumed present baryon density (Peebles 2014, eq. [24]).} about 5\,K. This is a fair point. But the considerations in Gamow (1948) are elegant too, they are about as simple, and they are  based on a specific physical condition, that thermonuclear reactions in the early universe produce a significant but not excessive mass fraction in atomic weights greater than the proton. It is difficult to understand why Gamow turned to the much less well motivated theory he presented in the 1950s.  

One may also wonder why there were such different community reactions to two announcements of evidence of fossils from a hot early universe. The first, by Osterbrock and Rogerson (1961), that the unexpectedly large abundance of helium may be a remnant from the hot early universe, received little notice. The second, by Dicke, Peebles, Roll, and Wilkinson (1965), that the unexpectedly large noise in Bell communications detectors may be a remnant from the hot early universe, received abundant attention. The latter was the more interesting to physicists, because they could set about measuring the radiation spectrum and angular distribution. But the former certainly could interest astronomers, who could have been motivated to more closely examine helium abundances in stars and nebulae, and the processes of stellar helium production and dispersal. Perhaps those who study the sociology of science are best positioned to explore why the presence of more microwave radiation than expected from known radio sources received so much more attention than the presence of more helium than expected from production in known stars. The rest of us might bear in mind that nature is quite capable of surprising us.

\subsection{Speculative and programmatic experiments}\label{classesofexpts}
This history presents us with examples of experiments that were purely speculative, done simply because they were  possible;  experiments that were inspired by ideas such as those just discussed that were speculative according to accepted ideas but attractive in other philosophies; and experiments that may be termed programmatic: designed to find what was expected from standard and accepted ideas. 

A good illustration of the first, speculative, class is the experimental test of equivalence of active and passive gravitational masses (Sec.~\ref{Sec:masses}). Discovery of a violation would be shocking, but one must consider that during the naissance little was known about empirical gravity physics, and such purely speculative experiments were well worth doing to help improve the situation. It would be difficult now to find support for a more precise laboratory test, which is regrettable; physics should be challenged.

The second class includes the tests for sensitivity of local physics to what the rest of the universe is doing. Some of these experiments were inspired by thoughts of a luminiferous aether of some sort, others by thoughts of Mach's Principle and Dirac's Large Numbers Hypothesis. Explorations of these ideas motivated experiments that have informed us about the nature of gravity. Consider, for example, Dicke's fascination with the idea that the strength of the gravitational interaction may be evolving, and that precision tracking of the orbit of the Moon might reveal the effect. Without Dicke's persistence and influence, would NASA, and its counterpart in the Soviet Union, have gone to the trouble of placing corner reflectors on the Moon?  We cannot answer, of course, but we do see how elegant ideas may lead to  great results, as in the demanding tests of gravity physics from the Lunar Laser Ranging Experiment.

I place in the programmatic class of experiments during the naissance the measurements of the gravitational  redshift and  deflection of light, and the search for tensor gravitational waves. This is the delicate art of finding what one expects to find (Hetherington 1980). One might argue that when these experiments were done during the naissance they belonged to the second class, because they were inspired by a theory that was not empirically well supported. The distinction is that the theory was broadly accepted, on nonempirical grounds. But empirical evidence is far better, in the philosophy of science as  exploration of the world around us. The programmatic searches for what general relativity theory predicts have been deeply important to this empirical establishment. 

As gravity physics grew the distribution of experiments among the three classes evolved, from considerable activity on the purely speculative side during the naissance to the present emphasis on the programmatic side, now that we have a well-established theory that tells us what to look for. This programmatic side is essential to the experimental gravity physics program, but it is not minor pedantry now to pursue experiments designed to be sensitive to departures from the standard model, such as checks of equivalence of the four masses defined in equation~(\ref{eq:kindsofmass}), or tests of the gravitational inverse square law (e.g. Adelberger, Gundlach, Heckel, et al. 2009), or searches for cosmic evolution of physical parameters such as ratios of elementary particle masses and the strengths of the fundamental interactions. All our physical theories are incomplete, the world is large, and it has ample opportunities to surprise us yet again. 

\subsection{General purpose and purpose-built instruments}
The pendulums usually seen in teaching laboratories for measurements of the acceleration, $g$, of gravity look very different from the ones illustrated in Figure~\ref{Fig:pendulum} and used in the Hoffmann (1962) and Curott (1965) experiments to probe possible evolution of $g$. They look  very different again from the sketch in Figure~\ref{Fig:Faller} of Faller's (1963) falling corner reflector experiment to measure the absolute value of $g$. These Princeton experiments, and the two versions of the E\"otv\"os experiment (Liebes 1963; Roll, Krotkov, and Dicke 1964), were designed, or we may say purpose-built, to be optimum for a specific measurement. The Pound and Rebka (1959) laboratory measurement of the gravitational redshift was a purpose-built experiment too, but it was inspired by the appearance of a new tool, the M{\"o}ssbauer effect, which made the experiment possible. The Princeton experiments certainly made heavy use of new tools as they became available, but many could have been done a decade earlier, though, I am informed, distinctly less well, or might have been done instead a decade later, and better. These are examples of experiments that awaited someone to act on the idea that a closer examination of a particular issue is worth doing, perhaps by design of a purpose-built experiment. 

At Gravity Research Group meetings Dicke told us about his preference for purpose-built experiments, and his dislike of sharing raw experimental data that others could reanalyze in foolish ways. The closest he came to Big Science was in the Lunar Laser Ranging Experiment, but it is to be observed that as that project grew he withdrew. The community has adapted to the working conditions of Big Science; shining examples are the precision measurements of the CMB, the  precision statistical measures of the natures and distributions of the galaxies, and the LIGO detection of gravitational waves. Dicke perhaps could not have anticipated the power of  ``data mining'' in modern observational surveys that can stimulate thinking  about  unexamined ideas. But there still is a good case for his preference for an experiment designed for optimum examination of a particular issue. For example, the Sloan Digital Sky Survey (Alam, Albareti, Allende, et al. 2015 and references therein) has made wonderfully broad contributions to our knowledge of the statistical properties of galaxies, but of course it could not have been designed for optimum exploration of all issues. An example is the spiral galaxies in which only a small fraction of the observed stars rise well above the disk (Kormendy, Drory, Bender, and Cornell  2010). Such galaxies are common nearby, and they are a fascinating challenge for theories of galaxy formation based on the $\rm\Lambda$CDM cosmology discussed in Section~\ref{Sec:GR-CMB}, because the predicted tendency of galaxies to grow by merging would tend to place stars in orbits that rise well above the disk.  Further exploration of this interesting phenomenon might best be served by an observational program designed for optimum examination of the properties of this particular class of galaxies.

\subsection{Support for curiosity-driven  research}\label{sec:militaryfunding}
The financial support for speculative curiosity-driven research during the naissance in experimental gravity physics is worth recalling, because science and society have changed since then. Research by Dicke and his Gravity Research Group was supported in part by the United States National Science Foundation, and in part also by the United States Army Signal Corps in the late 1950s, by the Office of Naval Research of the United States Navy through the 1960s, and in some papers one also sees acknowledgement of support by the U. S. Atomic Energy Commission. The Hoffmann, Krotkov,  and Dicke  (1960) proposal for precision tracking of satellites was published in the journal {\it IRE Transactions on  Military Electronics}. The first paper in the issue, by Rear Admiral Rawson Bennett, USN, opened with the sentence ``This issue \ldots is devoted to the United States Navy's interest and effort in space electronics.'' The military certainly had reason to be   interested in space electronics, but surely had little interest in the Hoffmann et al. proposal to test the idea that the strength of gravity may be changing by about a part in $10^{10}$ per year. But the military seemed to be comfortable supporting what Dicke and his group wanted to investigate.

This situation was not unusual. High energy physics papers often acknowledged support from military  agencies. In experimental gravity physics the development of the maser used in the Cedarholm et al. (1958) aether drift test (Sec.~\ref{sec:preferredmotion}) was supported ``jointly by the Signal Corps, the Office of Naval Research, and the Air Research and Development Command.'' The Pound and Rebka (1960) laboratory detection of the gravitational redshift (Sec.~\ref{sec:gravredshift}) acknowledged support ``in part by the joint program of the office of Naval Research and the U. S. Atomic Energy Commission and by a grant from the Higgins Scientific Trust.'' Irwin Shapiro, then at the MIT Lincoln Laboratory, recalled (Shapiro 2015) that application of his new test of general relativity by planetary radar ranging (Sec.~\ref{sec:gravredshift}) required a more powerful transmitter, and that after due deliberation the director of Lincoln Laboratory ``called up an Air Force general, who provided Lincoln Laboratory with funding, and asked him for \$500,000, which he then granted, for a new transmitter to enable Lincoln to carry out the experiment.'' This substantial financial contribution allowed a substantial advance in gravity physics. Wilson and Kaiser (2014) analyzed how the planetary radar experiments by Shapiro and colleagues may have been related to the US military research effort to detect and perhaps somehow learn to deal with an incoming USSR intercontinental ballistic missile. But Shapiro (2015) recalled that the military did not make any attempt to guide directions of the research by him and his colleagues on testing general relativity. That also was the experience in Dicke's Gravity Research Group.

The thinking by military funding agencies at the time may have combined the thoughts that this curiosity-driven research did not cost much, perhaps apart from exceptional cases such as Shapiro's transmitter, could do no harm, might lead to something of eventual value to society and the military, and perhaps also might help the military stay in contact with people whose expertise and advice they might seek on occasion. Weiss added that
\begin{quotation}
\noindent The military was absolutely the most wonderful way to get money. Their
mission at that time---and that's something that's grossly misunderstood by all the people that
got into trouble with Vietnam and everything else---the military was in the business of training
scientists. They wanted not to get caught again the next time there was a [need for a] Manhattan
Project or a Rad Lab [MIT Radiation Laboratory]. 
\end{quotation}
This permissive attitude of military funding agencies to speculative ideas had the result that in Dicke's group new directions of research often were pursued and the results then reported to the funding agencies, without prior approval of a well-reasoned motivation. This has changed. Compare Dicke's invitation to his graduate student Jim Faller to make an absolute measurement of ``little $g$'' (Sec.~\ref{sec:abs-g}) to the recent US National Science Foundation invitation to propose an absolute measurement of ``Big $G$''.\footnote{\url{http://www.nsf.gov/pubs/2016/nsf16520/nsf16520.htm}}  This evolution of thinking, from bottom-up toward top-down, is natural in a maturing science, though disturbing from the standpoint of innovation. During the naissance the scant empirical basis for gravity physics made it very appropriate to pursue the speculative curiosity-driven experiments that the military agencies were inclined to support. But the 1970 Mansfield Amendment ended this by prohibiting the Defense Department from funding ``any research project or study unless such project or study has a direct and apparent relationship to a specific military function.'' The last of the Gravity Research Group papers to acknowledge support from the ONR was published in 1972. And the arteries of natural science hardened a little.

\subsection{Unintended consequences of curiosity-driven research}
Pure curiosity-driven research has had such great unplanned consequences that custom may obscure recognition. So let us remember Dicke's invitation to Jim Faller to measure the acceleration of gravity by dropping a corner reflector (Sec.~\ref{sec:abs-g}), which grew into technology to measure changes in water table levels and the continental rebound from the last Ice Age. Remember also  Dicke's interest in probing the physics of temporal and spatial variation of the strength of gravity, and probes for gravitational waves, which led him to invite Bill Hoffmann and David Curott to design pendulums suited for the purpose, and Barry Block, Bob Moore, and Rai Weiss to do the same with spring-type (LaCoste) gravimeters. Section~\ref{sec:gravitationalwaves} recalls how this contributed to the Global Seismographic Network that monitors phenomena of practical interest to us all:  earthquakes, tsunamis, storm surges, and underground explosions. 

Of course, a good deal of curiosity-driven research serves only to satisfy curiosity. The Global Seismographic Network offers measures of the internal structure of the Earth, which certainly is interesting, though perhaps most satisfying to the curiosity of specialists. John Wheeler was fascinated by ``thought experiments'' that illustrate the curious apparently acausal nature of quantum physics. He discussed this at an Einstein Centennial Address delivered at the University of Maryland in 1979, on the centenary of Einstein's birth. Bill Wickes and Carroll Alley were in the audience. Wickes (1972) had completed his thesis with Dicke toward the end of the naissance; Alley's research with Dicke is reviewed in Sections~\ref{sec:gravredshift} and~\ref{sec:LLR}. Wickes (2016) recalled that 
\begin{quote}
\noindent Carroll and I independently realized that we could actually do Wheeler's double-slit delayed choice experiment, using some of the lasers and fast-switching methods that Carroll had pioneered for the lunar ranging work.  So we joined forces and recruited Oleg Jakubowicz to do the work for his Ph.D. thesis \ldots I like to think of the whole project as an elegant intersection of Wheeler's imagination and Dicke's practical tutelage.
\end{quote} 
This story of how a Wheeler ``thought experiment'' became a real experiment is told in Wheeler and Ford (1998, pp 336-338). Quantum physics is not really acausal: It does not allow you to foresee movements of the stock market, even in principle. But it is satisfying to see real experimental demonstrations of the well-advertised non-intuitive nature of quantum physics, even to non-specialists who take an interest in the world around us. And it is to be observed that many who are not involved in research in natural science find it satisfying to know that Einstein's general relativity theory of gravity, which he discovered a century ago, has  been experimentally shown to be a good approximation to how gravity actually operates.

\subsection{Establishing and disestablishing elements of natural science}\label{sec:establishment}
I offer some concluding thoughts about the empirical establishment of general relativity theory discussed in Section~\ref{theempiricalcase}. It is challenging if not impossible to define our working scientific philosophy. This attempt is meant to explain why the empirical case for general relativity theory is to me about as persuasive as it gets in natural science, even though the theory is manifestly incomplete. 

Common thinking in physical science, and adopted here, is that a theory is empirically established if it produces substantially more successful predictions than it allows adjustable parameters. The latter must take account of the fact that a sensible theorist will choose to work on the ideas that look most promising from the evidence at hand. Thus tests of gravity physics on the scale of the Hubble length depend on the $\rm\Lambda$CDM theory of cosmic structure formation, which grew out of several under discussion in the 1990s (Peebles and Silk 1990). The choice, initially CDM, was informed by the phenomena as well as simplicity. In effect, this was a free parameter to be added to the other adjustments made to fit the theory to the evidence. 

Also to be considered is the hazard Hetherington (1980) termed ``finding too facilely what they expected to find.'' Hetherington's example is the incorrect observational confirmation of an incorrect prediction of the gravitational redshift of the white dwarf star Sirius~B (reviewed in Sec.~\ref{sec:gravredshift}). Another example is Dawid (2015) comment that Einstein's general relativity ``had been confirmed  by Eddington's measurement of starlight bending in 1919.'' The measurement was greeted by the media as a great triumph for Einstein's theory, which perhaps is what Dawid had in mind. But the measurement was dubious, and if correct it only added ``very flimsy evidence on which to hang a theory'' (Dicke 1957a, as discussed in Sec.~\ref{Sec:WheelerDicke}). A third example is the BICEP2 measurement of the pattern of polarization of the CMB. The measurement was rightly welcomed as an important experimental advance. The initial interpretation of the BICEP2 measurement was that it had detected the effect of gravitational waves to be expected if inflation in the very early universe set our initial conditions, and if the expansion rate during inflation had been rapid enough to have produced waves of the proposed amplitude. This was greeted initially as a great triumph, a demonstration that inflation really happened. But it was soon understood that the experiment was not a credible detection, but rather an upper bound on the possible effect of inflation, if inflation actually happened. 

There could be undetected examples of misleading ``finding the expected'' in the tests of gravity physics summarized in Section~\ref{Sec:GR-CMB}. Consider for example the SNeIa redshift-magnitude relation marked in Figure~\ref{Fig:lengthscale}. It was a brilliant completion of Sandage's (1961) great goal for a test of cosmology.  But if this had been the only evidence for the detection of Einstein's cosmological constant, $\rm\Lambda$, the community would have had a choice: accept $\rm\Lambda$ in the standard model, or argue that that the evidence of its detection is misleading because, despite very careful tests that argue to the contrary, SNeIa at high redshift had lower luminosities than apparently identical ones nearby. In this hypothetical situation, with no other relevant evidence, I expect the community would have argued for the latter,  because a nonzero value of $\rm\Lambda$ is awkward from a theoretical point of view, and it destroys the elegant simplicity of the Einstein-de Sitter model that was so influential in the 1990s. This would have been an example of willful disregard of the unexpected. In fact this evidence for $\rm\Lambda$ was not disregarded, perhaps in part because the community was too sensible, but certainly also because the CMB anisotropy measurements, with the astronomers' value for the extragalactic distance scale, $H_o^{-1}$, also  were seen to be pointing to nonzero $\rm\Lambda$. (The situation was reviewed in Peebles, Page, and Partridge 2009, pp. 447-477.) And, at the time, measurements of stellar evolution ages with the astronomers  $H_o^{-1}$ were found to be difficult to understand in the absence of a nonzero $\rm\Lambda$. This litany of crosschecks for other tests may be continued at considerable length. And as discussed in Section~\ref{Sec:GR-CMB}, it is this abundance of crosschecks that argues against excessive willful misinterpretation of the evidence, and empirically establishes general relativity theory as a good approximation to reality.

The result is not what Dicke anticipated in the mid-1950s when he decided to remedy the scant attention to experimental gravity physics. I do not recall his mentioning the tightening evidence against the scalar-tensor theory; Parkinson's disease might have slowed his recognition of these developments. But his big questions remain. Is local physics really not related to the rest of the universe, apart from the tantalizing effect on inertial motion? Is the enormous difference of strengths of gravity and electromagnetism really only a result of anthropic selection? Is the classical theory of gravity really in a satisfactory state?

\subsection{Future developments}
Among open issues in gravity physics, the one most immediately relevant to the empirical theme of this paper may be the question of what happened in the very early universe, before classical general relativity theory could have been a good approximation. It may be instructive to compare thinking now and in the 1950s. General relativity theory then was widely accepted as  logically compelling. Now the inflation picture for the very early universe is accepted by many to be promising, even logically compelling. There is the serious difference that general relativity is a theory, while inflation is a framework on which to hang a theory. But there is a serious similarity. In the 1950s the empirical basis for general relativity was generally considered to be necessarily schematic, because better experiments were not feasible. Now the empirical basis for inflation, or other ideas about the very early universe, is considered to be necessarily schematic, because better experiments are not feasible. The community was surprised by the abundance of evidence that has grown out of the naissance of experimental gravity physics. If the community will not be surprised by another harvest of tests it will have to be content with evidence from elegance and logical completeness. But the communities of natural science have been surprised by many great empirical advances. This experience suggests there may be more surprising empirical developments to come, perhaps even some that postpone the need for nonempirical assessment of some future deeper theory of the nature of gravity and how the world began.

\begin{acknowledgement}
I have greatly benefitted from recollections, guidance to the literature, and advice on the preparation of this paper, from Eric Adelberger, Pete Bender, Jon Berger, Bart Bernstein, Steve Boughn, Paul Boynton, Richard Dawid, Dieter Brill, David Curott, Jim Faller, Masataka Fukugita, Richard Garwin, Henry Hill, Bill Hoffmann, David Kaiser, Helge Kragh, Bob Krotkov, Jeff Kuhn, Adele La Rana, Sid Liebes, Ed McDonald, Martin Mchugh, Charlie Misner, Lyman Page, Bruce Partridge, Peter Roll, Irwin Shapiro, Joe Taylor, Scott Tremaine, Virginia Trimble, Rainer Weiss, Bill Wickes, and Clifford Will. I must make special mention of Masataka Fukugita, Bill Hoffmann, Bruce Partridge, Irwin Shapiro, Virginia Trimble, Rainer Weiss, and the editors of this journal, for their careful readings and annotated commentaries on how to improve the paper; Dieter Brill, for his notes of what was said at Gravity Group meetings; Sid Liebes, who made the drawing of the interferometer on the left side of Figure~\ref{Fig:EotvosDetectors}; and Bill Hoffmann, who had in his possession the drawing of the interferometer in Figure~\ref{Fig:Faller}, and  made the first draft of the tables in  Appendix A. I must also recognize with gratitude Linda Chamberlin's help in guiding me through the vast but not always well organized resources of the Princeton University Library. This research was supported through the good offices of the Department of Physics, Princeton University. 
 \end{acknowledgement}
 
 \bigskip\bigskip

{\noindent\bf\large Appendix A: Dicke's Gravity Research Group}\medskip

\noindent Examples of research in Dicke's Gravity Research Group were discussed in Sections~\ref{sec:GravityGroup} and~\ref{sec:tests}. This Appendix is meant to give a broader picture by listing all of Dicke's PhD graduate students and all the post-PhD members of his Gravity Research Group. The  records I have may be incomplete; I would be grateful for information about anyone I have overlooked or misidentified. The dissertation titles serve to illustrate Dicke's quite abrupt mid-career change of direction of  research.

Table 1A lists the pre-gravity PhD theses Dicke directed, with title and year of acceptance of the thesis by the Princeton Department of Physics. The compilation of Table~2A, on research in the Gravity Research Group, required somewhat more creative rules. A graduate student who completed a PhD under Dicke's direction is marked as PhD in the second column, and the third column gives the thesis title and date of acceptance. Some graduate students published papers in addition to their dissertations while they were group members, and some stayed in the group as postdocs after completion of the PhD,  but none of this information is entered. Those who joined the group as a postdoc, instructor, or assistant professor are marked as post-PhDs in the second column. The third column in this case lists the title and reference for the first publication reporting research done in the group  (but excluding multiple-author papers that are already in the table.) Some titles are shortened to fit.  A more complete picture would list visitors, some of whom stayed for a day or two, perhaps to present a seminar, and others stayed longer, but I do know of reasonably complete records. Dicke's publications with colleagues outside the group, and his single-author papers, are not tabulated, but almost all of his papers on gravity are discussed in the  text. Apart from special cases to be noted the undergraduate papers and theses Dicke directed are not entered.

Some entries in Table 2A require special explanation. Dicke advised Lawrence Cathles's 1965 Princeton  undergraduate paper on {\it The physics of glacial uplift}. Cathles wrote his PhD thesis on deeper exploration of the same subject, in the Princeton Department of Geological and Geophysical Sciences, under the direction of Jason Morgan, who wrote his PhD thesis under Dicke's direction in the Department of Physics.  While Henry Hill was in the gravity group he directed Carl Zanoni's 1957 PhD research, which is entered. Curtis Callan was most closely associated with Sam Treiman's  research in elementary particle theory, but Dicke proposed and supervised research on his dissertation. Wheeler supervised Dieter Brill's 1959 PhD thesis, but Brill often attended Gravity Group meetings, and he wrote a 1962 review of experimental tests of relativity with Bob Krotkov, who was in the Gravity Group, and Bruno Bertotti, who was visiting the group. 

\begin{table}[t]
\centering
\begin{tabular}{ l l  }
\multicolumn{2}{c}{Table 1A: Graduate student research with Dicke, pre-gravity}\\
\noalign{\medskip}
\toprule
\toprule
 Name & Research\\
\noalign{\smallskip}
\toprule
Alexander Pond & A Experimental Investigation of Positronium (1952)\\
George Newell & A Method for Reducing the Doppler Width  \\
& \qquad of Microwave Spectrum Lines (1953)\\
Bruce Hawkins& The Orientation and Alignment of Sodium Atoms \\
& \qquad by Means of Polarized Resonance Radiation (1954)\\
Robert Romer & A Method for the Reduction of the Doppler \\
& \qquad Width of Microwave Spectral Lines (1955)\\
James Wittke & A Redetermination of the Hyperfine Splitting \\
& \qquad in the Ground State of Atomic Hydrogen (1955) \\
Christopher Sherman & Nuclear Induction with Separate \\
& \qquad Regions of Excitation and Detection (1955) \\
Lowell White & The Gyromagnetic Ratio of the Electron in the \\
& \qquad Metastable State of Hydrogen (1956) \\
Peter Bender & The Effect of a Buffer Gas on the Optical \\
& \qquad Orientation Process in Sodium Vapor  (1956)\\
Edward Lambe & A Measurement of the g-Value of the Electron\\
& \qquad in the Ground State of the Hydrogen Atom (1959) \\
\noalign{\smallskip} 
\bottomrule
\end{tabular}
\end{table}

\begin{table}[t]
\centering
\begin{tabular}{ l l l }
\multicolumn{3}{c}{Table 2A: Research with Dicke in the Gravity Group}\\
\noalign{\medskip}
\toprule
\toprule
 Name & Status & Research\\
\noalign{\smallskip}
\toprule
Robert Krotkov & post-PhD & Comparison between theory and observation \\
&&  \qquad for the outer planets (Krotkov and Dicke 1959)\\
Carl Brans & PhD & Mach's Principle \& Varying Gravitational Constant (1961)\\
James Peebles & PhD & Observational Tests and Theoretical Problems with Variable\\
&&  \qquad   Strength of the Electromagnetic Interaction (1961) \\
Carroll Alley & PhD & Optical Pumping and Optical Detection Involving \\
&&  \qquad Microwave \& Radio Frequency Coherence Effects (1962) \\
William Hoffmann &  PhD & A Pendulum Gravimeter for Measurement of Periodic\\
&&  \qquad  Annual Variations in the Gravitational Constant (1962) \\
Kenneth Turner & PhD & New Limit on Velocity Dependent Interaction\\
&&   \qquad  Between Natural Clocks and Distant Matter (1962)  \\
James Brault & PhD & The Gravitational Red Shift in the Solar Spectrum (1962)\\
Dieter Brill & post-PhD & Experiments on Gravitation (Bertotti, Brill, Krotkov 1962§¨)\\
James Faller & PhD & An Absolute Interferometric Determination of \\
&&  \qquad the Acceleration of Gravity (1963) \\
John Stoner & PhD & Production of narrow balmer spectrum lines in an \\
&&  \qquad electron-bombarded atomic hydrogen beam (1963)\\
Henry Hill & post-PhD &  Experimental Limit on Velocity-Dependent Interactions\\
&&  \qquad  of Clocks and Distant Matter (Turner and Hill 1964)\\
Sidney Liebes & post-PhD & Gravitational Lenses (Liebes 1964)\\
Curtis Callan & PhD & Sphericaly Symmetric Cosmological models (1964)\\
Lawrence Jordan & PhD & The velocities of 4 BeV/c pions and 8 BeV/c pions, \\
&& \qquad  kaons, and protons (1964)\\
Jason Morgan & PhD & An Astronomical and Geophysical Search for \\
&&  \qquad Scalar Gravitational Waves (1964)\\
William Hildreth & PhD & The interaction of scalar gravitational waves \\
&&  \qquad with the Schwarzschild Metric (1964)\\
Peter Roll & post-PhD & The equivalence of inertial and passive gravitational\\  
&&  \qquad  mass (Roll, Krotkov, Dicke 1964)\\
Rainer Weiss & post-PhD & A  Gravimeter  to  Monitor  the  $_0{\rm S}_0$ Dilational\\
&&   \qquad  Mode  of  the  Earth (Weiss and Block 1965) \\
David Curott & PhD & A Pendulum Gravimeter for Precision Detection of \\
&&  \qquad Scalar Gravitational Radiation (1965)\\
David Wilkinson & post-PhD & Cosmic Black-Body Radiation \\
&&  \qquad (Dicke, Peebles, Roll, and Wilkinson 1965)\\
Robert Moore & PhD & Study of Low Frequency Earth Noise and New Upper Limit\\
&&  \qquad   to the Intensity of Scalar Gravitational Waves (1966) \\
Lloyd Kreuzer & PhD & The Equivalence of Active and Gravitational Mass (1966)\\
 Barry Block & post-PhD & Measurements in Earth mode frequency, electrostatic \\
&&  \qquad  sensing \& feedback gravimeter (Block and Moore 1966) \\
Mark Goldenberg & post-PhD & Solar Oblateness and General Relativity\\
&&  \qquad (Dicke and Goldenberg 1967)\\
Carl Zanoni & PhD & Development of Daytime Astrometry to Measure \\
&&  \qquad the Gravitational Deflection of Light (1967)\\
Dennis Heygi & PhD & The Primordial Helium Abundance as Determined\\
&&  \qquad  from the Binary Star System $\mu$\,Cassiopeiae (1968) \\
B. Edward McDonald & PhD & Meridian Circulation in Rotating Stars (1970)\\
Lawrence Cathles & PhD & The viscosity of the Earth's mantle (1971) \\
William Wickes & PhD & Primordial Helium Abundance and Population-II \\
&&  \qquad Binary Stars: Measurement Technique (1972)\\
Jeffrey Kuhn & PhD & Global scale photospheric velocity fields: \\
&&  \qquad Probes of the solar interior (1980) \\
Ken Libbrecht & PhD & The shape of the Sun (1984)\\
\noalign{\smallskip} 
\bottomrule
\end{tabular}
\end{table}

\end{document}